\documentclass[pra,superscriptaddress,twocolumn,nobibnotes]{revtex4-2}
\usepackage{amsmath}
\usepackage{amsfonts}
\usepackage{braket}
\usepackage{graphicx}
\usepackage{color}
\usepackage{float}

\DeclareMathOperator{\Tr}{Tr}

\begin{document}
	
\title{Proposal for distribution of multi-photon entanglement\\ with optimal rate-distance scaling}

\author{Monika E. Mycroft}
\affiliation{Faculty of Physics, University of Warsaw, ul.~Pasteura 5, 02-093 Warsaw, Poland}

\author{Thomas McDermott}
\affiliation{levelQuantum s.r.l., Piazzale Luigi Cadorna 4, 20123 Milan, Italy}

\author{Adam Buraczewski}
\affiliation{levelQuantum s.r.l., Piazzale Luigi Cadorna 4, 20123 Milan, Italy}

\author{Magdalena Stobi\'nska}
\email{magdalena.stobinska@gmail.com}
\thanks{Corresponding author}
\affiliation{Faculty of Mathematics, Informatics and Mechanics, University of Warsaw, ul.~Banacha 2, 02-097 Warsaw, Poland}
\affiliation{levelQuantum s.r.l., Piazzale Luigi Cadorna 4, 20123 Milan, Italy}

\date{\today}

\begin{abstract}
We propose a protocol to perform long-range distribution of near-maximally entangled multiphoton states, allowing versatile applications such as quantum key distribution (QKD) and quantum metrology which can provide alternatives to state-of-the-art protocols. Our scheme uses resources available within the current integrated quantum photonic technology: squeezed vacuum states and photon-number-resolving detectors. The distributed entanglement can be certified by Bell tests which have the potential to be loophole free, and may be directly used in well established QKD protocols. Generally, this provides measurement-device-independent (MDI) levels of security, which may be upgraded to fully device-independent (DI) security if the Bell test is loophole-free. In both cases, the protocol is robust to extremely high transmission losses, matching the optimal $O(\sqrt{\eta})$ scaling of key rate with channel transmittance.
\end{abstract}

\maketitle

\section{Introduction}

\begin{figure*}[t]\centering
	\includegraphics[width=15cm]{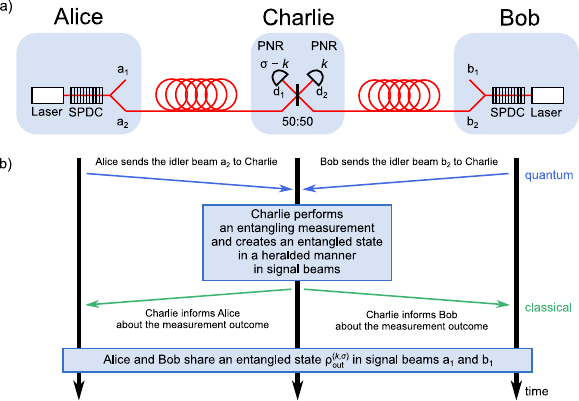}
	\caption{(Color online) Distribution of multiphoton bipartite entanglement. An explanation of the protocol as shown by 
		a)~a schematic of the setup and 
		b)~a sequence diagram.
		Alice and Bob locally generate a pulsed, photon-number entangled two-mode squeezed vacuum state each, obtained by spontaneous parametric down-conversion (SPDC) -- the most ubiquitous source of quantum light. They send their idler modes $a_2$ and $b_2$ to a remote station Charlie, who interferes them on a $50:50$ beam splitter (BS) and performs a photon-number-resolving (PNR) detection on the BS output modes $d_1$ and $d_2$. This is an entangling measurement by which Charlie creates the heralded mixed state $\rho^{(k,\sigma)}_{\text{out}}$ (or, in the lossless case, the pure state $\ket{\Psi_{\text{out}}^{(k,S=\sigma)}}$) in Alice and Bob's signal modes $a_1$ and $b_1$. The state and the amount of shared entanglement is parameterized by the outcomes of Charlie's measurement, $k$ and $\sigma-k$. In general, this entanglement is multiphoton, with $S \geq \sigma$ ($S = \sigma$ for a pure state) photons in total distributed between Alice and Bob. After Charlie informs the parties classically of the measurement outcome, Alice and Bob know which state $\rho^{(k,\sigma)}_{\text{out}}$ they possess and may employ it for quantum applications.}
	\label{fig:setup}
\end{figure*}

Distribution of photonic entanglement is key to building quantum networks which facilitate secure long-distance quantum communications~\cite{Gisin2007}, distributed quantum computation~\cite{Cirac1999} and sensing~\cite{Eldredge2018, Zhao_2021}, leading finally to a full realization of the quantum internet~\cite{Wehner_2018}. A fundamental problem is that entanglement becomes easily corrupted by losses in the transmitting channels, which results in low transmission rates of entangled states. Since amplification of quantum signals is impossible~\cite{Wootters1982}, alternative remedies are in high demand. A popular option is to use quantum repeaters~\cite{Briegel_PRL_1998, Rozpedek2019}. However, this approach necessitates numerous intermediate stations, quantum memories and multiple two-photon Bell pairs, a resource that is often created non-deterministically. Despite rapid advances in the field~\cite{Munro_NatPhot_2012, Muralidharan_PRL_2014, Azuma_NatComm_2015, Humphreys_2018, Zwerger_PRL_2018, Hasegawa_NatComm_2019, Li_NatPhot_2019, Yu_Nature_2020, Pu_NatPhot_2021, Wallnofer2021}, experimental realizations of quantum repeaters currently show deterioration with distance which is insufficient for most quantum communications applications~\cite{Khatri_npj_2021}. Another possibility is to use satellites to distribute two-photon Bell pairs, which is advantageous thanks to photons traveling the majority of the distance in free space and thus experiencing less loss than in atmospheric links or optical fibers, while connecting arbitrarily distant locations. This approach has seen much progress recently~\cite{Aspelmeyer_2003, Tang_PRA_2016, Bedington_EPJ_2016, Takenaka_NatPhot_2017, Ren_Nature_2017, Oi_EPJ_2017, Yin_Science_2017, Kerstel_EPJ_2018, Dai_NatPhys_2020, Yin_Nature_2020, Cao_PRL_2020, Khatri_npj_2021, Chen_Nature_2021, Pirandola_review_2021, Chen2021a, Zhang2022}, although the losses are still detrimental, with typically only approximately 1~in $10^6$ of produced Bell pairs surviving the trip to the detection stations on the ground~\cite{Yin_Science_2017}. Thus the distribution of entanglement in a way that is resource-efficient, verifiable and well suited to the existing quantum-photonic technologies continues to be an open problem.

Quantum key distribution (QKD) uses quantum correlations to distribute cryptographic keys between remote parties which enables secure communication. In the long distance regime, the current state-of-the-art is twin-field (TF)-QKD~\cite{Lucamarini2018}, which displays a ground breaking improvement over previous methods, exemplified by the key-rate scaling $O(\sqrt{\eta})$ with channel transmittance $\eta$, allowing efficient demonstrations of QKD over 600\kern.25emkm~\cite{Chen2020, Pittaluga2021, Liu2021}. There, weak coherent pulses are phase-encoded and sent to an untrusted node for measurement. This is an example of measurement-device-independent (MDI)-QKD~\cite{Lo2012}, which closes all detector side channels but remains susceptible to eavesdropping attacks at the source. For example, photon-number-splitting attacks~\cite{Brassard2000} which must be carefully avoided via decoy-state methods~\cite{Hwang2003}. Entanglement-based versions, which benefit from the same key-rate scaling, were soon theorized~\cite{Kamaruddin2015, Li2019}. Generally, entanglement-based approaches make security easier to verify and have the potential for loophole-free Bell tests which could close all possible side channels. This is called device-independent (DI)-QKD~\cite{Pironio2009,Zhang2022}. It is clear that for entanglement distribution protocols to remain competitive for QKD at long distances, they must achieve the same optimal $O(\sqrt{\eta})$ key-rate scaling.

Here, we propose a protocol that uses multiphoton bipartite entanglement and photon-number-resolving (PNR) detectors to establish long-distance near-maximal entanglement distribution, allowing entanglement-based QKD protocols that match the $O(\sqrt{\eta})$ scaling. It is robust to high transmission losses which, remarkably, deteriorate only its efficiency, but not the amount of generated entanglement, a feature until now seen only in protocols based on Bell pairs. As a result, the protocol has the potential to outperform other methods of entanglement distribution which have transmission rates scaling as $O(\eta)$. This is particularly true in high loss scenarios, for example satellite-based Earth-space channels and long fiber networks. Once distributed, the multiphoton entanglement of the states may be verified by a CHSH Bell inequality~\cite{CHSH1969}, forming a basis for establishing a cryptographic key. Depending on the detection efficiencies available, this can be achieved either by DI-QKD~\cite{Kamaruddin2015}, or entanglement-based phase-matching QKD~\cite{Li2019}. The higher photon number states generated are additionally useful for quantum metrology which has been extensively covered in our previous work~\cite{Thekkadath_npj_2020}.

Bipartite entanglement between two physical systems such as, e.g.\ two photons, can be revealed by their Bell nonlocality, i.e.\ the fact that outcomes of local measurements on these two subsystems exhibit correlations which cannot be described by local hidden variable models~\cite{Horodecki2009}. In this paper we focus on multiphoton bipartite entanglement, engineered from two-mode squeezed-vacuum (SV) states, which are used as our primary resource. This Gaussian quantum state is naturally and deterministically produced  from a pulsed coherent light via the spontaneous parametric down-conversion process (SPDC), the most ubiquitous source of quantum light~\cite{Pan2012}. It carries entanglement between the quadratures of the electric field (continuous variables, CVs) but also between photon numbers in the two modes (discrete variables, DVs), which are unbounded~\cite{Chekhova2015}. The latest advances in experimental integrated quantum photonics facilitate precise generation and manipulation of near-single-mode SV states in optical chips~\cite{Lenzini2019}, as well as detection of their photon number statistics using state-of-the-art PNR detectors~\cite{Gerrits2011}. The most successful PNR detectors are transition edge sensors (TES) which have negligible dark count rates and quantum efficiencies which can exceed $95\%$~\cite{Lita2008, Fukuda_2011}. Integrated photonic devices have also contributed to the development of practical quantum communications scenarios~\cite{integrated_QKD, integrated_qcomm}.

In our protocol, a verifiable DV entanglement of high local dimension is created from two sources of an SV state using an entangling measurement at a remote station. The measurement is realized by multiphoton quantum interference on a beam splitter (BS) followed by PNR detectors. The protocol can also be carried out in a delayed-choice scheme which frees the parties from using quantum memories and allows them to share near-maximally entangled states in realistic implementations. These states have been proven to be useful not only for QKD, as specified above, but also find an immediate application in quantum metrology. These are multiphoton generalized Holland--Burnett (HB) states, which have been experimentally proven to allow near-optimal quantum-enhanced optical phase estimation in a lossy environment~\cite{Thekkadath_npj_2020}.

\section{Results}

\begin{figure*}[t]\centering
	\includegraphics[width=13cm]{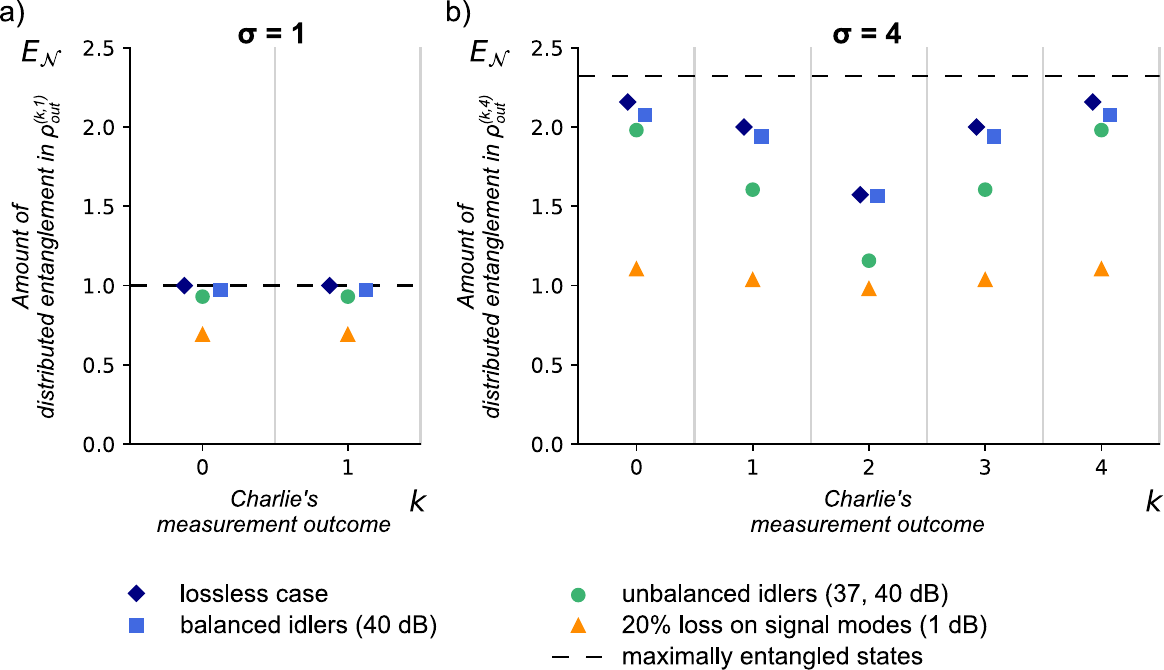}
	\caption{(Color online)  The amount of distributed entanglement in the shared state $\rho^{(k, \sigma)}_{\mathrm{out}}$. 
		The entanglement is quantified by the value of the logarithmic negativity $E_{\mathcal{N}}$ and is calculated for $g=0.1$ and a) $\sigma = 1$ and b) $\sigma = 4$ photons detected in total by Charlie. The value of $E_{\mathcal{N}} = \log_2(\sigma + 1)$ for the maximally entangled state possible with the same number of photons is shown by the dashed black line. The dark blue diamonds depict the lossless case, Eq.~(\ref{EN_ideal}). At $k=\sigma / 2$, as a result of the multiphoton Hong--Ou--Mandel (HOM) effect, all odd photon-number components disappear, halving the basis and reducing the maximal entanglement to $\log_2(\sigma/2 + 1)$. This is reflected by the dip at $k=2$. The blue squares represent the numerically computed values for a system where 40\kern.25emdB losses were experienced by the photons in the idler modes $a_2$ and $b_2$. It is remarkable that these values match almost exactly the ideal case. This is one of the two main results of this work. For comparison, the results for two other scenarios of losses in the system are also shown. The green circles indicate the situation where the photons in the idler modes $a_2$ and $b_2$ experience different losses: 37\kern.25emdB in one mode and 40\kern.25emdB in the other (50\% higher attenuation in one arm). The orange triangles illustrate the situation where a small amount of losses (20\%) are present in the signal modes $a_1$ and $b_1$ only. Losses in signal modes are detrimental to the amount of distributed entanglement but can be overcome using the delayed-choice scheme. All non-ideal cases were computed for 50\% detection efficiency at the PNR detectors. Vertical grid lines separate the photon-number bins.}
	\label{fig:logneg}
\end{figure*}

\subsection{Resources}

We employ two copies of an SV state $\ket{\Psi}$ coming from a pulsed source as our input, $\ket{\Psi_{\text{in}}}=\ket{\Psi}_a\otimes\ket{\Psi}_b$. In its Schmidt basis, $\ket{\Psi}$ takes the form of a superposition of $n$-photon pairs with real-valued probability amplitudes $\sqrt{\lambda_{n}} = \frac{\tanh^{n}g}{\cosh g}$
\begin{equation} 
\ket{\Psi} = \sum_{n=0}^{\infty}\sqrt{\lambda_{n}}\ket{n}_1 \ket{n}_2,
\label{SV}
\end{equation}
where $g$ is the parametric gain, which is the key parameter characterizing an SPDC source, setting the mean photon number in $\ket{\Psi}$ to $2\sinh^2g$ for each pulse. Quantum correlations in $\ket{\Psi}$ are manifested by equal photon numbers in modes 1 and 2, called the signal and idler, which can be spatially resolved. The subsequent photon number contributions to the SV state: the vacuum, single-photon and higher-order ($n>1$) emissions, occur with a probability which follows a geometric progression with common ratio $\tfrac{\lambda_{n+1}}{\lambda_n} = \tanh^2g$. Thus, the SPDC generates a considerable amount of multiphoton events which are utilized to our benefit in our protocol.

A schematic explanation of the proposed entanglement distribution protocol is shown in Fig.~\ref{fig:setup}. The setup consists of two pulsed SPDC sources, one held by Alice and one by Bob, each generating $\ket{\Psi}$. The idler beams emitted into modes $a_2$ and $b_2$ are sent to Charlie at a remote station, who interferes them on a balanced ($50:50$) BS and then detects them with PNR detectors. This is an entangling measurement, whereby the signal modes $a_1$ and $b_1$ become photon-number correlated.

\subsection{Protocol in ideal circumstances} 

Let us consider the lossless case first. The detection of $\sigma$ photons in total in the entangling measurement means that $\sigma$ photons were distributed between the two idler beams entering the BS and thus that the total number of photons, $S$, in the output state shared by Alice and Bob is $S = \sigma$. The BS performs a linear operation on the input idler creation operators, $U_{\text{BS}}^{\dagger}\, a_2^{\dagger}\, U_{\text{BS}} =  \tfrac{1}{\sqrt{2}} \, (a_2^\dagger - i b_2^\dagger)$, $U_{\text{BS}}^{\dagger}\, b_2^{\dagger}\, U_{\text{BS}} =  \tfrac{1}{\sqrt{2}} \, (-i a_2^\dagger + b_2^\dagger)$, while the signal modes $a_1$ and $b_1$ are intact. Applying this operation to Fock states
\begin{equation}
\ket{n}_{a_2} = \frac{(a_2^{\dagger})^{n}}{\sqrt{n!}}\ket{0},\qquad
\ket{\sigma\!-\!n}_{b_2} = \frac{(b_2^{\dagger})^{\sigma-n}}{\sqrt{(\sigma-n)!}}\ket{0},
\end{equation}
requires taking powers of the transformed operators. This results in a transformation on $a_2^{\dagger}$ and $b_2^{\dagger}$ governed by a binomial distribution and an output state described by an arcsine probability distribution. Through such a multiphoton Hong--Ou--Mandel (HOM) effect, entanglement between the BS output modes is generated~\cite{Kim2002}. The  probability amplitudes of detecting $k$ and $\sigma\!-\!k$ photons behind the BS are equal to~\cite{Magda-Kravchuk}
\begin{equation}
\begin{aligned}
\mathcal{A}_{\sigma}(k,n) ={}& \langle k, \sigma\!-\!k \vert U_{\text{BS}} \vert n, \sigma\!-\!n \rangle \nonumber \\
{}={}& i^{k-n}\,
\phi_k(n-\tfrac{\sigma}{2}, \sigma),
\end{aligned}
\end{equation}
where $k=0, \dots, \sigma$ and $\phi_k$ are symmetric Kravchuk functions -- orthonormal discrete polynomials which converge to Hermite--Gauss polynomials for large $\sigma$~\cite{Atakishiyev_1997}. The output state in our protocol is therefore
\begin{equation}
\begin{aligned}[c] \ket{\Psi_{\text{out}}^{(k,\sigma)}} ={}& \mathcal{N} \bra{k,\sigma\!-\!k}U_{\text{BS}}\ket{\Psi}^{\otimes 2}\\
{}={}&\mathcal{N}  \sum_{n,m=0}^{\infty}\sqrt{\lambda_{n}\lambda_{m}}\ket{n,m}_{a_1,b_1}\times{}\\
&\qquad\qquad{}\times \langle k, \sigma\!-\!k \vert U_{\text{BS}}  \ket{n,m}_{a_2,b_2},
\end{aligned}
\end{equation}
where $\mathcal{N}$ is the normalization and $\ket{\Psi}$ is defined in Eq.~(\ref{SV}). Since $n+m=\sigma$ must hold true,
\begin{equation}
\ket{\Psi_{\text{out}}^{(k,\sigma)}} = \sum_{n=0}^{\sigma} \mathcal{A}_{\sigma}(k,n)\ket{n,\sigma-n}_{a_1,b_1}.
\end{equation}
We now see that $\ket{\Psi_{\text{out}}^{(k,\sigma)}}$ lives in a Hilbert space of dimension $(\sigma\!+\!1)$ and is parameterized by the entangling measurement readouts $k$ and $\sigma\!-\!k$ which define its photon number statistics. The probability of detecting $n$ photons in mode $a_1$ (and $\sigma-n$ in mode $b_1$) is $\lvert\mathcal{A}_{\sigma}(k,n)\rvert^2=\lvert\phi_k(n-\frac{\sigma}{2}, \sigma)\rvert^2$. Example photon number distributions, and details of the derivation, are given in Figs.~\ref{fig:As24}-\ref{fig:As210} and in Appendix~\ref{app:snote1}.

This protocol can also be realized in a delayed-choice scheme. Then, the measurements taken by Alice and Bob on the signal modes $a_1$ and $b_1$ precede the ones performed on the idler modes $a_2$ and $b_2$ at the remote station. By the no-signaling principle, the photon-number statistics observed by Alice and Bob is again determined by $\lvert\mathcal{A}_{\sigma}(k,n)\rvert^2$.

To quantify entanglement in $\ket{\Psi_{\text{out}}^{(k,\sigma)}}$ we employ the logarithmic negativity $E_{\mathcal{N}} = \log_2 \lvert \lvert\rho^{\Gamma}\rvert \rvert_1$, where $\rho$ denotes a density operator, $\Gamma$ is the partial transpose operation and $\lvert \lvert \cdot \rvert \rvert_1$ is the trace norm. This entanglement measure is easily computable and gives an upper bound to the distillable entanglement, reflecting the maximal number of Bell states that can be extracted~\cite{Vidal_PRA_2002, Plenio_PRL_2005}. Inserting $\rho = \ket{\Psi_{\text{out}}^{(k,\sigma)}}\bra{\Psi_{\text{out}}^{(k,\sigma)}}$ we obtain  
\begin{align}
&E_{\mathcal{N}} \big(\ket{\Psi_{\text{out}}^{(k,\sigma)}}\bra{\Psi_{\text{out}}^{(k,\sigma)}}\big)={}
\nonumber\\&\qquad{}=2\log_{2} \Bigg\{\sum_{n = 0}^{\sigma} \Big| \phi_k (n-\tfrac{\sigma}{2}, \sigma) \Big| \Bigg\},
\label{EN_ideal}
\end{align}
see Appendix~\ref{app:snote2} for the full derivation.
Since the readouts $k$ and $\sigma\!-\!k$ uniquely define the state $\ket{\Psi_{\text{out}}^{(k,\sigma)}}$, they also determine the amount of entanglement in it. The maximum amount of entanglement which can be created in Hilbert space of dimension $(\sigma+1)$ is $E_{\mathcal{N}_\text{max}} = \log_2(\sigma+1)$. Our protocol allows one to achieve values close to this maximum, as shown in the lossless case for $\sigma = 4$ in Fig.~\ref{fig:logneg}, with the maximal value $E_{\mathcal{N}_\text{max}} \approx 2.3$. The decrease in entanglement for $k=\frac{\sigma}{2}$ results from the multiphoton HOM effect, whereupon all odd photon-number components disappear. This leads to an effective reduction in the Hilbert space dimension to $(\frac{\sigma}{2} + 1)$ where the maximal entanglement is $\log_2(\frac{\sigma}{2}+1)$, or $\approx 1.6$ at $\sigma = 4$. For comparison, Bell states have logarithmic negativity, $E_{\mathcal{N}_\text{Bell}}=1$. This is the same as we obtain for $\sigma = 1$.

\subsection{Operation in lossy conditions}

Depending on the application of our protocol, the amount of losses in the photon transmission may vary, ranging from 0.2\kern.25emdB/km for optical fibers to approximately 40\kern.25emdB for an uplink ground-to-satellite channel~\cite{Jennewein_NJP_2013}. In addition, the efficiency of a PNR detection system may drop to 50--60\% (by 2--3\kern.25emdB) if one includes uncorrelated background counts and inefficient coupling~\cite{Thekkadath_npj_2020}.

\begin{figure*}[t]\centering
	\includegraphics[width=13cm]{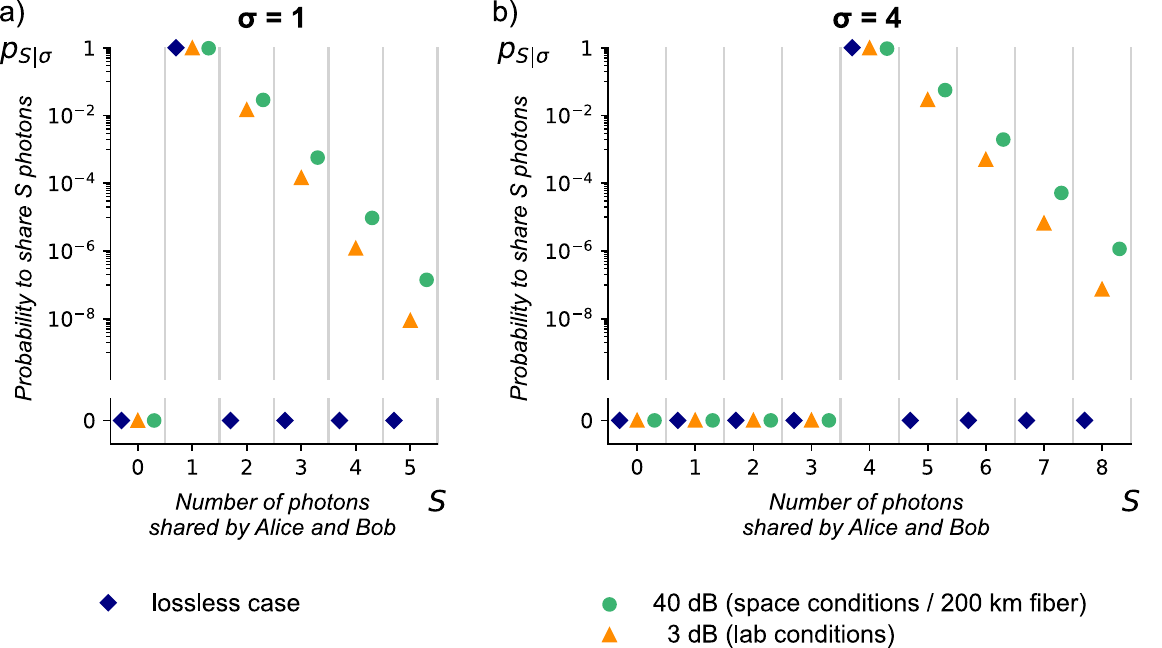}
	\caption{(Color online) Multiphoton character of the state $\rho^{(k, \sigma)}_{\mathrm{out}}$ shared by Alice and Bob. The probability, $p_{S|\sigma}$, that Alice and Bob share $S$ total photons given a)~$\sigma = 1$ and b)~$\sigma = 4$ photons were measured by Charlie, plotted for a realistic value of parametric gain, $g = 0.1$. The blue diamonds depict the case of lossless idler modes ($r=0$), here $S$ is always equal to $\sigma$ because the signal and idler modes have equal photon numbers and Charlie's entangling measurement is photon number preserving. The orange triangles depict idler losses of 3\kern.25emdB ($r=0.5$), as might be present in lab conditions, while green circles depict 40\kern.25emdB ($r=0.9999$), as in an Earth-space channel or 200\kern.25emkm of optical fiber. Idler losses introduce higher-order contributions $S > \sigma$ but the primary component and lower-bound is still $S = \sigma$ since, in the absence of dark counts and background (stray) photons, Charlie cannot measure more photons than were sent to him in Alice and Bob's idler modes. Losses in the signal modes may introduce components $S < \sigma$ but these can be significantly reduced by using a delayed-choice scheme where Alice and Bob's measurements precede that of Charlie. Vertical grid lines separate the photon-number bins.}
	\label{fig:chi}
\end{figure*}

Losses are modeled by inserting additional beam splitters into the pathways of the photons, whose reflectivity $r$ quantifies the amount of loss. In the case of symmetric idler mode losses, the state shared by Alice and Bob is no longer the pure state $\ket{\Psi_{\text{out}}^{(k,\sigma)}}$, but a mixed state described by the density operator  
\begin{equation} 
\rho^{(k,\sigma)}_{\text{out}}=\sum_{S=\sigma}^{\infty}p_{S|\sigma}\,\rho^{(\sigma,k,S)}_{\text{int}},
\end{equation}
where $\rho^{(\sigma,k,S)}_{\text{int}}$ is a density operator of a state with $S$ photons in total in the signal modes which are assumed lossless. The full representation can be found in Appendix~\ref{app:disc1}. Unlike in the lossless case, $S$ is not always equal to $\sigma$, and instead the probability for Alice and Bob to detect $S$ photons between them is given by $p_{S|\sigma} = \tilde{\mathcal{N}}^2 \chi_{\sigma,S}$ where $\chi_{\sigma,S} = r^{S-\sigma} \lambda_S\,\binom{S+1}{\sigma+1}$ and $\tilde{\mathcal{N}}$ is a normalization constant independent of $S$. As shown in Fig.~\ref{fig:chi}, in the limit $r \tanh^2(g) \ll 1$, the most likely event is $S=\sigma$, with additional components $S > \sigma$ becoming increasingly less relevant as $\chi_{\sigma,S}$ drops rapidly towards zero. The primary component of the mixed state is then $\rho^{(\sigma,k,\sigma)}_{\text{int}}$, which is identical to the lossless state $\ket{\Psi_{\text{out}}^{(k,\sigma)}}\bra{\Psi_{\text{out}}^{(k,\sigma)}}$. Provided that the gain $g$ is at typical low levels, $g^2 \ll1$, the limit $r \tanh^2(g) \ll 1$ holds even as $r \to 1$, so that the entanglement of the final states remains high even in the presence of arbitrarily high losses in the idler modes.

To understand how realistic conditions affect the distributed entanglement we consider different scenarios of losses. Fig.~\ref{fig:logneg} shows numerically calculated logarithmic negativity at $g = 0.1$ in the ideal case, the case of large symmetric losses in the idler beams ($r_{a_2, b_2}\!=\!99.99\%$, i.e., 40\kern.25emdB), the case of unbalanced losses in the idler beams with relative transmittance $\epsilon = \frac{1-r_{b_2}}{1-r_{a_2}}=0.5$ ($r_{a_2}\!=\!99.98\%$, $r_{b_2}\!=\!99.99\%$, i.e., 37, 40\kern.25emdB, respectively) and the case of non-ideal signal modes with 20\% losses in those arms. Losses in dB are given by $-10 \log_{10}(1-r)$. We do not assume perfect entangling measurements, instead setting realistic losses at the detectors ($r_{d_1,d_2}\!=\!50\%$), however, we found that their influence is negligible since they can be viewed as a minor addition to the already large idler losses. For more examples of scenarios involving losses see Figs.~\ref{fig:log_neg_loss}-\ref{fig:log_neg_signal} and \ref{fig:unsymlogneganalysis}-\ref{fig:log_neg_all}, and Appendices~\ref{app:disc1}-\ref{app:disc3}. As can be seen in Fig.~\ref{fig:logneg}, the amount of entanglement obtained in the case of large symmetric losses is almost exactly the same as in the ideal case. In fact, the entangling measurement can be viewed as a $g^{(2)}$ correlation function measurement, which is known to be immune to losses~\cite{Stevens_OpticsExpress_2014}.

As for the asymmetrical transmission losses, the logarithmic negativity stays within 90\% of the lossless values provided that the relative transmittance of the channels $\epsilon$ stays above 0.4, see Fig.~\ref{fig:unsymscalinganalysis} and Appendix~\ref{app:disc2} for an explanation of this effect. One should note that relative transmittance $\epsilon < 0.4$ is unlikely to occur in reality~\cite{Yin_Nature_2020, Jennewein_NJP_2013} and artificial losses may always be introduced to balance the channels. Another consideration is that losses for each channel will vary with time~\cite{Jennewein_NJP_2013}, especially with atmospheric losses when applied to a satellite-based scheme. However, losses fluctuate on time-scales an order of magnitude higher than the time of flight of photons between the ground and satellite, therefore it suffices to take the average of Alice and Bob's output logarithmic negativity. The effect of fluctuations on the average is not detrimental, as shown in Fig.~\ref{fig:simulation_time}.

In contrast to idler mode losses, detection inefficiencies or losses in signal modes $a_1$ and $b_1$ are critical since they spoil generated entanglement, see Fig.~\ref{fig:log_neg_signal} for more examples of how signal losses affect the entanglement. However, these losses may be significantly reduced by using a delayed-choice scheme. In such a scheme, Alice and Bob would perform their measurements immediately before waiting for the detection by Charlie at the remote station. By the no-signaling principle, all measurement outcomes are independent of this modification. If Alice, Bob and Charlie's measurements have space-like separation from each other, this can allow Bell tests which simultaneously close the locality and detection loopholes, as in ``event-ready" Bell tests~\cite{Hensen2015}. In our case the event-ready signal is Charlie's successful measurement. The delayed-choice scheme also mitigates the effect of phase-damping. After Alice and Bob perform their near-immediate measurement of the signal beams, from their reference frame, the idlers are projected onto Fock states which are unaffected by phase-damping. Remaining synchronization errors and phase-damping may be further reduced by using two reference pulses, one to measure the delay in the arrival times of the two idler beams, and one to measure the rate of phase-shifting. These errors can then be corrected in postprocessing~\cite{Chen2020, Pittaluga2021, Liu2021}.

\subsection{Protocol performance}

The probability to generate the state $\rho^{(k,\sigma)}_{\text{out}}$ is equal to the probability for Charlie to obtain the measurement outcomes $k$ and $\sigma-k$ photons after the BS
\begin{equation}
p_{(k, \sigma)} = \frac{\lambda_\sigma}{\cosh^2 g} \cdot \frac{(1-r)^\sigma}{(1-r\tanh^2 g)^{\sigma+2}},
\label{efficiencyfull}
\end{equation}
which is independent of $k$, see Appendix~\ref{app:disc4} for the calculation. This probability is useful to calculate the efficiency of our protocol, which we define as the probability of Charlie receiving at least one photon. When that happens, we are certain that Alice and Bob share entangled states, which may be multiphoton. The protocol's efficiency, $1-p_{(0, 0)} \approx 2(1-r)\lambda_1/\cosh^2(g)$, increases with the parametric gain $g$ and decreases with idler losses $r$. Investigating the dependence on transmission losses more closely, we see that it is proportional to the transmittance $t = 1-r$ between Alice/Bob and Charlie. It is customary to instead use the transmittance $\eta = t^2$ of the summed distance from Alice to Bob~\cite{Lucamarini2018}, assuming that Charlie is approximately at the mid-point between them. In this context, it is clear that our protocol's efficiency scales as $O(\sqrt{\eta})$. This should be contrasted to schemes where Charlie directly distributes Bell pairs to Alice and Bob~\cite{Yin_Science_2017, Yin_Nature_2020}, which have efficiencies $O(\eta)$. The reason for the quadratic improvement is that only a single photon has to survive the trip between Alice/Bob and Charlie, since an entangled state is established even with $\sigma=1$. In contrast, distributing a Bell pair requires two photons to survive the trip, therefore losses are more significant.

\begin{figure*}
	\centering
	\includegraphics[width=10cm]{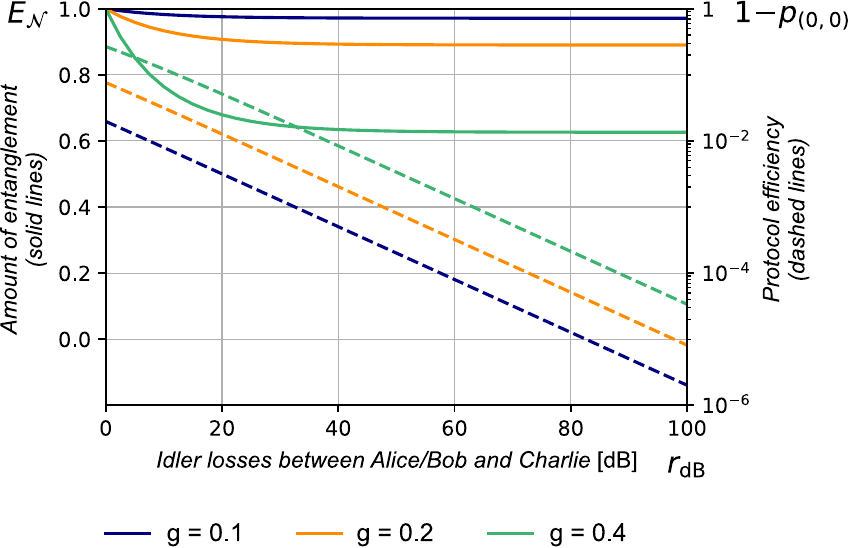}
	\caption{(Color online) Protocol performance. The performance of the entanglement distribution is quantified by the amount of generated entanglement, as measured by the logarithmic negativity $E_{\mathcal{N}}$ (solid lines) and the protocol efficiency, $1-p_{(0, 0)}$ (dashed lines), plotted here for the example state $\rho^{(0, 1)}_{\mathrm{out}}$. The idler losses between Alice/Bob and Charlie are expressed in dB, $r_\textrm{dB} = -10\log_{10}(1-r)$. The protocol efficiency decreases gradually with idler losses, having the favorable scaling $O(\sqrt{\eta})$ where $\eta = (1-r)^2$ is the effective transmittance between Alice and Bob. Charlie is assumed to be approximately at the mid-point between them. The amount of generated entanglement also decreases with idler losses, but quickly saturates to a finite value. Provided that the parametric gain $g$ is at realistically low levels ($g^2 \ll 1$), this decrease is negligible and the protocol is robust to extremely high losses (blue lines, $g=0.1$). Increasing the parametric gain (orange lines, $g=0.2$ and green lines, $g=0.4$) leads to an increase of protocol efficiency at the cost of reduced entanglement of the generated states.}
	\label{fig:success_logneg}
\end{figure*}

Fig.~\ref{fig:success_logneg} depicts the protocol efficiency and generated entanglement as a function of the idler losses for three values of the parametric gain: $g=\{0.1, 0.2, 0.4\}$. Provided that $g^2 \ll 1$, as can be seen for the case $g=0.1$, the idler losses influence only the efficiency of the protocol without changing its principle of operation or the generated entanglement. This feature is characteristic to protocols based on two-photon Bell pair entanglement, and now we have shown it for all multiphoton schemes based on our setup. If $g$ is increased, the efficiency of the protocol is improved but at the cost of reducing the generated entanglement, as can be seen for $g=0.2$ and $g=0.4$. 

\subsection{Applications}

\subsubsection{Quantum key distribution}

Depending on the detection efficiencies available, two different Bell tests are possible for Alice and Bob to verify the amount of entanglement in their shared state, with each test being associated with a well-established QKD protocol: single-photon entanglement-based DI-QKD~\cite{Banaszek2009, Kamaruddin2015}, and single-photon entanglement-based phase-matching (SEPM)-QKD~\cite{Li2019}. Both of these protocols display the optimal $O(\sqrt{\eta})$ scaling of key rate with channel transmittance $\eta$.  They rely on Alice and Bob interfering their signal modes with coherent states $\ket{\alpha}$ and $\ket{\beta}$ on local beam splitters with reflectivities $r_a$ and $r_b$ respectively, and then measuring the output photon numbers.

The first test is specifically suited to cases where Alice and Bob's detector efficiencies and signal mode transmittances are high enough for the test to be performed without any postselection. Alice and Bob interfere their signal modes with coherent states to perform the displacement operations $D(-\delta_\alpha) =  \exp(\delta_\alpha^* a_1 -\delta_\alpha a_1^\dagger)$ and $D(-\delta_\beta) = \exp(\delta_\beta^* b_1 -\delta_\beta b_1^\dagger)$, where $\delta_\alpha = i \alpha \sqrt{r_a}$ and $\delta_\beta = i \beta \sqrt{r_b}$, and finally measure their transmitted photon numbers with PNR detectors. They evaluate a CHSH inequality by assigning ${+}1/{-}1$ either to vacuum/non-vacuum events, or to even/odd photon numbers. By choosing randomly between two local measurement settings each, $\delta_{\alpha 1}, \delta_{\alpha 2}$ and $\delta_{\beta 1}, \delta_{\beta 2}$, they can then observe violation of the CHSH inequality $|B| \leq 2$ where $B = E(\delta_{\alpha 1}, \delta_{\beta 1}) + E(\delta_{\alpha 1}, \delta_{\beta 2}) + E(\delta_{\alpha 2}, \delta_{\beta 1}) - E(\delta_{\alpha 2}, \delta_{\beta 2})$, and $E(\delta_\alpha, \delta_\beta)$ is the correlation between their dichotomized variables. The greatest violation $|B| = 2.63$ is found for the $\sigma = 1$ states, even with large transmission losses in excess of 40\kern0.25emdB, where they comprise the vast majority (99.9999\%) of the generated states. For these states we found the optimal choice of measurement settings to be $\delta_{\alpha 1} = +0.17, \delta_{\alpha 2} = -0.56, \delta_{\beta 1} = \pm0.17i, \delta_{\beta 2}=\mp0.56i$, with the upper/lower signs corresponding to the $k=0/1$ states. This test remains effective $|B| > 2$ as long as Alice and Bob's local detector efficiencies, including signal mode transmittances, are above 85\%. It does not employ any postselection to filter out any vacuum or multiphoton events, closing the detection loophole and opening up the possibility of performing a loophole-free Bell test~\cite{Stobinska2011,Peres1997,ZeilingerTES}. Bell inequality violation is also observed for the states $\sigma \geq 2$, e.g., for $\sigma = 2$ one obtains $|B| =  2.35$ in the ideal case, but it is not as tolerant to losses. 

\begin{figure*}[t]\centering
	\includegraphics[width=10cm]{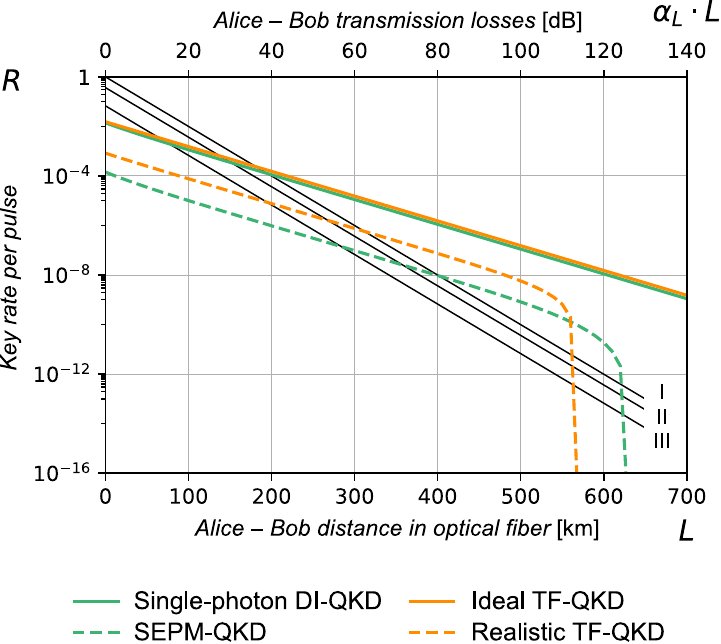}
	\caption{(Color online) Key rate comparison. The key rates of the two quantum key distribution (QKD) protocols performed with our distributed entanglement, single-photon DI-QKD~\cite{Kamaruddin2015} and single-photon entanglement-based phase-matching (SEPM)-QKD~\cite{Li2019}, are compared to the state-of-the-art twin-field (TF)-QKD scheme~\cite{Lucamarini2018}. All three of these protocols display the rate-distance scaling $O(\sqrt{\eta})$, showing improvement at long distances compared to protocols with scaling $O(\eta)$ (I -- ideal BB84, II -- decoy-state QKD and III -- decoy-state MDI-QKD~\cite{Lucamarini2018}). When high detection efficiencies are available, single-photon DI-QKD can be performed. In ideal conditions (100\% detection efficiency, 100\% error correction efficiency, no dark counts, no background or stray photons) the key rates (green solid line) match almost exactly the ideal TF-QKD rate (orange solid line). In realistic conditions (here 30\% detector efficiencies, error correction coefficient 1.15, false detection probability $10^{-8}$ per gate interval) SEPM-QKD may be performed (green dashed line), with similar performance as realistic TF-QKD (orange dashed line). This is the second main result of this work. While TES detectors have negligible dark count rate, the false detections could be due to background photons, leading to a maximum distance of SEPM-QKD when the false detection rate and protocol efficiency become comparable. The conversion between Alice-Bob losses in dB, $2\,r_\textrm{dB} = -10 \log_{10}(\eta) = \alpha_L L$, and distance in optical fiber, $L$, is given by the standard attenuation rate $\alpha_L = $ 0.2\kern.25emdB/km. The key rate per pulse, $R$, is the number of secure key bits generated per unit time, divided by the repetition rate of the source.}
	\label{fig:keyrate}
\end{figure*}

If a loophole-free Bell test can be established, DI-QKD schemes become possible which are immune to all possible side channels and eavesdropping strategies~\cite{Pironio2009}. The $\sigma = 1$ states are particularly well suited for this due to their tolerance to losses. These states are generally multiphoton (see Fig.~\ref{fig:chi}a) but have a dominant single-photon contribution $\frac{1}{\sqrt{2}}(\ket{0}_A\ket{1}_B\mp i\ket{1}_A\ket{0}_B)$. DI-QKD using these states can then be viewed as a form of single-photon DI-QKD~\cite{Kamaruddin2015}. Once a state has been successfully distributed to Alice and Bob, they perform the same procedure as described in the Bell test above but now Alice has an additional choice for her measurement setting, $\delta_{\alpha 0} = -0.17$. The pair $\{\delta_{\alpha 0}$, $\delta_{\beta 1}\}$ is chosen such that Alice and Bob's transmitted photon numbers are highly correlated for these measurements, and these results are kept secret and used to generate the raw key. Results for the combination of settings $\{\delta_{\alpha 1}, \delta_{\alpha 2}, \delta_{\beta 1}, \delta_{\beta 2}\}$ are communicated publicly and used to violate the CHSH inequality as described above, which bounds the information that is leaked to Eve~\cite{Pironio2009}. The key rate is proportional to the efficiency of our entanglement distribution protocol, which we have seen scales as $O(\sqrt{\eta})$. Thus single-photon DI-QKD based on our distributed entanglement can achieve the optimal scaling with channel transmittance. In fact, as shown in Fig.~\ref{fig:keyrate}, the key rates are almost identical to those obtained in TF-QKD if both protocols are run in ideal conditions, i.e., 100\% detector efficiencies, 100\% error correction efficiency, no dark counts and no background or stray photons. The key rate is calculated assuming collective attacks, following the general procedure outlined in \cite{Pironio2009}, although the security may be improved to protect against completely general/coherent attacks~\cite{Vazirani2014, Miller2016}. See Appendix~\ref{app:QKD} for details of the calculation and a sketch of the protocol in Fig.~\ref{fig:QKD}.

The higher photon-number states $\sigma \geq 2$ also violate Bell inequalities and thus can be used for DI-QKD. However, while the entanglement distribution protocol as a whole has an efficiency scaling as $O(\sqrt{\eta})$, from Eq.~(\ref{efficiencyfull}) the efficiency of extracting a specific state $\rho^{(k,\sigma)}_{\text{out}}$ scales as $O(\eta^{\sigma/2})$. Thus these higher states don't benefit from such scaling but could still be useful in shorter range networks with lower losses.

If high detection efficiencies and signal mode transmittances ($>85\%$) are not available, a different Bell test may be carried out using postselection and the fair-sampling assumption~\cite{Li2019}. Alice and Bob now set $r_a = r_b = \frac{1}{2}$, and assign a value ${+}1/{-}1$ when photons are detected in the transmitted/reflected channels, ignoring outcomes where neither or both detectors click. Near maximum violation of the CHSH inequality may be obtained if Alice chooses the phase of her coherent state, $\theta_a$, randomly from the set $\{0, \frac{\pi}{2}\}$, while Bob chooses his phase, $\theta_b$, randomly from $\{\frac{\pi}{4} - \theta, -\frac{\pi}{4} - \theta\}$, where $\theta = -\frac{\pi}{2}$ when $k=0$ and $\theta = \frac{\pi}{2}$ when $k=1$. The amplitudes of the coherent states are chosen to be equal $|\alpha|=|\beta|$. This Bell inequality is no longer loophole-free so it cannot provide DI levels of security, but it can nevertheless be used to perform SEPM-QKD, first described in \cite{Li2019}. If the phase-matching condition $\theta_a - \theta_b - \theta = \pm \pi$ is satisfied, their results are highly correlated and can be used to extract a shared key, while results for other phases are used to evaluate the CHSH inequality, placing a limit on the information leaked to an eavesdropper Eve. Since the fair-sampling assumption is used, Alice and Bob must now trust their local sources and measurement apparatus, but still need not trust the measurement at the remote station performed by Charlie/Eve, placing the security within the framework of measurement-device-independent (MDI)-QKD~\cite{Lo2012}. The protocol differs slightly in our case since the entanglement is now heralded by Charlie's measurement rather than being sent to Alice and Bob directly, but is otherwise identical in operation. The key rate in this case is once again proportional to the efficiency of the entanglement distribution, and thus has the same scaling $O(\sqrt{\eta})$. To compare SEPM-QKD based upon our entanglement distribution to TF-QKD we have considered both in the same realistic scenario discussed in \cite{Lucamarini2018}, i.e., 30\% detector efficiencies and an error correction coefficient of 1.15. Although TES detectors have a negligible dark count rate~\cite{Miller2003, Lita2008}, for a fairer comparison we have also added a small probability of false detection events, $10^{-8}$ per gate interval, which could be due to either background (stray) photons or dark counts if the TES were replaced with single-photon detectors. The results are presented in Fig.~\ref{fig:keyrate}, and we see that the two key rates remain within an order of magnitude, with SEPM-QKD achieving a slightly larger maximum distance. In the case of SEPM-QKD, this maximum distance results when the false detection probability becomes comparable with the protocol efficiency, and can in principle be increased by using extremely low dark count TES detectors and sufficiently shielding from background photons~\cite{Yang2019}. Our version of SEPM-QKD may also be viewed as a realistic implementation of the ideal entanglement-based protocol presented in \cite{Curty2019}. See Appendix~\ref{app:SEPM-QKD} for more information.

\subsubsection{Earth-space communications with $O(\sqrt{\eta})$ scaling}

The robustness to extreme losses of our protocol makes it ideal for deployment in a satellite-based quantum communications scheme where entanglement is distributed between two remote locations on Earth and employed for quantum key distribution. Our protocol has the potential to outperform state-of-the-art demonstrations based on the distribution of two-photon Bell pairs, such as those reported in \cite{Yin_Science_2017} and \cite{Yin_Nature_2020}. There, an SPDC source pumped by a continuous wave laser generated space-borne Bell pairs at random times, which were next sent by a downlink channel to two stations on Earth separated by nearly 1200\kern.25emkm. The entanglement between the stations was verified by a CHSH Bell test under the fair sampling hypothesis. Moreover, only approximately two Bell pairs out of $5.9\times10^6$ reached the stations per second. The receiving stations had to be synchronized by the beacon laser and classical two-way communication between Alice and Bob was necessary to discard the cases when only one of the stations received a photon.

In contrast, in our protocol both Alice and Bob create locally an SV state. Their sources can be synchronized offline similarly to entanglement swapping schemes. A third party, Charlie, is located on a satellite and performs conditional state preparation. He acts as a central authority who broadcasts via a classical downlink if Alice and Bob share an entangled state. Thus, direct communication between Alice and Bob is not required to generate the entanglement, although for QKD some one-way communication is still necessary to perform information reconciliation. Most importantly, in a satellite-based approach, the improved scaling $O(\sqrt{\eta})$ is more significant. For example, in \cite{Yin_Nature_2020}, two Bell pairs reach the ground stations per second, whereas for similar atmospheric losses (40 dB between ground and satellite, 80 dB summed losses) and $g=0.1$ in our scenario, we estimate 160 successful entangled states to be generated per second.

Our protocol operating in an Earth-space environment is also original in that it uses an uplink rather than a downlink to send the quantum optical signals. Although downlinks have been shown to perform better than uplinks~\cite{Jennewein_NJP_2013}, from an experimental point of view uplinks provide better control and flexibility of the quantum source when it is located on the ground rather than in space, and thus are the better choice to study global-scale quantum communications implementations~\cite{Jennewein_NJP_2013}. 
Turbulence and background radiation affect uplinks more significantly, however, the pointing accuracy is a more consequential factor in our scheme, due to the necessity of time synchronization of the ground sources. In this aspect uplinks have in fact been shown to be more precise~\cite{Jennewein_NJP_2013}. Further analysis on the feasibility of uplink channels has been done in \cite{Toyoshima_uplink_2004} and \cite{Ding_uplink_2013}.

\subsubsection{Quantum metrology}

The entanglement distribution protocol described above can be used to generate states with applications to quantum metrology. 
$\ket{\Psi^{(k,\sigma)}_{\text{out}}}$ is a generalized HB state, which becomes an exact HB state for $k=\sigma/2$~\cite{Holland1993}. It offers quantum-enhanced optical phase estimation even in the presence of significant optical loss and approximates the performance of an optimal probe.
Details of this analysis together with an experimental demonstration can be found in \cite{Thekkadath_npj_2020}.

\section{Discussion}

We have proposed an entanglement distribution protocol based on an optimal use of the existing integrated quantum optical components, namely SPDC sources and PNR detectors. 
An experiment in which our family of states was successfully produced has been reported in \cite{Magda-Kravchuk} and \cite{Thekkadath_npj_2020}, manifesting the feasibility of our protocol. Other distinctive features of our protocol include robustness to arbitrarily high transmission losses, ability of choosing the local dimension of the generated multiphoton entangled state, the possibility of performing a loophole-free Bell test by adopting a preselection instead of a postselection procedure and the flexibility in the choice of the security framework depending on the available detector efficiencies. 
These allow our protocol to be applicable to quantum metrology, as already shown in \cite{Thekkadath_npj_2020}, Earth-space quantum communications and QKD. This would require space-ready TES detection, which is already being developed~\cite{space_TES}. We demonstrated that our protocol is capable of achieving a quadratic improvement in transmission rates compared to the distribution of polarization-entangled photon pairs, such as in the recently deployed Earth-space setup~\cite{Yin_Nature_2020}. This result enabled us to employ our distributed entangled states in well-established QKD protocols that match the key rates of state-of-the-art protocols. At smaller distances, such as in metropolitan networks, the generation of robust and near-maximally entangled multiphoton states also opens the possibility of performing high-dimensional QKD, which has recently produced record breaking results in other platforms~\cite{Hu2021}.

\section{Acknowledgments}

We would like to thank Stefanie Barz, Christine Silberhorn and Tim Bartley for discussions. M.E.M., T.M., A.B. and M.S. were supported by the Foundation for Polish Science ``First Team'' project No.\ POIR.04.04.00-00-220E/16-00 (originally FIRST\ TEAM/2016-2/17) and the National Science Centre `Sonata Bis' project No.\ 2019/34/E/ST2/00273. Numerical computations were performed in the ACK ``Cyfronet'' AGH computer centre (Zeus cluster). 

\vfill

\clearpage

\begin{table}[t]
	\begin{center}
		\begin{tabular}{ |c|c||c|c|c|c|c| } 
			\hline
			$k$ & $\sigma$ & $\ket{\Psi_{\text{out}}^{(k,\sigma)}}$\\
			\hline
			0 & 1 & $\frac{1}{\sqrt{2}}(\ket{0,1} - i\ket{1,0})$\\	
			1 & 1 & $\frac{1}{\sqrt{2}}(\ket{0,1} + i\ket{1,0})$\\	
			\hline
			0 & 2 & $\frac{1}{2}\ket{0,2} - \frac{i}{\sqrt{2}}\ket{1,1} - \frac{1}{2}\ket{2,0}$\\
			1 & 2 & $\frac{1}{\sqrt{2}}(\ket{0,2}+\ket{2,0})$\\
			2 & 2 & $\frac{1}{2}\ket{0,2} + \frac{i}{\sqrt{2}}\ket{1,1} - \frac{1}{2}\ket{2,0}$\\
			\hline
			0 & 3 & $\frac{1}{\sqrt{8}}(\ket{0,3}-i\sqrt{3}\ket{1,2}-\sqrt{3}\ket{2,1}+i\ket{3,0})$\\
			1 & 3 & $\frac{1}{\sqrt{8}}(\sqrt{3}\ket{0,3}-i\ket{1,2}+\ket{2,1}-i\sqrt{3}\ket{3,0})$\\
			2 & 3 & $\frac{1}{\sqrt{8}}(\sqrt{3}\ket{0,3}+i\ket{1,2}+\ket{2,1}+i\sqrt{3}\ket{3,0})$\\\
			3 & 3 & $\frac{1}{\sqrt{8}}(\ket{0,3}+i\sqrt{3}\ket{1,2}-\sqrt{3}\ket{2,1}-i\ket{3,0})$\\
			\hline
		\end{tabular}
		\label{states_table}
		\caption{Example states $\ket{\Psi_{\text{out}}^{(k,\sigma)}}$ for $\sigma \leq 3$.}
	\end{center}
\end{table}

\appendix

\section{Photon number statistics\\ of the output state $\ket{\Psi^{(k,\sigma)}_{\text{out}}}$}
\label{app:snote1}

In the lossless case, the family of output states generated by our protocol is given by
\begin{equation}
	\ket{\Psi_{\text{out}}^{(k,\sigma)}} = \sum_{n=0}^{\sigma} \mathcal{A}_\sigma(k,n)\ket{n,\sigma-n}_{a_1,b_1},
\end{equation}
where $\sigma$ is the total number of photons distributed between the signal modes and the probability amplitudes $\mathcal{A}_\sigma(k,n)$ are expressed using symmetric Kravchuk functions $\phi_k$~\cite{Magda-Kravchuk, Atakishiyev_1997, Hakioglu2000}
\begin{equation}
	\mathcal{A}_\sigma(k,n) = \bra{k, \sigma-k} U_{\text{BS}} \ket{n, \sigma-n} = i^{k-n}
	\phi_k(n-\sigma/2, \sigma).
	\label{eq:A}
\end{equation}
Here $U_\text{BS} = U_\text{BS}^{(1/2)}$ is the unitary for a beam splitter (BS) of reflectivity $r=\frac{1}{2}$ where $U_\text{BS}^{(r)} = \exp(-iH_\text{BS}\theta)$ with $H_\text{BS} = \frac{1}{2}(a^\dagger b + ab^\dagger)$ and $\theta =  2\arcsin(\sqrt{r})$. The Kravchuk functions $\phi_k$ are defined in terms of the Gaussian hypergeometric function ${}_2F_1$ as
\begin{equation}
	\phi_k(n-\sigma/2,\sigma) = (-1)^k \sqrt{\frac{1}{2^\sigma}\binom{\sigma}{n}\binom{\sigma}{k}} \, {}_2 F_1(-k,-n;-\sigma;2),
\end{equation}
and form an orthonormal set
\begin{equation}
	\sum_{n=0}^\sigma \phi_k(n-\sigma/2,\sigma) \phi_{k'}(n-\sigma/2,\sigma) = \delta_{k,k'}.
\end{equation}
The states $\ket{\Psi_{\text{out}}^{(k,\sigma)}}$ then form a complete orthonormal set in the $(\sigma+1)$-dimensional Hilbert space of fixed photon number $\sigma$
\begin{equation}
	\braket{\Psi_{\text{out}}^{(k,\sigma)}|\Psi_{\text{out}}^{(k',\sigma)}} = \delta_{k,k'}.
\end{equation}

\begin{figure*}[t]\centering
	\includegraphics[height=3.5cm]{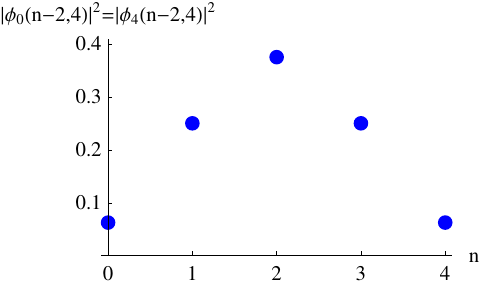}\hskip0.5cm
	\includegraphics[height=3.5cm]{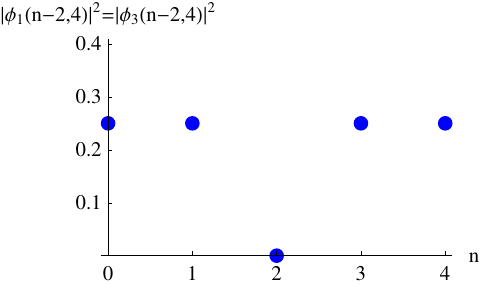}\hskip0.5cm
	\includegraphics[height=3.5cm]{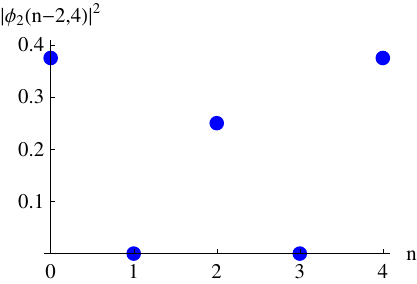}
	\caption{(Color online) Photon number distributions
			$\lvert\mathcal{A}_\sigma(k,n)\rvert^2=\lvert\phi_k(n-\sigma/2, \sigma)\rvert^2$
			computed for $\sigma=4$ and $k=0,1,2$. Note how the odd photon number components vanish for $ k=\sigma / 2 $.}
	\label{fig:As24}	
\end{figure*}

\begin{figure*}[t]\centering
	\includegraphics[height=3.3cm]{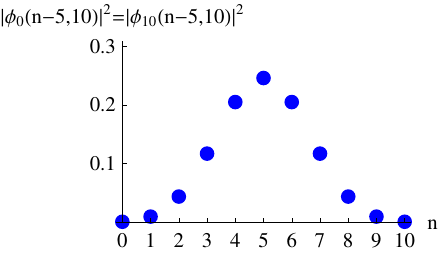}\hfill
	\includegraphics[height=3.3cm]{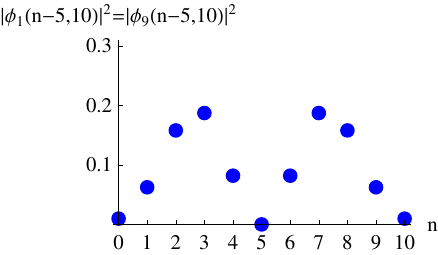}\hfill
	\includegraphics[height=3.3cm]{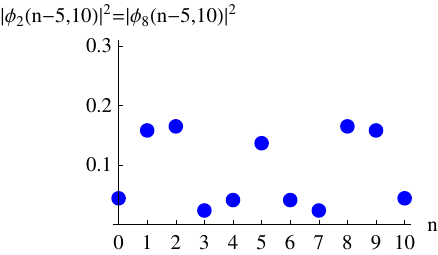}\\[2mm]
	\includegraphics[height=3.3cm]{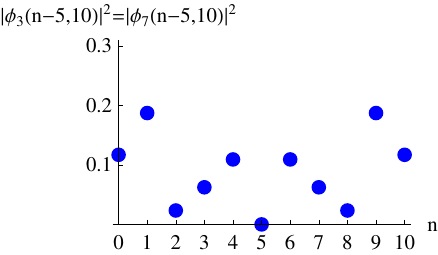}\hfill
	\includegraphics[height=3.3cm]{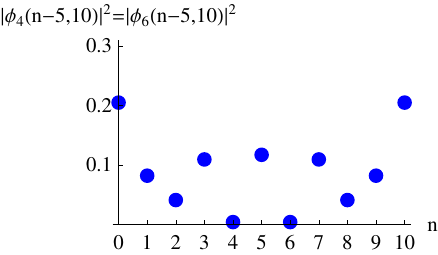}\hfill
	\includegraphics[height=3.3cm]{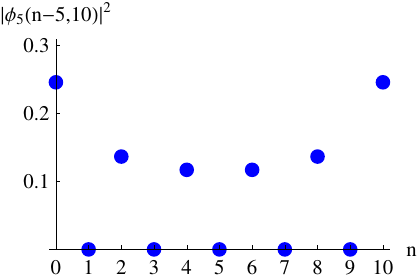}
	\caption{Photon number distributions $\lvert\mathcal{A}_\sigma(k,n)\rvert^2=\lvert\phi_k(n-\sigma/2, \sigma)\rvert^2$
			computed for $\sigma=10$ and $k=0,\dots,5$. Note how the odd photon number components vanish for $ k=\sigma / 2 $.}
	\label{fig:As210}	
\end{figure*}

These states are listed in Table I below for $\sigma = 1,2,3$, while Figs.~\ref{fig:As24} and \ref{fig:As210} depict photon number distributions $\lvert\mathcal{A}_\sigma(k,n)\rvert^2=\lvert\phi_k(n-\sigma/2, \sigma)\rvert^2$ for some higher states, $\sigma=4$ and $\sigma=10$ respectively. Notice that for $k=\sigma/2$, the odd photon number components vanish due to multi-photon Hong-Ou-Mandel interference, effectively reducing the Hilbert space dimension to $(\frac{\sigma}{2}+1)$. Additionally, the probability distribution $\lvert\mathcal{A}_\sigma(\sigma/2,n)\rvert^2$ possesses an envelope $f(n)=\frac{4}{\pi \sigma\sqrt{1-(2n/\sigma-1)^2}}$ which is a probability density function of an arcsine distribution~\cite{Nakazato}.

\section{Logarithmic negativity of $\ket{\Psi^{(k,\sigma)}_{\text{out}}}$} 
\label{app:snote2}

The logarithmic negativity of a physical state is a measure of its amount of entanglement. Given a state described by a density operator $\rho$, it is defined by $E_{\mathcal{N}}(\rho)=\log_2\left\lVert\rho^{\Gamma}\right\lVert_1$, where $\Gamma$ denotes the partial transposition operation and $\left\lVert\rho^{\Gamma}\right\lVert_1$ is the trace norm of $\rho^{\Gamma}$. In the lossless case our output state is $\rho_{\text{out}}^{(k,\sigma)} = \ket{\Psi_{\text{out}}^{(k,\sigma)}}\bra{\Psi_{\text{out}}^{(k,\sigma)}}$ and the logarithmic negativity is
\begin{equation}
	E_{\mathcal{N}}(\rho_{\text{out}}^{(k,\sigma)}) = 2\log_2 \left(\sum_{n=0}^\sigma 
	\bigl\lvert\mathcal{A}_\sigma(k,n)\bigr\rvert\right).
	\label{logneg}
\end{equation}

This formula can be derived in the following way
\begin{align}
	\rho_{\text{out}}^{(k,\sigma)}&{}= \ket{\Psi_{\text{out}}^{(k,\sigma)}} \bra{\Psi_{\text{out}}^{(k,\sigma)}} \nonumber\\
	&{}= \sum_{n,m=0}^\sigma \begin{aligned}[t]&\underbrace{[\rho]_{n,\sigma-n;m,\sigma-m}}_{\mathcal{A}_\sigma(k,n) \mathcal{A}^{*}_\sigma(k,m)}\times{}\\&\qquad\times\ket{n,\sigma-n}\bra{m,\sigma-m},\end{aligned}
	\\
	\left(\rho_{\text{out}}^{(k,\sigma)}\right)^{\Gamma_2}&{}= \sum_{n,m=0}^\sigma \begin{aligned}[t]&[\rho]_{n,\sigma-n;m,\sigma-m}\times{}\\&\qquad\times\ket{n,\sigma-m}\bra{m,\sigma-n},\end{aligned}
	\\
	\left(\left(\rho_{\text{out}}^{(k,\sigma)}\right)^{\Gamma_2}\right)^\dagger&{}= \sum_{n,m=0}^\sigma \begin{aligned}[t]&[\rho]^{*}_{n,\sigma-n;m,\sigma-m}\times\\&\qquad\times\ket{m,\sigma-n}\bra{n,\sigma-m},\end{aligned}
\end{align}

\begin{widetext}
\begin{align}
	\left(\left(\rho_{\text{out}}^{(k,\sigma)}\right)^{\Gamma_2}\right)^\dagger
	\cdot
	\left(\rho_{\text{out}}^{(k,\sigma)}\right)^{\Gamma_2}&
	{}=
	\sum_{n,m,p,q=0}^\sigma [\rho]^{*}_{n,\sigma-n;m,\sigma-m}
	[\rho]_{p,\sigma-p;q,\sigma-q}
	\ket{m,\sigma-n}\braket{n,\sigma-m\vert p,\sigma-q}\bra{q,\sigma-p}
	\\
	&{}=
	\sum_{n,m,p,q=0}^\sigma [\rho]^{*}_{n,\sigma-n;m,\sigma-m}
	[\rho]_{p,\sigma-p;q,\sigma-q}
	\delta_{n,p}\delta_{m,q}
	\ket{m,\sigma-n}\bra{q,\sigma-p}
	\\
	&
	{}=
	\sum_{n,m=0}^\sigma [\rho]^{*}_{n,\sigma-n;m,\sigma-m}
	[\rho]_{n,\sigma-n;m,\sigma-m}
	\ket{m,\sigma-n}\bra{m,\sigma-n}
	\\
	&{}=
	\sum_{n,m=0}^\sigma 
	\bigl\lvert[\rho]_{n,\sigma-n;m,\sigma-m}\bigr\rvert^2
	\ket{m}\bra{m}\otimes\ket{\sigma-n}\bra{\sigma-n},
	\\
	\left(\left(\rho_{\text{out}}^{(k,\sigma)}\right)^{\Gamma_2}\right)^\dagger
	\cdot
	\left(\rho_{\text{out}}^{(k,\sigma)}\right)^{\Gamma_2}
	&{}=
	\sum_{n,m=0}^\sigma 
	\bigl\lvert\mathcal{A}_\sigma(k,n)\bigr\rvert^2\,
	\bigl\lvert\mathcal{A}_\sigma(k,m)\bigr\rvert^2
	\ket{m}\bra{m}\otimes\ket{\sigma-n}\bra{\sigma-n}
	\\
	&{}=
	\left(\sum_{m=0}^\sigma 
	\bigl\lvert\mathcal{A}_\sigma(k,m)\bigr\rvert^2\,
	\ket{m}\bra{m}\right)
	\otimes
	\left(\sum_{n=0}^\sigma 
	\bigl\lvert\mathcal{A}_\sigma(k,n)\bigr\rvert^2\,
	\ket{\sigma-n}\bra{\sigma-n}\right).
\end{align}
The above operator is diagonal and therefore, it is easy to compute the square root of it by taking the square root of its eigenvalues
\begin{align}
	\sqrt{\left(\left(\rho_{\text{out}}^{(k,\sigma)}\right)^{\Gamma_2}\right)^\dagger
		\cdot
		\left(\rho_{\text{out}}^{(k,\sigma)}\right)^{\Gamma_2}}&
	{}=\begin{aligned}[t]&
	\left(\sum_{m=0}^\sigma 
	\bigl\lvert\mathcal{A}_\sigma(k,m)\bigr\rvert\,
	\ket{m}\bra{m}\right)
	\otimes\\&\qquad\otimes
	\left(\sum_{n=0}^\sigma 
	\bigl\lvert\mathcal{A}_\sigma(k,n)\bigr\rvert\,
	\ket{\sigma-n}\bra{\sigma-n}\right),\end{aligned}
	\\
	\left\lVert\left(\rho_{\text{out}}^{(k,\sigma)}\right)^{\Gamma_2}\right\lVert_1 = \Tr
	\sqrt{\left(\left(\rho_{\text{out}}^{(k,\sigma)}\right)^{\Gamma_2}\right)^\dagger
		\cdot
		\left(\rho_{\text{out}}^{(k,\sigma)}\right)^{\Gamma_2}}&
	{}=
	\sum_{a,b=0}^\sigma\begin{aligned}[t]
	&\left(\sum_{m=0}^\sigma 
	\bigl\lvert\mathcal{A}_\sigma(k,m)\bigr\rvert\,
	\braket{a\vert m}\braket{m\vert a}\right)\times{}\\&\qquad\times
	\left(\sum_{n=0}^\sigma 
	\bigl\lvert\mathcal{A}_\sigma(k,n)\bigr\rvert\,
	\braket{b\vert \sigma-n}\braket{\sigma-n\vert b}\right)
	\end{aligned}
	\\
	&{}=
	\left(\sum_{n=0}^\sigma 
	\bigl\lvert\mathcal{A}_\sigma(k,n)\bigr\rvert\right)^2.
\end{align}
In these formulas, $\Gamma_2$ denotes the partial transpose with respect to the second subsystem, and $\delta_{n,m}$ denotes the Kronecker delta, equal to $1$ for $n=m$ and $0$ otherwise.
\end{widetext}

\section{Protocol performance in case of symmetric losses in idler modes $a_2$ and $b_2$}
\label{app:disc1}

\subsection{Ideal case}

A two-mode squeezed vacuum state is given by
\begin{equation}
	\ket{\Psi} = \sum_{n=0}^{\infty}\sqrt{\lambda_{n}}\ket{n}_1 \ket{n}_2,
\end{equation}
so that the density operator for the input state $\ket{\Psi_{\text{in}}}=\ket{\Psi}^{\otimes2}$ is as follows
\begin{align}
	\rho_{\text{in}} ={}&\!\!\!\sum_{n,n',m,m'=0}^{\infty}\!\!\! \begin{aligned}[t]&\sqrt{\lambda_n\,\lambda_{n'}\,\lambda_m\,\lambda_{m'}}
	\ket{n,m}\bra{n',m'}_{a_1,b_1}\otimes{}\\&\quad\otimes \ket{n,m}\bra{n',m'}_{a_2,b_2}.
	\end{aligned}
\end{align}
Modes $a_2$ and $b_2$ interfere on a beam splitter and the resulting output modes are projected onto $\ket{k,\sigma-k}_{d_2,d_1}$. In the ideal case, the total number of photons registered behind the BS, $\sigma$, is equal to the total number of photons, $S$, in the remaining modes $a_1$ and $b_1$. The state created in modes $a_1$ and $b_1$ takes the form
\begin{equation}
	\rho^{(k,\sigma)}_{\text{out}} =
	\mathcal{N}^2\!\!\!\sum_{n,n',m,m'=0}^{\infty}\!\!\! 
	\begin{aligned}[t]	
	&\delta_{\sigma,n+m}\,\delta_{\sigma,n'+m'}\times{}\\
	&\qquad\times\sqrt{\lambda_n\,\lambda_{n'}\,\lambda_m\,\lambda_{m'}}\times{}\\
	&\qquad\times\mathcal{A}_{\sigma}(k,n)\,\bigl(\mathcal{A}_{\sigma}(k,n')\bigr)^{*}\times{}\\
	&\qquad\times\ket{n,m}\bra{n',m'}_{a_1,b_1},
	\end{aligned}
	\label{rhoABnolosses}
\end{equation}
where $\mathcal{N}^2=\frac{\cosh^4g}{\tanh^{2\sigma}g}$. This simplifies to
\begin{align}
	\rho^{(k,\sigma)}_{\text{out}}&=
	\sum_{n,n'=0}^{\sigma}
	\begin{aligned}[t]
	&\mathcal{A}_{\sigma}(k,n)\,
	\bigl(\mathcal{A}_{\sigma}(k,n')\bigr)^{*}\times{}
	\\&\qquad\times
	\ket{n,\sigma-n}\bra{n',\sigma-n'}_{a_1,b_1}
	\end{aligned}
	\label{rhoABnolossesFinal}\\
	&= \ket{\Psi^{(k,\sigma)}_{\text{out}}}\bra{\Psi^{(k,\sigma)}_{\text{out}}}.\nonumber
\end{align}

\subsection{Symmetric idler losses}
\label{symmetric_derivation}

\begin{figure*}[t]\centering
	\includegraphics[width=12cm]{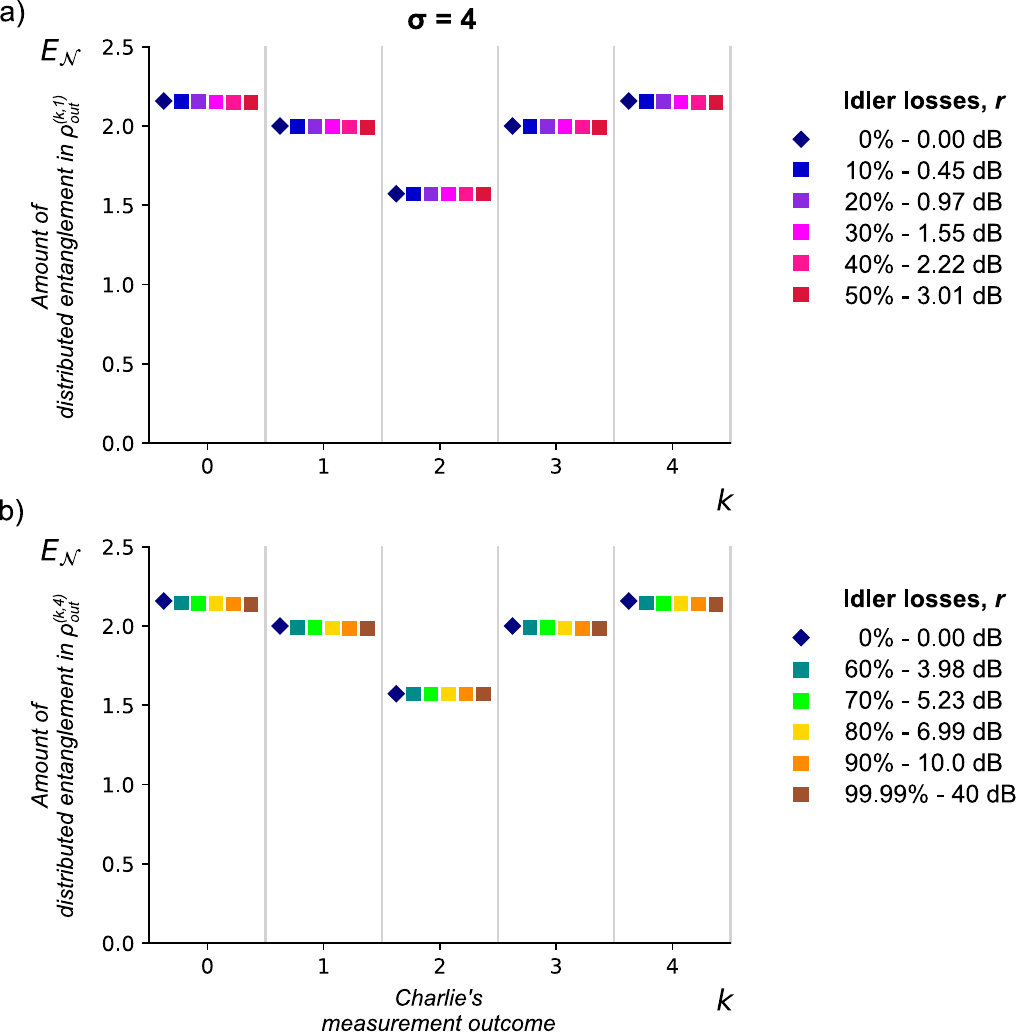}
	\caption{(Color online) Logarithmic negativity of the output state
		$\rho^{(k,\sigma)}_{\text{out}}$ assuming symmetric losses $r$ in the idler modes $a_2$ and $b_2$. The results were computed using Eq.~\eqref{rhoABsymmetric_fullform} for $g=0.1$ and $\sigma=4$ 
		with losses all the other modes and at the detectors set to
		zero.}
	\label{fig:log_neg_loss}
\end{figure*}

\begin{figure*}[t]\centering
	\includegraphics[width=12cm]{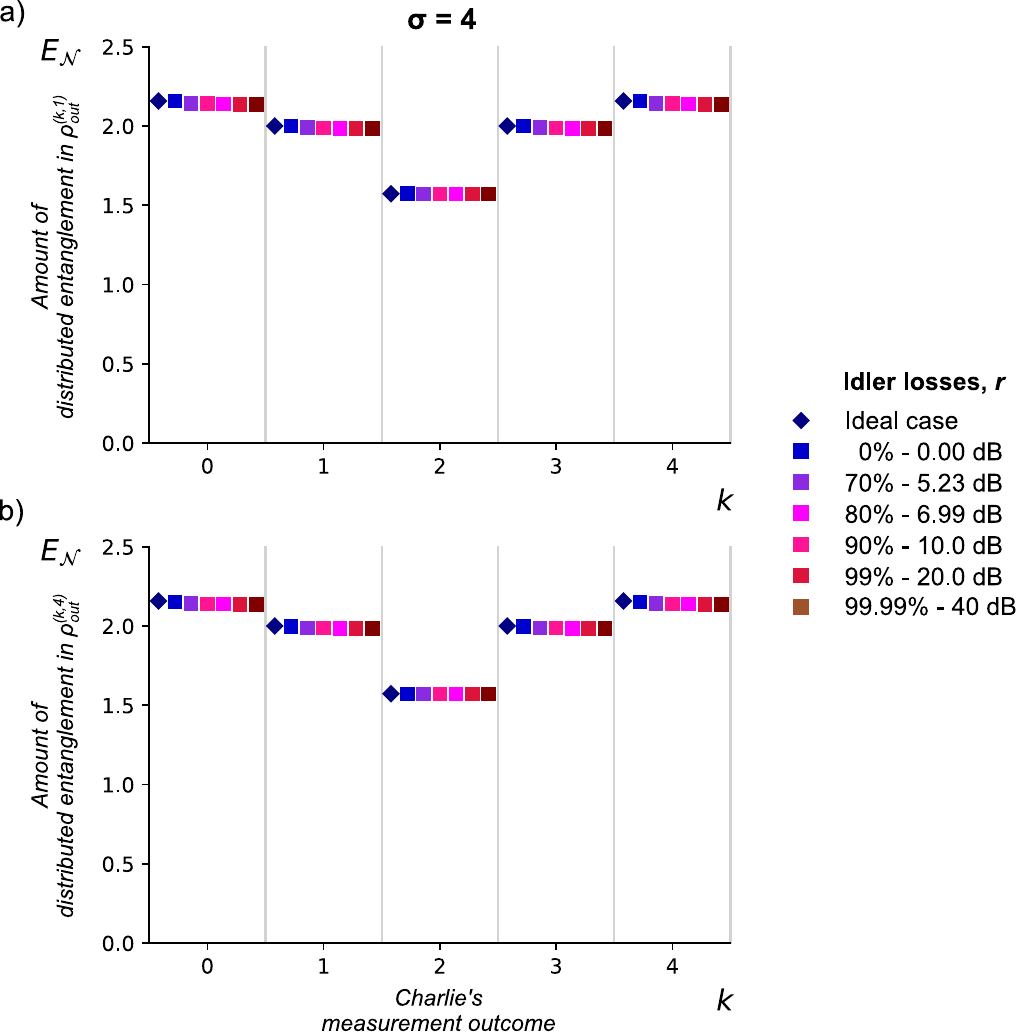}
	\caption{(Color online) Logarithmic negativity of the output state
		$\rho^{(k,\sigma)}_{\text{out}}$ assuming symmetric losses $r$ in the idler modes
		$a_2$ and $b_2$ and losses at the detectors. The results were computed using Eq.~(\eqref{rhoABsymmetric_fullform}) for $g=0.1$ and $\sigma=4$. Losses at detectors located behind the beam splitter are set
		to a)~$r_d = 5\%$, b)~$r_d = 21\%$. Losses at signal modes are set to zero.
		There is a negligible difference compared to the ideal case $r = r_s = r_d =
		0$.}
	\label{fig:log_neg_lossy_TESs}
\end{figure*}

Here we assume no losses in signal modes $a_1$ and $b_1$, as well as ideal detection performed by photon-number-resolving (PNR) detectors. More general cases including losses in idler and signal modes are studied in subsequent sections.

Losses can be modelled by a beam splitter with reflectivity $r\in[0,1]$ which quantifies the amount of loss, and then tracing over the reflected mode. For a Fock state $\ket{n}$ it reduces to
\begin{align}
	&\mathrm{Tr}_r\left\{U_{\text{BS}}^{(r)}\,\ket{n,0}\bra{n,0} \left(U_{\text{BS}}^{(r)}\right)^{\dagger}\right\}\nonumber\\&\qquad{}= 
	\sum_{p=0}^n \binom{n}{p}\,(1-r)^{n-p}\,r^p\,\ket{n-p}\bra{n-p}_t,
\end{align}
where $t$ and $r$ denote the transmitted and reflected mode, respectively, $r+t=1$, and
\begin{equation}
	U_{\text{BS}}^{(r)}\,\ket{n,0} = \sum_{p=0}^n (-i)^p \sqrt{\binom{n}{p}\,(1-r)^{n-p}\,r^p}\,\ket{n-p,p}_{t,r}.
\end{equation}
For $r=0$ it reduces to
\begin{equation}
\mathrm{Tr}_r\left\{U_{\text{BS}}^{(0)}\,\ket{n,0}\bra{n,0} \left(U_{\text{BS}}^{(0)}\right)^{\dagger}\right\}=\ket{n}\bra{n}_t.
\end{equation}

\begin{widetext}
Applying this procedure to the idler modes of the input state $\ket{\Psi_{\text{in}}}$ leads to
\begin{align}
	\rho_{\text{in}} ={}&
	\sum_{n,n',m,m'=0}^{\infty} \sqrt{\lambda_n\,\lambda_{n'}\,\lambda_m\,\lambda_{m'}}
	\sum_{p=0}^{\min(n,n')}\, \sum_{q=0}^{\min(m,m')}
	\sqrt{\binom{n}{p}\binom{n'}{p}\,(1-r_a)^{n+n'-2p}\,r_a^{2p}}
	\times{}\nonumber\\&\qquad{}\times
	\sqrt{\binom{m}{q}\binom{m'}{q}\,(1-r_b)^{m+m'-2q}\,r_b^{2q}}
	\times{}\nonumber\\&\qquad{}\times
	\ket{n,m}\bra{n',m'}_{a_1,b_1}\otimes\ket{n-p,m-q}\bra{n'-p,m'-q}_{a_2,b_2},
	\label{rho_in}
\end{align}
where $r_a$ and $r_b$ denote losses at Alice's and Bob's idler modes. As we assume that there are no losses in modes $a_1$ and $b_1$, the total number of photons in these modes equals $S=n+m=n'+m'$. The lossy modes $a_2$ and $b_2$ interfere on a balanced beam splitter and are projected onto $\ket{k,\sigma-k}_{d_2,d_1}$, where $\sigma \le S$. This results in the creation of the following state
\begin{align}
	\rho^{(k,\sigma)}_{\text{out}} ={}&\tilde{\mathcal{N}}^2\,
	\sum_{n,n',m,m'=0}^{\infty} \sqrt{\lambda_n\,\lambda_{n'}\,\lambda_m\,\lambda_{m'}}\ket{n,m}\bra{n',m'}_{a_1,b_1}
	\times{}\nonumber\\&\qquad{}\times
	\sum_{p=0}^{\min(n,n')}\, \sum_{q=0}^{\min(m,m')}
	\sqrt{\binom{n}{p}\binom{n'}{p}\,(1-r_a)^{n+n'-2p}\,r_a^{2p}}
	\sqrt{\binom{m}{q}\binom{m'}{q}\,(1-r_b)^{m+m'-2q}\,r_b^{2q}}
	\times{}\nonumber\\&\qquad{}\times
	\mathcal{A}_\sigma(k,n-p)\,\left(\mathcal{A}_\sigma(k,n'-p)\right)^{*}\,\delta_{\sigma,n+m-p-q}\,\delta_{\sigma,n'+m'-p-q},
	\label{rhoABlosses}
\end{align}
where
\begin{align}
	\frac{1}{\tilde{\mathcal{N}}^2}={}&
	\sum_{n,m=0}^{\infty} \lambda_n\,\lambda_m\kern-1em
	\sum_{p=\max(0,n-\sigma)}^{\min(n,n+m-\sigma)}
	\binom{n}{p}\binom{m}{n+m-\sigma-p}\,(1-r_a)^{n-p}\,r_a^{p}\,(1-r_b)^{p+\sigma-n}\,r_b^{n+m-p-\sigma}
	\left\lvert\mathcal{A}_\sigma(k,n-p)\right\rvert^2.
	\label{N2orig}
\end{align}

We now assume that the losses are equal, i.e.\ $r_a=r_b=r$. Then,
\begin{equation}
	(1-r)^{n+n'-2p}\,r^{2p}\,(1-r)^{m+m'-2q}\,r^{2q}=(1-r)^{2\sigma}\,r^{n+m+n'+m'-2\sigma}.
\end{equation}
This allows us to write
\begin{align}
	\rho^{(k,\sigma)}_{\text{out}} ={}&\tilde{\mathcal{N}}^2
	\sum_{n,n',m,m'=0}^{\infty} r^{-\sigma}\sqrt{r^n\lambda_n\,r^{n'}\lambda_{n'}\,r^m\lambda_m\,r^{m'}\lambda_{m'}}\ket{n,m}\bra{n',m'}_{a_1,b_1}
	\times{}\nonumber\\&\qquad{}\times
	\sum_{p=0}^{\min(n,n')}\, \sum_{q=0}^{\min(m,m')}
	\sqrt{\binom{n}{p}\binom{n'}{p}\binom{m}{q}\binom{m'}{q}}
	\times{}\nonumber\\&\qquad{}\times
	\mathcal{A}_\sigma(k,n-p)\,\left(\mathcal{A}_\sigma(k,n'-p)\right)^{*}\,\delta_{\sigma,n+m-p-q}\,\delta_{\sigma,n'+m'-p-q},
\end{align}
where we have absorbed the factor $(1-r)^\sigma$ into $\tilde{\mathcal{N}}^2$. Since $S=n+m=n'+m'$, we notice that $n,n'\leq S$ and $\sigma=S-p-q$. Since $q=S-\sigma-p$, we are able to remove the sum over $q$, which creates constraints on $p$: $p\leq S-\sigma$ and $p\geq n-\sigma$ as well as $p\geq n'-\sigma$. The state takes the form
\begin{align}
	\rho^{(k,\sigma)}_{\text{out}} ={}&\tilde{\mathcal{N}}^2\sum_{S=\sigma}^{\infty}\,
	\sum_{n,n'=0}^{S} r^{-\sigma} \sqrt{r^n\lambda_n\,r^{n'}\lambda_{n'}\,r^{S-n}\lambda_{S-n}\,r^{S-n'}\lambda_{S-n'}}\ket{n,S-n}\bra{n',S-n'}_{a_1,b_1}
	\times{}\nonumber\\&\qquad{}\times
	\sum_{p=0}^{\min(n,n')}\, \sum_{q=0}^{\min(S-n,S-n')}
	\sqrt{\binom{n}{p}\binom{n'}{p}\binom{S-n}{q}\binom{S-n'}{q}}
	\times{}\nonumber\\&\qquad{}\times
	\mathcal{A}_\sigma(k,n-p)\,
	\left(\mathcal{A}_\sigma(k,n'-p)\right)^{*}\,\delta_{\sigma,S-p-q}
	\\{}={}&
	\tilde{\mathcal{N}}^2\sum_{S=\sigma}^{\infty} r^{S-\sigma}\,\lambda_S
	\sum_{n,n'=0}^{S} \ket{n,S-n}\bra{n',S-n'}_{a_1,b_1}
	\times{}\nonumber\\&\qquad{}\times
	\sum_{p=\max(0,n-\sigma,n'-\sigma)}^{\min(S-\sigma,n,n')}
	\sqrt{\binom{n}{p}\binom{n'}{p}\binom{S-n}{S-\sigma-p}\binom{S-n'}{S-\sigma-p}}
	\times{}\nonumber\\&\qquad{}\times
	\mathcal{A}_\sigma(k,n-p)\,
	\left(\mathcal{A}_\sigma(k,n'-p)\right)^{*}
	\label{rhoABsymmetric_fullform}
	\\{}={}&
	\tilde{\mathcal{N}}^2\sum_{S=\sigma}^{\infty} \frac{r^{S-\sigma}\lambda_S}{\tilde{\mathcal{N}}_{\text{int}}^2}\,\rho^{(\sigma,k,S)}_{\text{int}}
	=
	\tilde{\mathcal{N}}^2\sum_{S=\sigma}^{\infty} \chi_{\sigma,S}\,\rho^{(\sigma,k,S)}_{\text{int}},
	\label{rhoABsymmetric}
\end{align}
where $\chi_{\sigma,S}=\frac{r^{S-\sigma}\lambda_S}{\tilde{\mathcal{N}}^2_{\text{int}}}$. Here we have used $\lambda_n \lambda_{S-n} = \frac{\lambda_S}{\cosh^2 g}$ and absorbed the factor $\frac{1}{\cosh^2 g}$ into $\tilde{\mathcal{N}}^2$. Eq.~(\ref{N2orig}) then simplifies to
\begin{align}
	\frac{1}{\tilde{\mathcal{N}}^2}={}&
	\sum_{S=\sigma}^{\infty}
	r^{S-\sigma}
	\lambda_S
	\sum_{n=0}^{S} 
	\sum_{p=\max(0,n-\sigma)}^{\min(S-\sigma,n)}
	\binom{n}{p}\binom{S-n}{S-\sigma-p}
	\left\lvert\mathcal{A}_\sigma(k,n-p)\right\rvert^2.
	\label{ntilde}
\end{align}
The internal matrix is as follows
\begin{align}
	\rho^{(\sigma,k,S)}_{\text{int}} ={}& \tilde{\mathcal{N}}^2_{\text{int}}
	\sum_{n,n'=0}^{S} \ket{n,S-n}\bra{n',S-n'}_{a_1,b_1}
	\times{}\nonumber\\&\qquad{}\times
	\sum_{p=\max(0,n-\sigma,n'-\sigma)}^{\min(S-\sigma,n,n')}
	\sqrt{\binom{n}{p}\binom{n'}{p}\binom{S-n}{S-\sigma-p}\binom{S-n'}{S-\sigma-p}}
	\times{}\nonumber\\&\qquad{}\times
	\mathcal{A}_\sigma(k,n-p)\,
	\left(\mathcal{A}_\sigma(k,n'-p)\right)^{*},
\end{align}
where
\begin{equation}
	\frac{1}{\tilde{\mathcal{N}}^2_{\text{int}}} =
	\sum_{n=0}^{S}\,\sum_{p=\max(0,n-\sigma)}^{\min(S-\sigma,n)}
	\binom{n}{p}\binom{S-n}{S-\sigma-p}
	\bigl\lvert\mathcal{A}_\sigma(k,n-p)\bigr\rvert^2
\end{equation}
By shifting the summation index $p\to n-p$ and then using the Chu-Vandermonde identity $\sum_{n=0}^S \binom{n}{p}\binom{S-n}{\sigma-p} = \binom{S+1}{\sigma+1}$, this may be evaluated as $\frac{1}{\tilde{\mathcal{N}}^2_{\text{int}}}=\binom{S+1}{\sigma+1}$. This allows one to simplify the form of $\chi_{\sigma,S}$ to
\begin{equation}
	\chi_{\sigma,S} = r^{S-\sigma}\,\lambda_S\,\binom{S+1}{\sigma+1}.
\end{equation} 
Furthermore, Eq.~(\ref{ntilde}) now simplifies to
\begin{align}
	\frac{1}{\tilde{\mathcal{N}}^2}&={}
	\sum_{S=\sigma}^{\infty}
	r^{S-\sigma}
	\lambda_S
	\binom{S+1}{\sigma+1}\\
	&= \frac{\lambda_\sigma}{(1-r \tanh^2 g)^{\sigma+2}},
\end{align}
where we have used the identity $\sum_{S=\sigma}^\infty x^S \binom{S+1}{\sigma+1} = (1-x)^{-\sigma-2} x^\sigma$. The probability to obtain $S$ photons, given a measurement $\ket{k, \sigma-k}$ at the remote station, is then equal to
\begin{equation}
	p_{S|\sigma} = \tilde{\mathcal{N}}^2 \chi_{\sigma,S} = (r \tanh^2 g)^{S-\sigma} (1- r \tanh^2 g)^{\sigma+2} \binom{S+1}{\sigma+1}.
\end{equation}
and Eq.~(\ref{rhoABsymmetric}) may be rewritten as
\begin{equation} 
	\rho^{(k,\sigma)}_{\text{out}}=\sum_{S=\sigma}^{\infty}p_{S|\sigma}\,\rho^{(\sigma,k,S)}_{\text{int}}.
\end{equation}
For $S=\sigma$, as expected, $\rho^{(\sigma,k,S)}_{\text{int}}$ reproduces the density operator for the lossless case
\begin{align}
	\rho^{(\sigma,k,\sigma)}_{\text{int}} ={}&
	\sum_{n,n'=0}^{\sigma} \ket{n,\sigma-n}\bra{n',\sigma-n'}_{a_1,b_1}
	\times{}\nonumber\\&\qquad{}\times
	\sum_{p=\max(0,n-\sigma,n'-\sigma)}^{\min(\sigma-\sigma,n,n')}
	\sqrt{\binom{n}{p}\binom{n'}{p}\binom{\sigma-n}{\sigma-\sigma-p}\binom{\sigma-n'}{\sigma-\sigma-p}}
	\times{}\nonumber\\&\qquad{}\times
	\mathcal{A}_\sigma(k,n-p)\,
	\left(\mathcal{A}_\sigma(k,n'-p)\right)^{*}
	\\
	{}={}&
	\sum_{n,n'=0}^{\sigma}
	\mathcal{A}_\sigma(k,n)\,
	\left(\mathcal{A}_\sigma(k,n')\right)^{*}\,
	\ket{n,\sigma-n}\bra{n',\sigma-n'}_{a_1,b_1}
	=
	\ket{\Psi_{\text{out}}^{(k,\sigma)}}\bra{\Psi_{\text{out}}^{(k,\sigma)}}.
\end{align}
The probability $p_{S|\sigma}$ quantifies the contribution of $\rho^{(\sigma,k,S)}_{\text{int}}$ to the final density operator. Provided we are in the limit $r \tanh^2 g \ll 1$, the term $(r \tanh^2 g)^{S-\sigma}$ decreases with $S$ more rapidly than $\binom{S+1}{\sigma+1}$ increases, so that $p_{S|\sigma}$ is largest for $S=\sigma$ and rapidly decreases. In the realistic limit $g^2 \ll 1$, this is satisfied for arbitrarily high idler losses $r\to 1$, leading to the output state being largely unchanged by the losses $\rho^{(k,\sigma)}_{\text{out}} \approx \ket{\Psi_{\text{out}}^{(k,\sigma)}}\bra{\Psi_{\text{out}}^{(k,\sigma)}}$.

\subsection{Numerical computations}

We have computed the logarithmic negativity for the density operator in Eq.~(\ref{rhoABsymmetric_fullform}) assuming losses $r$ in idler modes $a_2$ and $b_2$ to be symmetric. The signal mode and detector losses, $r_s$ and $r_d$, have been set to zero. As is customary, the reflectivities in the results below and in the following subsections are displayed in percentages (\%). 

We have changed $r$ between $0$ and $90\%$. The case of $r=99.99\%$ ($40\kern.25em\mathrm{dB}$), which is typical of Earth-to-space scenarios, has been also examined. The results are shown in Fig.~\ref{fig:log_neg_loss}. The logarithmic negativity is not deteriorated by increasing values of $r$ and even for $r=99.99\%$ it is close to the ideal case $r=0$.

\subsection{Numerical computations including lossy detection}

We have repeated the above computations for the density operator in Eq.~(\ref{rhoABsymmetric_fullform}) with non-zero losses $r_d$ at the PNRs located behind the beam splitter included in the numerical programme. Fig.~\ref{fig:log_neg_lossy_TESs} depicts the values of $E_{\mathcal{N}}$ for $r_d$ equal to $5\%$ and $21\%$, which correspond to detector efficiency of $95\%$ and $79\%$ in case of on-chip detection scheme \cite{Calkins_2013}. We have also set the range of $r\in[0,99.99\%]$. Similarly to the previous case, the results are not deteriorated by the losses.

\subsection{Comparison with the case of symmetric losses in signal modes $a_1$ and $b_1$}

In order to check the influence of losses $r_s$ in signal modes $a_1$ and $b_1$ we have set  
them to symmetric values in the range between $0$ and $50\%$. The losses at the detectors have been set to $5\%$ and $21\%$ while we assumed no losses at idler modes. The results are depicted in Fig.~\ref{fig:log_neg_signal}. Non-zero values of $r_s$ significantly lower the obtained logarithmic negativity, which drops below $1.0$ for $r_s=30\%$, in the case of $g=0.1$, $\sigma=4$. This is not a major problem since the signal beams only travel a short distance to Alice / Bob's local detectors, compared to the large distance travelled by the idler beams. The signal beams may be measured before the idler beams (the delayed-choice scheme) without affecting the measurement statistics, by the no-signaling principle.

\begin{figure*}[t]\centering
	\includegraphics[width=12cm]{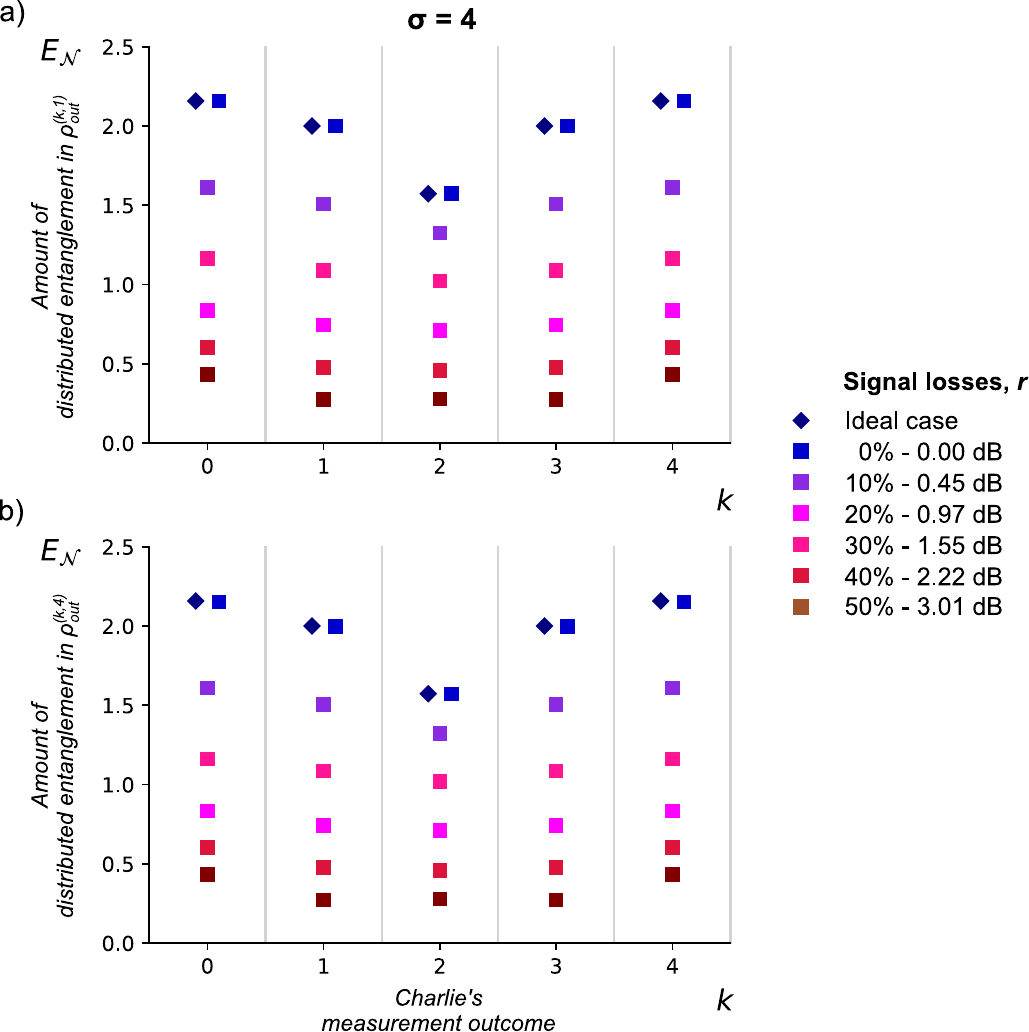}
	\caption{(Color online) Logarithmic negativity of the output state
		$\rho^{(k,\sigma)}_{\text{out}}$ assuming
		symmetric losses in signal modes $a_1$ and $b_1$ and losses at the detectors. The results were computed for $g=0.1$ and $\sigma=4$. Losses at detectors located behind the beam splitter are set to a) $5\%$, b) $21\%$. Losses at the idler modes are set to zero.}
	\label{fig:log_neg_signal}
\end{figure*}

\section{Protocol performance for asymmetric losses in idler modes $a_2$ and $b_2$}
\label{app:disc2}
	
\subsection{Analytic derivations} 

In this subsection we assume no losses in signal modes as well as ideal detection performed by all PNR detectors. Moreover, since analytical derivations become quite complex for a general case of losses occurring in our setup, we will first analyse the case when Bob's idler mode, $b_2$, is lossless ($t_b = 1$ or $r_b = 0$) and Alice's idler mode's transmittance, $t_a$, is a fraction $\epsilon$ of Bob's mode's transmittance ($t_a = \epsilon$). This approach will allow us to understand the impact of asymmetric losses on the protocol's performance.

We repeat the steps as in Appendix~\ref{app:disc1}, but with $r_a  = 1 - \epsilon$ to arrive at the form equivalent to Eq.~(\ref{rhoABlosses})
\begin{align}
	\rho^{(k,\sigma)}_{\text{out}} ={}&\tilde{\mathcal{N}}^2\,
	\sum_{n,n',m,m'=0}^{\infty} \sqrt{\lambda_n\,\lambda_{n'}\,\lambda_m\,\lambda_{m'}}\ket{n,m}\bra{n',m'}_{a_1,b_1}
	\times{}\nonumber\\&\qquad{}\times
	\sum_{p=0}^{\min(n,n')}\, \sum_{q=0}^{\min(m,m')}
	\sqrt{\binom{n}{p}\binom{n'}{p}\,t_a^{n+n'-2p}\,(1-t_a)^{2p}}
	\sqrt{\binom{m}{q}\binom{m'}{q}\,t_b^{m+m'-2q}\,(1-t_b)^{2q}}
	\times{}\nonumber\\&\qquad{}\times
	\mathcal{A}_\sigma(k,n-p)\,\left(\mathcal{A}_\sigma(k,n'-p)\right)^{*}\,\delta_{\sigma,n+m-p-q}\,\delta_{\sigma,n'+m'-p-q},
\end{align}
\begin{align}
	\rho^{(k,\sigma)}_{\text{out}} ={}&\tilde{\mathcal{N}}^2\,
	\sum_{n,n',m,m'=0}^{\infty} \sqrt{\lambda_n\,\lambda_{n'}\,\lambda_m\,\lambda_{m'}}\ket{n,m}\bra{n',m'}_{a_1,b_1}
	\times{}\nonumber\\&\qquad{}\times
	\sum_{p=0}^{\min(n,n')}\, 
	\sqrt{\binom{n}{p}\binom{n'}{p}\,\epsilon^{n+n'-2p}\,(1-\epsilon)^{2p}}
	\times{}\nonumber\\&\qquad{}\times
	\mathcal{A}_\sigma(k,n-p)\,\left(\mathcal{A}_\sigma(k,n'-p)\right)^{*}\,\delta_{\sigma,n+m-p}\,\delta_{\sigma,n'+m'-p},
\end{align}        
where we could remove the sum over $q$, as only the $q=0$ term would contribute to the sum for $t_b = 1$.

The Kronecker delta functions tell us that $n+m = n' + m' = S$, we can then set $m=S-n, m'=S-n'$ and replace the sum over $m, m'$ with a single sum over $S$. They also tell us that $p = S-\sigma$, so that the sum over $p$ can also be removed. Thus, the density matrix takes the following form
\begin{align}
	\rho^{(k,\sigma)}_{\text{out}} ={}& \tilde{\mathcal{N}}^2\,
	\sum_{S=\sigma}^{\infty}\sum_{n,n'=S-\sigma}^{S} \sqrt{\lambda_n\,\lambda_{n'}\,\lambda_{S-n}\,\lambda_{S-n'}}\ket{n,S-n}\bra{n',S-n'}_{a_1,b_1}
	\times{}\nonumber\\&\qquad{}\times
	\sqrt{\binom{n}{S-\sigma}\binom{n'}{S-\sigma}\,\epsilon^{n+n'+2(\sigma-S)}\,(1-\epsilon)^{2(S-\sigma)}}
	\times{}\nonumber\\&\qquad{}\times
	\mathcal{A}_\sigma(k,n-S+\sigma)\,\left(\mathcal{A}_\sigma(k,n'-S+\sigma)\right)^{*}. 
\end{align}  
Finally, we notice that $\lambda_n\lambda_{S-n}=\frac{\lambda_S}{\cosh^2 g}$, and again absorb $\frac{1}{\cosh^2 g}$ into $\tilde{\mathcal{N}}^2$ 
\begin{align}
	\rho^{(k,\sigma)}_{\text{out}} ={}& \tilde{\mathcal{N}}^2\,
	\sum_{S=\sigma}^{\infty}\lambda_S \big(\tfrac{1-\epsilon}{\epsilon}\big)^{S-\sigma} \sum_{n,n'=S-\sigma}^{S} \epsilon^{\frac{n+n'}{2}}\ket{n,S-n}\bra{n',S-n'}_{a_1,b_1}
	\times{}\nonumber\\&\qquad{}\times
	\sqrt{\binom{n}{S-\sigma}\binom{n'}{S-\sigma}}
	\mathcal{A}_\sigma(k,n-S+\sigma)\,\left(\mathcal{A}_\sigma(k,n'-S+\sigma)\right)^{*},
	\label{rhoAlossesBideal_fullform}
\end{align}
where the normalization factor equals
\begin{align}
	\frac{1}{\tilde{\mathcal{N}}^2}={}&
	\sum_{S=\sigma}^{\infty} \lambda_S\big(\tfrac{1-\epsilon}{\epsilon}\big)^{S-\sigma}
	\sum_{n=S-\sigma}^{S}\epsilon^n \binom{n}{S-\sigma}  
	\left\lvert\mathcal{A}_\sigma(k,n-S+\sigma)\right\rvert^2.
\end{align}
We notice that, unlike in Eq.~(\ref{rhoABsymmetric}), it is impossible to decompose the matrix into $\epsilon$-independent components. Instead we obtain
\begin{align}
	\rho^{(k,\sigma)}_{\text{out}} ={}& \tilde{\mathcal{N}}^2 \sum_{S=\sigma}^{\infty} \chi_{\sigma,S}(\epsilon) \rho^{(\sigma,k,S)}_{\text{int}}(\epsilon), 
\end{align}
where $\chi_{\sigma,S}(\epsilon) = \tfrac{\lambda_S}{\tilde{\mathcal{N}}(\epsilon)^2} \big(\tfrac{1-\epsilon}{\epsilon}\big)^{S-\sigma}$ and the density matrix components are
\begin{align}
	\rho^{(\sigma,k,S)}_{\text{int}}(\epsilon) ={}& \tilde{\mathcal{N}}(\epsilon)^2\,
	\sum_{n,n'=S-\sigma}^{S} \epsilon^{\frac{n+n'}{2}}\ket{n,S-n}\bra{n',S-n'}_{a_1,b_1}
	\times{}\nonumber\\&\qquad{}\times
	\sqrt{\binom{n}{S-\sigma}\binom{n'}{S-\sigma}}
	\mathcal{A}_\sigma(k,n-S+\sigma)\,\left(\mathcal{A}_\sigma(k,n'-S+\sigma)\right)^{*},
\end{align}
\end{widetext}
with
\begin{align}
	\frac{1}{\tilde{\mathcal{N}}(\epsilon)^2}={}&
	\sum_{n=S-\sigma}^{S}\epsilon^n \binom{n}{S-\sigma}  
	\left\lvert\mathcal{A}_\sigma(k,n-S+\sigma)\right\rvert^2.
\end{align}

\begin{figure}[htbp]\centering
	\includegraphics[width=\columnwidth]{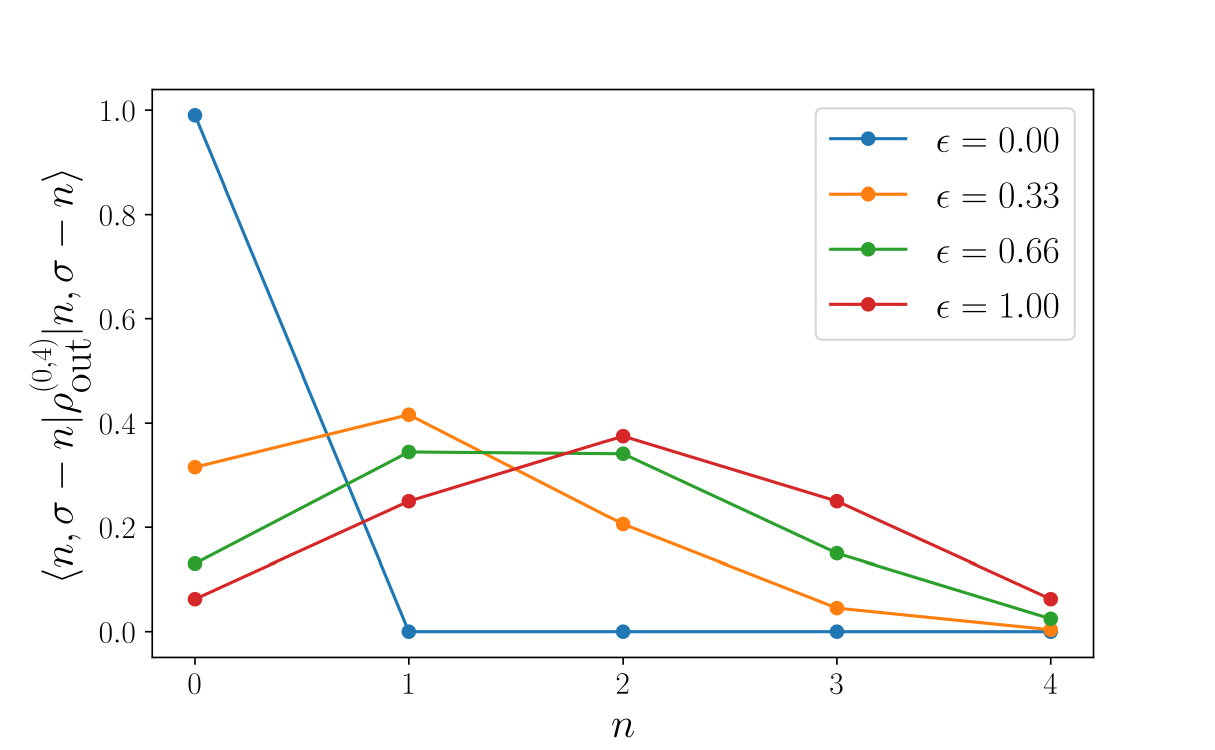}
	\caption{(Color online) Effect of asymmetry on the probability for Alice and Bob to measure $ n, \sigma -n $ photons. Given a state $\rho_\textrm{out}^{(0,4)}$ generated with asymmetric
		losses and $g=0.1$, the probability for Alice, Bob to measure $n, \sigma-n$
		photons is given by $\langle
		n,\sigma-n|\rho_\textrm{out}^{(k,\sigma)}|n,\sigma-n\rangle$ and plotted for
		$\epsilon \in \{0, 0.33, 0.66, 1\}$. Asymmetry in losses leads to asymmetry of
		the shared state, lowering entanglement. The probabilities sum to $\sim0.99$
		rather than one due to a small probability to detect $S \geq \sigma$ photons.}
	\label{fig:asymm}
\end{figure}

We find that, similar to the case of symmetric losses, the probability $\tilde{\mathcal{N}}^2 \chi_{\sigma,S}(\epsilon)$ is strongly peaked at $S=\sigma$ when $g^2 \ll 1$, so that $\rho^{(k,\sigma)}_{\text{out}} \approx \rho^{(\sigma,k,\sigma)}_{\text{int}}$. However $\rho^{(\sigma,k,\sigma)}_{\text{int}}$ is no longer equal to the lossless state $\ket{\Psi_{\text{out}}^{(k,\sigma)}}\bra{\Psi_{\text{out}}^{(k,\sigma)}}$ but is instead given by $\rho^{(\sigma,k,\sigma)}_{\text{int}} = \ket{\Psi_\epsilon^{(k,\sigma)}}\bra{\Psi_\epsilon^{(k,\sigma)}}$ where
\begin{equation}
	\ket{\Psi_\epsilon^{(k,\sigma)}} = \tilde{\mathcal{N}}(\epsilon)^2\sum_{n=0}^\sigma \epsilon^{n/2} \mathcal{A}_\sigma(k,n)\ket{n,\sigma-n}.
\end{equation}
The asymmetry lowers the number of photons, $n$, measured by Alice, breaking the $n \to \sigma-n$ symmetry of the state. This is shown for the state $k=0, \sigma=4$ and $g=0.1$ in Fig.~\ref{fig:asymm}, where the probability for Alice, Bob to measure $n, \sigma-n$ photons is plotted for $\epsilon \in \{0, 0.33, 0.66, 1\}$. The suppression of high $n$ states leads to an effective reduction of the Hilbert space dimension, lowering the entanglement. Following the same steps as in Note 2, the logarithmic negativity of $\ket{\Psi_\epsilon^{(k,\sigma)}}$ is
\begin{equation}
	E_{\mathcal{N}}(\ket{\Psi_\epsilon^{(k,\sigma)}}\bra{\Psi_\epsilon^{(k,\sigma)}}) = 2\log_2 \left(\frac{\sum_{n=0}^\sigma \epsilon^{n/2} \bigl\lvert\mathcal{A}_\sigma(k,n)\bigr\rvert}{\sqrt{\sum_{n=0}^\sigma \epsilon^{n} \bigl\lvert\mathcal{A}_\sigma(k,n)\bigr\rvert^2}}\right).
	\label{lognegasymm}
\end{equation}
This correctly reduces to Eq.~(\ref{logneg}) in the symmetric limit $\epsilon \to 1$, while in the limit of extreme asymmetry $\epsilon \to 0$ it becomes $E_{\mathcal{N}} = 0$, since the state reduces to the unentangled $\ket{0, \sigma}$.

\begin{figure*}[t]
	\centering
	\includegraphics[width=12cm]{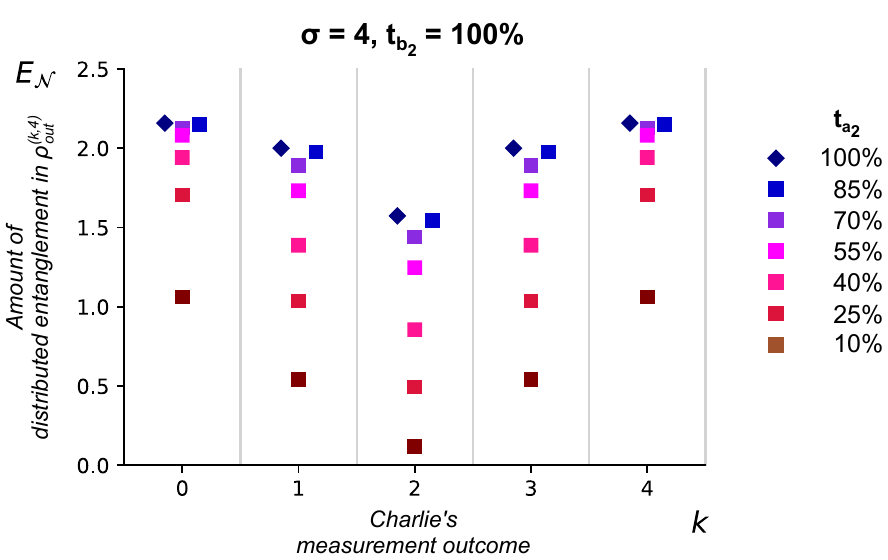}
	\caption{(Color online) Logarithmic negativity computed for the asymmetric state in
		Eq.~(\ref{rhoAlossesBideal_fullform}). The results are computed for $ g0.1 $ and $ \sigma = 4 $. The trasnmissivity of Bob's idler modes is assumed ideal and Alice's drops from ideal to 10\%.}	
	\label{fig:unsymlogneganalysis}
\end{figure*}

\subsection{Numerical computations}

To back up our analytic derivations we have computed the logarithmic negativity for the full state given in Eq.~(\ref{rhoAlossesBideal_fullform}) for various values of transmittances $t_a=\epsilon \cdot 100\%$, while keeping Bob's mode ideal ($t_b=100\%$). As before, we take $g=0.1$ and display transmittances in percentages (\%) in the following results. A representative figure showing the behavior of logarithmic negativity as the transmittance of idler modes is made progressively more unequal is shown in Fig.~\ref{fig:unsymlogneganalysis}. We find that a small asymmetry $\epsilon \gtrsim 70\%$ has almost no effect but thereafter there is a significant reduction in the amount of entanglement.

\begin{figure}[hb]
	\centering
	\includegraphics[width=\columnwidth]{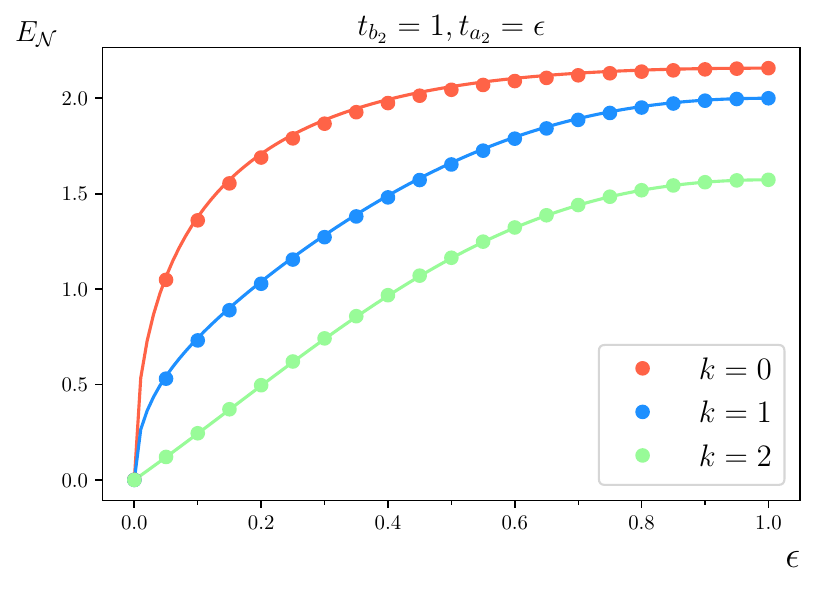}
	\caption{(Color online) Dependence of logarithmic negativity shown in
			Fig.~\ref{fig:unsymlogneganalysis} on imbalance of losses in the two idler modes. The imbalance is quantified by
		$\epsilon = \tfrac{t_a}{t_b}$, and the results are computed for $g=0.1$ and $\sigma=4$. The points are the
		exactly calculated numerical results, while the solid lines are calculated from
		the analytical approximation Eq.~(\ref{lognegasymm}).}
	\label{fig:unsymscalinganalysis}
\end{figure}

Further analysis reveals that the entanglement is best preserved for the case where all photons emerge from the same output of the beam splitter, i.e. for $k=0$ and $k=\sigma$. The entanglement deteriorates quickest when the outputs of the beam splitter are equally populated, $k=\sigma/2$. These results are shown in Fig.~\ref{fig:unsymscalinganalysis} where the dependence of the logarithmic negativity on $\epsilon = \tfrac{t_a}{t_b}$ is plotted for $\sigma=4$. The numerically calculated values (points) match almost exactly the approximate analytical result (lines) calculated from Eq.~(\ref{lognegasymm}). From these calculations it can be inferred that the protocol maintains its quality of output (up to 90\% of the maximal value of entanglement) down to $\epsilon=0.7$ for $k=2$ and $\epsilon=0.4$ for $k=0$.

\begin{figure*}[tp]
	\centering
	\includegraphics[width=\textwidth]{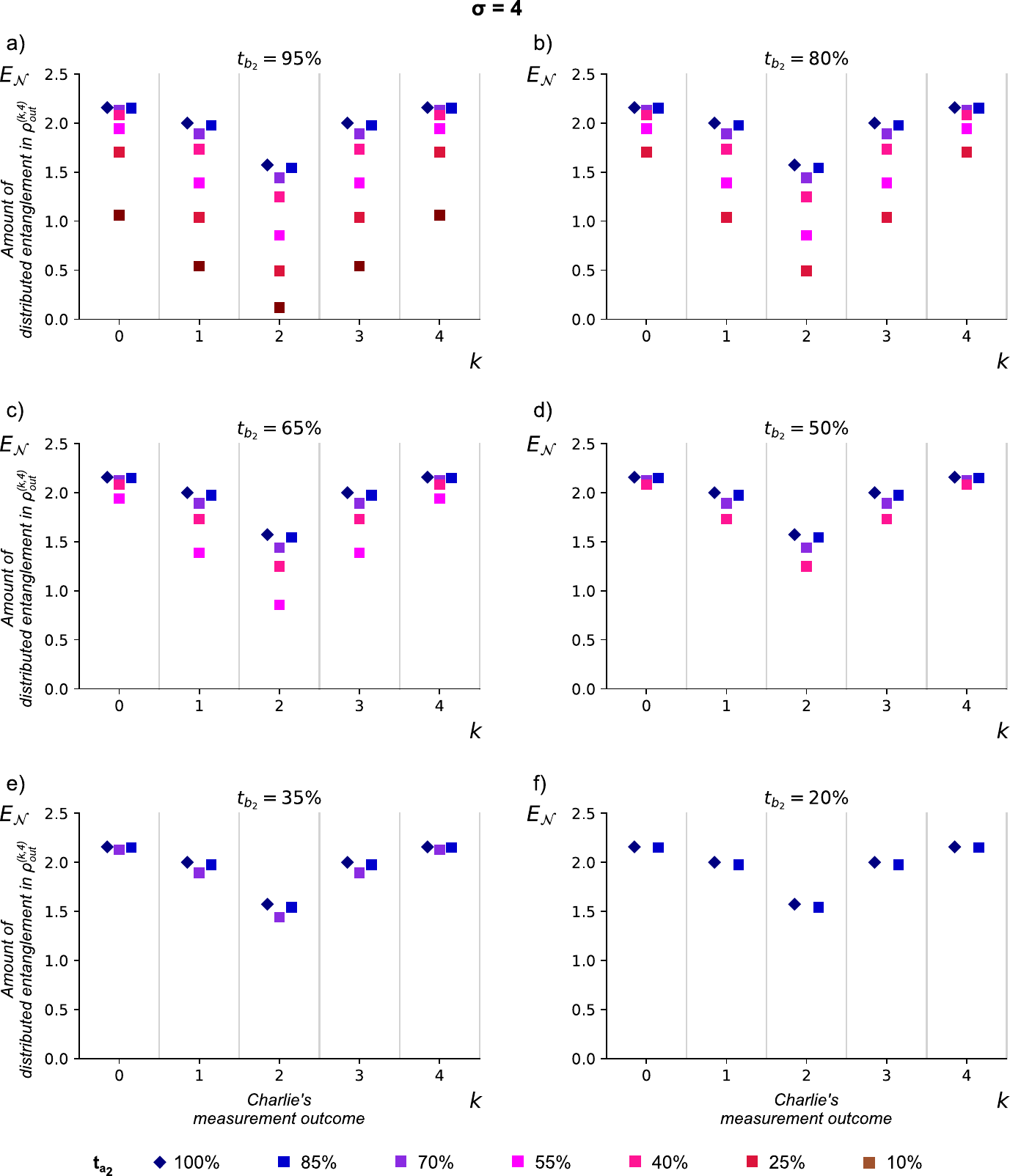}
	\caption{Logarithmic negativity computed for imbalanced idler modes with non-ideal transmissivity in Bob's mode. The results were computed for $g=0.1$ and $\sigma=4$. The transmissivity of Bob's mode drops from 95\% to 20\%. The transmissivity of Alice's mode is a fraction of Bob's transmissivity as specified by the legend below the figures.}
	\label{fig:logneg_wholespectrum}
\end{figure*}

\subsection{Numerical computations with both idler modes lossy}

We include computations for the more general cases, when Bob's mode is no longer assumed lossless. The results for $\sigma = 4$ are shown in Fig.~\ref{fig:logneg_wholespectrum}. $t_{b_2}$ takes values from 90\% to 10\% and for each $t_{b_2}$, $t_{a_2}$ goes from 5\% to $t_{b_2}$ in steps of 5\%. When the transmittances of the two idler modes are equal, the amount of entanglement is close to maximal, given by $\log_2(5) \approx 2.32$. As the symmetry is broken, entanglement decreases. In fact, Eq.~(\ref{lognegasymm}) gives a very good analytical approximation even in this more general case, showing that the entanglement depends only on $\epsilon = \frac{t_a}{t_b}$.

\subsection{General case which assumes losses in all modes and in detection}

\begin{figure}[htbp]\centering
	\includegraphics[width=\columnwidth]{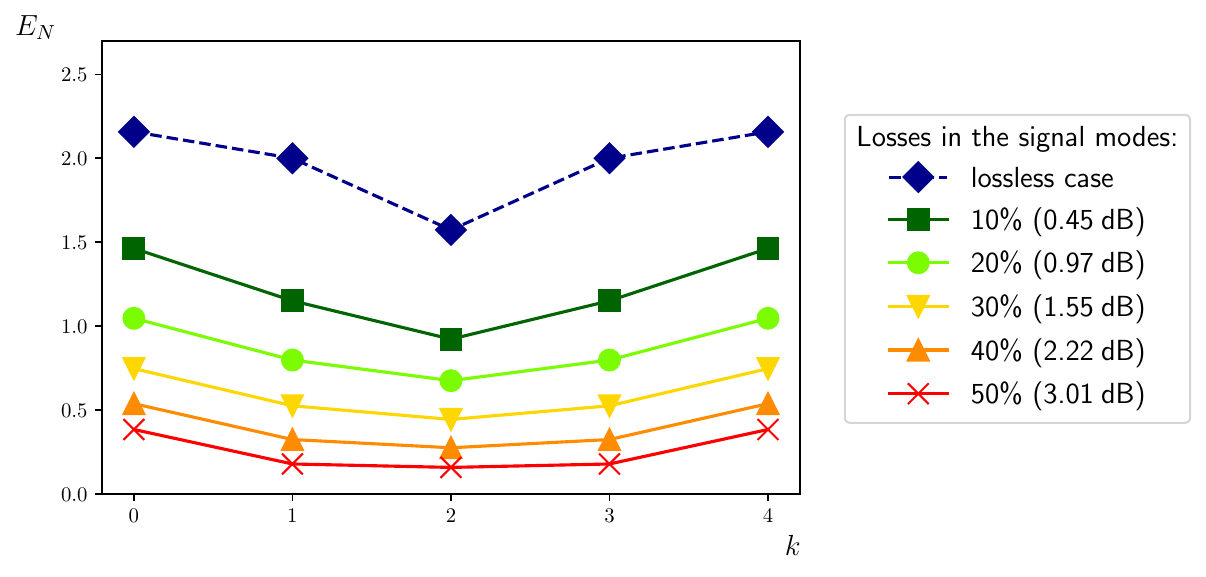}
	\caption{(Color online) Logarithmic negativity for the most general case which assumes losses in all modes and in detection. The results were computed for $g=0.1$ and $\sigma=4$ assuming
		unbalanced losses in modes $a_2$ and $b_2$ (37 and 40\kern.25emdB,
		respectively). Losses in signal modes $a_1$ and $b_1$ vary between $0$ and
		$50\%$. Losses at the detectors are set to $50\%$.}
	\label{fig:log_neg_all}
\end{figure}

Finally, we include computations for losses present in all modes and detectors, and asymmetry in idler losses. We have set the losses in the idler modes $a_2$ and $b_2$ to 37 and 40\kern.25emdB, the detector losses to 50\%, while changing losses in signal modes $a_1$ and $b_1$ between $0$ and $50\%$. The results are depicted in Fig.~\ref{fig:log_neg_all}, and we see that there is still a high amount of entanglement, even in these unfavourable conditions.

\section{Losses varying with time}
\label{app:disc3}

As has been discussed in ref.~\cite{Jennewein_NJP_2013}, in Earth-space scenarios, losses in each mode will vary over time on the timescale of 10-100\kern0.25em ms for losses coming from atmospheric turbulence and 0.1-1\kern0.25em s for losses deriving from jitter in the telescopes. A single pulse travelling towards a satellite on a low-Earth orbit will take approximately 3\kern0.25em ms to reach the satellite. Therefore any fluctuations will only affect the time-averaged logarithmic negativity Alice and Bob compute for all successful events.

\begin{figure}[ht]
	\centering
	\includegraphics[width=0.8\columnwidth]{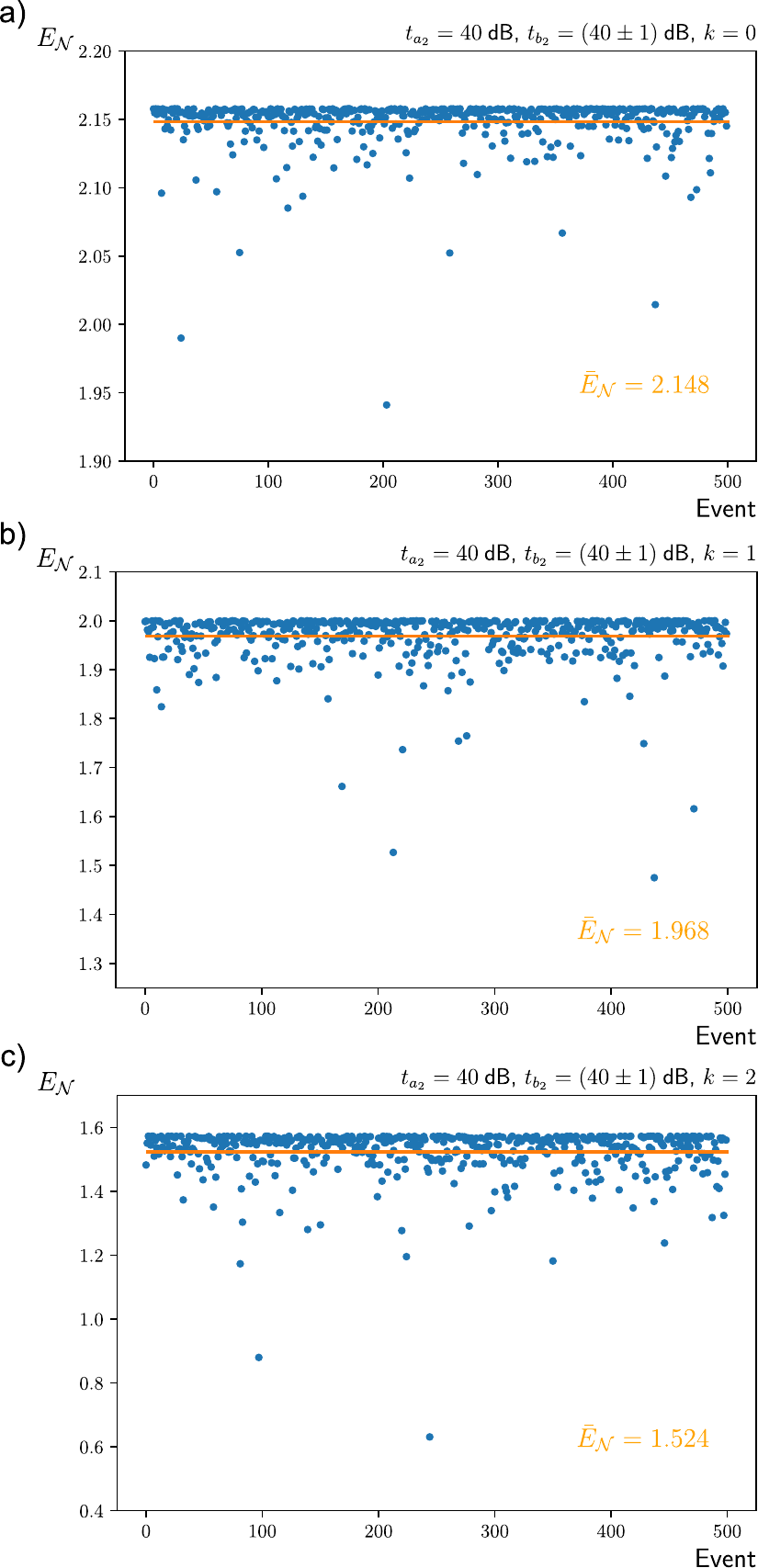}
	\caption{(Color online) Effect of fluctuations on the logarithmic negativity. The results were obtained by randomly generating 500 values of logarithmic negativity for the state
		$\rho_\textrm{out}^{(k,4)}$. We assumed 40\kern0.25emdB
		attenuation and variance 1\kern0.25emdB. a)~$k=0$, b) $k=1$, and c) $k=2$.}
	\label{fig:simulation_time}
\end{figure}

Let us assume that Bob's modes's transmittance is fixed and the Alice's has a normal distribution with variance 1\kern0.25emdB. We then generate 500 random values from that distribution and calculate the resulting logarithmic negativity. We repeat the process for different values of transmittance $t_{b_2}$. An example is shown in Fig.~\ref{fig:simulation_time} for $\sigma=4$ and typical attenuation of 40\kern0.25em dB. 

\section{Efficiencies and success rates}
\label{app:disc4}

\subsection{Ideal case}

\begin{figure}[ht]\centering
	\includegraphics[height=5cm]{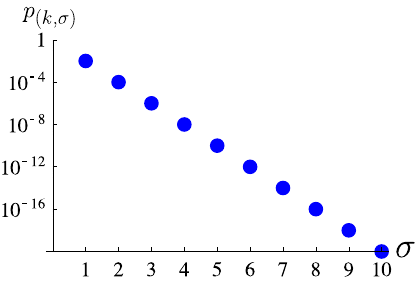}\hskip1cm
	\caption{(Color online) Probability to generate a specific state, $p_{(k,\sigma)}$ as a function of $\sigma$. The probability was computed for $g=0.1$ and no losses.}
	\label{fig:eff}	
\end{figure}

Interference of modes $a_2$ and $b_2$ on the BS alters the input state $\ket{\Psi_{\text{in}}}=\ket{\Psi}^{\otimes2}$ in the following way
\begin{align}
	\ket{\Psi_{\text{BS}}}&{} =
	\sum_{n,m=0}^{\infty} \sqrt{\lambda_n\,\lambda_m}
	\ket{n,m}_{a_1,b_1}
	U_{\text{BS}}\ket{n,m}_{a_2,b_2},
	\\
	\rho_{\text{BS}}&{} = \ket{\Psi_{\text{BS}}} \bra{\Psi_{\text{BS}}}\nonumber\\
	&{}=
	\sum_{n,m,p,q=0}^{\infty}\begin{aligned}[t]&\sqrt{\lambda_n\,\lambda_m\,\lambda_p\,\lambda_q}
	\ket{n,m}\bra{p,q}_{a_1,b_1}\otimes{}\\&\qquad\otimes	U_{\text{BS}}\ket{n,m}\bra{p,q}_{a_2,b_2}U_{\text{BS}}^\dagger.
	\end{aligned}
\end{align}
After tracing out modes $a_1$ and $b_1$ we obtain 
\begin{align}
	\rho_{a_2,b_2}&{}=\Tr_{a_1,b_1}\{\rho_{\text{out}}\}
	\nonumber\\&{}=
	\sum_{n,m=0}^{\infty} \lambda_n\,\lambda_m
	U_{\text{BS}}\ket{n,m}\bra{n,m}_{a_2,b_2}U_{\text{BS}}^\dagger.
\end{align}
Thus, the probability of detecting $\ket{k,\sigma-k}$ behind the BS, generating the state $\rho_\textrm{out}^{(k,\sigma)}$, equals
\begin{align}
	p_{(k,\sigma)}&{}=\Tr\left\{\ket{k,\sigma-k}\bra{k,\sigma-k}\rho_{a_2,b_2}\right\}
	\\
	&{}=
	\sum_{n,m=0}^{\infty}\begin{aligned}[t]&\lambda_n\,\lambda_m
	\bra{k,\sigma-k}U_{\text{BS}}\ket{n,m}\times{}\\
	&\qquad\times\bra{n,m}_{a_2,b_2}U_{\text{BS}}^\dagger\ket{k,\sigma-k}.
	\end{aligned}
\end{align}
Since the beam splitter interaction is particle number conserving, $\sigma=n+m$ must hold true. In addition, we note that $\lambda_a\lambda_b=\frac{\lambda_{a+b}}{\cosh^2 g}$ and therefore, 
\begin{align}
	p_{(k,\sigma)}
	&{}=
	\sum_{n=0}^{\infty} \lambda_n\,\lambda_{\sigma-n} \bigl\lvert\mathcal{A}_\sigma(k,n)\bigr\rvert^2
	\\
	&{}=
	\frac{\lambda_\sigma}{\cosh^2 g}\sum_{n=0}^{\infty} \bigl\lvert\mathcal{A}_\sigma(k,n)\bigr\rvert^2
	\\
	&{}=
	\frac{\lambda_\sigma}{\cosh^2 g}.
	\label{efficiency}
\end{align}
This probability depends solely on the parametric gain $g$ and $\sigma$. Since there are $\sigma+1$ values of $k$, the probability to generate any $\sigma$ photon state is $(\sigma+1)p_{(k,\sigma)}$. Fig.~\ref{fig:eff} shows $p_{(k,\sigma)}$ for $g=0.1$.

The success rate of obtaining a particular total photon number $\sigma$ is then calculated from multiplying $(\sigma+1) p_{(k,\sigma)}$ and the repetition rate of the pump, $f_{\mathrm{rep}}$. For example, assuming $g=0.1$ and $f_{\mathrm{rep}} = 80$\kern0.25emMHz, we would observe events having $\sigma=4$ photons in total with frequency 3.82\kern0.25emHz and events having $\sigma=2$ with frequency 23.2\kern0.25emkHz.   

\subsection{Symmetric idler losses}

Including idler losses, the probability $p_{(k,\sigma)}$ may be calculated similarly by
\begin{equation}
	p_{(k,\sigma)} = \Tr_{a_1,b_1}\{ \bra{k,\sigma-k}_{a_2,b_2} U_{\text{BS}} \rho_{\text{in}} U_{\text{BS}}^\dagger \ket{k,\sigma-k}_{a_2,b_2} \}
\end{equation}
where $\rho_{\text{in}}$ is given by Eq.~(\ref{rho_in}) and $U_{\text{BS}}$ is understood to act on the modes $a_2, b_2$. The state we are tracing over has already been calculated in Appendix~\ref{app:disc1}, for the case of symmetric losses $r_a = r_b = r$. The trace should then be given by the normalization constant $\frac{1}{\tilde{\mathcal{N}}^2}$, but we subsequently absorbed a factor $\frac{(1-r)^\sigma}{\cosh^2(g)}$ into $\tilde{\mathcal{N}}^2$. The probability is then
\begin{equation}
	p_{(k,\sigma)} = \frac{(1-r)^\sigma}{\cosh^2 g}\cdot\frac{1}{\tilde{\mathcal{N}}^2} = \frac{\lambda_\sigma}{\cosh^2 g} \cdot \frac{(1-r)^\sigma}{(1-r\tanh^2 g)^{\sigma+2}},
	\label{efficiencyfullapp}
\end{equation}
which correctly reduces to Eq.~(\ref{efficiency}) in the lossless case $r=0$.

The protocol efficiency is proportional to the probability of receiving at least one photon at the remote station, as then we are certain that Alice and Bob share entangled states. The complement of that event is that no photons reach the satellite, equivalent to $\sigma=0$. Setting $\sigma = 0$ in Eq.~(\ref{efficiencyfullapp}) we find
\begin{align}
	p_{(0,0)} &= \frac{1}{\cosh^4 g} \cdot \frac{1}{\big(1-r \tanh^2 g\big)^2}\nonumber\\
	&= \left(\frac{1-\tanh^2 g}{1-r \tanh^2 g}\right)^2.
	\label{efficiency2}
\end{align}
The equivalent expression in the case of asymmetric losses is calculated as follows
\begin{align}
	p_{(0,0)}&{}= \Tr_{a_1,b_1} \big\{\bra{0,0}_{a_2,b_2} \rho_{\mathrm{in}} \ket{0,0}_{a_2,b_2}\big\}\nonumber\\  
	&{}= \sum_{n,m=0}^{\infty} \lambda_n\lambda_m r_a^n r_b^m\nonumber\\
	&{}= \sum_{n=0}^{\infty} \lambda_n r_a^n \sum_{m=0}^{\infty} \lambda_m r_b^m\nonumber\\
	&{}= \frac{1}{\cosh^4 g} \cdot \frac{1}{1-r_a \tanh^2 g} \cdot \frac{1}{1-r_b \tanh^2 g}.
\end{align}
The overall success rate is then $f_{\mathrm{rep}}(1-p_{(0,0)})$. For a typical parametric gain ($g=0.1$) and Earth-space losses ($r_a=r_b=40$\kern0.25emdB) we obtain $1-p_{(0,0)}=2 \times 10^{-6}$. Assuming a pulse repetition rate of $80$\kern0.25emMHz, our success rate will be $160$\kern0.25emHz, i.e. 160 entangled states generated per second.

\section{Loophole-free Bell test and DI-QKD}
\label{app:QKD}

\subsection{Loophole-free Bell test}
The entanglement of Alice and Bob's state $\rho^{(k,\sigma)}_\textrm{out}$ can be verified by a Bell test which was first introduced for generic two-mode states of light in ref. \cite{Banaszek1999}. Alice and Bob interfere their individual modes with coherent states $\ket{\alpha}$ and $\ket{\beta}$, on beam splitters of reflectivities $r_a$ and $r_b$ (Alice and Bob's idler losses are now assumed equal and given by $r$). These operations can be approximated by displacement operations $D(-\delta_\alpha)$ and $D(-\delta_\beta)$, where $\delta_\alpha = i \alpha \sqrt{r_a}$, $\delta_\beta = i \beta \sqrt{r_b}$, which become exact in the limit $r_a, r_b \to 0$ and $|\alpha|^2, |\beta|^2 \to \infty$ such that $\delta_\alpha$ and $\delta_\beta$ are finite. They then measure their transmitted photon numbers $i$ and $j$, separating their results into two categories (dichotomized variables). This can be done by assigning $+1/-1$ either to vacuum/non-vacuum events, or to even/odd photon numbers. By choosing randomly between two local measurement settings each, $\delta_{\alpha 1}, \delta_{\alpha 2}$ and $\delta_{\beta 1}, \delta_{\beta 2}$, they can observe violation of the CHSH inequality $|B| \leq 2$ where $B = E(\delta_{\alpha 1}, \delta_{\beta 1}) + E(\delta_{\alpha 1}, \delta_{\beta 2}) + E(\delta_{\alpha 2}, \delta_{\beta 1}) - E(\delta_{\alpha 2}, \delta_{\beta 2})$, and $E(\delta_\alpha, \delta_\beta)$ is the correlation between their dichotomized variables.

First let us consider the zero/non-zero case. It is convenient to write $B$ in the Clauser--Horne form
\begin{equation}
	B = 2 + 4 \Big[\begin{aligned}[t]
		&{}+p(i=0,j=0|\delta_{\alpha1},\delta_{\beta1}) +{}\\
		&{}+p(i=0,j=0|\delta_{\alpha1},\delta_{\beta2}) +{}\\ &{}+p(i=0,j=0|\delta_{\alpha2},\delta_{\beta1}) +{}\\
		&{}-p(i=0,j=0|\delta_{\alpha2},\delta_{\beta2}) +{}\\
		&{}- p(i=0|\delta_{\alpha1}) - p(j=0|\delta_{\beta1})\Big].
		\end{aligned}
		\label{B}
\end{equation}
Here $p(i=0, j=0|\delta_\alpha, \delta_\beta)$ are the probabilities that they both detect zero photons, given measurement settings $\delta_\alpha$, $\delta_\beta$. The use of TES detectors (transition edge sensors) as PNRs make the detection of vacuum events much easier by significantly reducing the dark count rate. Alternatively, swapping the labelling of the outputs has no effect on $B$, so an equivalent form may be obtained by replacing $i = 0, j = 0$ with $i \neq 0, j \neq 0$ throughout this equation, which may be a more practical form for experiments. In the case of perfect detectors, the probability $p(i=0,j=0|\delta_{\alpha},\delta_{\beta})$ is equal to the Husimi Q function of the state
\begin{align}
	&p(i=0,j=0|\delta_{\alpha},\delta_{\beta})\nonumber\\
	&\qquad{}= \bra{0}_a\bra{0}_b D(-\delta_\alpha)D(-\delta_\beta) \rho^{(k,\sigma)}_\textrm{out}\otimes\nonumber\\
	&\qquad\qquad\otimes D^\dagger(-\delta_\alpha) D^\dagger(-\delta_\beta) \ket{0}_a \ket{0}_b\nonumber\\
	&\qquad{}= \bra{\delta_\alpha}_a\bra{\delta_\beta}_b \rho^{(k,\sigma)}_\textrm{out} \ket{\delta_\alpha}_a\ket{\delta_\beta}_b\nonumber\\
	&\qquad{}= Q(\delta_\alpha, \delta_\beta),
	\label{Q2}
\end{align}
where following \cite{Banaszek1999} we have dropped unnecessary factors of $\pi$ from the usual definition. Similarly, the marginal probabilities $p(i=0|\delta_{\alpha})$ and $p(j=0|\delta_{\beta})$ are given by the marginal Q functions
\begin{align}
	p(i=0|\delta_\alpha) &= \sum_j p(i=0, j|\delta_{\alpha},\delta_{\beta})\nonumber\\
	&= \sum_j \bra{0}_a\bra{j}_b D(-\delta_\alpha) \rho^{(k,\sigma)}_\textrm{out}  D^\dagger(-\delta_\alpha)\ket{0}_a\ket{j}_b\nonumber\\
	&= \bra{\delta_\alpha}_a \textrm{Tr}_b(\rho^{(k,\sigma)}_\textrm{out}) \ket{\delta_\alpha}_a \nonumber\\
	&= Q_a(\delta_\alpha),
	\label{Q}
\end{align}
and similarly for Bob, where $\textrm{Tr}_b$ indicates a partial trace over Bob's modes. With zero idler losses, the Q functions of the state $\ket{\Psi^{(k,\sigma)}_\textrm{out}}$ are
\begin{align}
	Q(\delta_\alpha, \delta_\beta) &{}= \begin{aligned}[t]&\frac{e^{-(|\delta_\alpha|^2+|\delta_\beta|^2)}}{\sigma !}\times{}\\&\qquad\times \left|\sum_{n=0}^\sigma (\delta_\alpha^*)^n (\delta_\beta^*)^{\sigma-n} \sqrt{\binom{\sigma}{n}} \mathcal{A}_\sigma(k,n) \right|^2,\end{aligned}
	\\
	Q_a(\delta_\alpha) &{}= e^{-|\delta_\alpha|^2} \sum_{n=0}^\sigma \frac{|\delta_\alpha|^{2n}}{n!}|\mathcal{A}_\sigma(k,n)|^2.
\end{align}
These forms are extremely useful for finding the optimal measurement settings that maximize the Bell inequality violation.

\begin{table*}[h]
	\begin{center}
		\begin{tabular}{ |c|c||c|c|c|c|c| } 
			\hline
			$k$ & $\sigma$ & variables & $[\delta_{\alpha 1}, \delta_{\alpha 2}], [\delta_{\beta 1}, \delta_{\beta 2}]$ & $|B|$ & $|B|$ (40\kern0.25emdB) & Alice/Bob detector efficiency (40\kern0.25emdB)\\
			\hline
			0 & 1 & zero/non-zero & [0.17, -0.56], [0.17$i$, -0.56$i$] & 2.69 & 2.63 & $\geq85$\%\\ 
			1 & 1 & zero/non-zero & [0.17, -0.56], [-0.17$i$, 0.56$i$] & 2.69 & 2.63 & $\geq85$\%\\
			\hline
			0 & 2 & even/odd & [0.06, -0.28], [0.06$i$, -0.28$i$] & 2.35 & 2.17 & $\geq98$\% \\
			1 & 2 & zero/non-zero & [0.00, 0.63], [0.00, 0.63] & 2.07 & 2.03 & $\geq96$\% \\
			2 & 2 & even/odd & [0.06, -0.28], [-0.06$i$, 0.28$i$] & 2.35 & 2.17 & $\geq98$\% \\
			\hline
			0 & 3 & even/odd & [0.06, -0.24], [0.06$i$, -0.24$i$] & 2.40 & 2.16 & $\geq99$\% \\
			1 & 3 & even/odd & [0.02, -0.22], [0.02$i$, -0.22$i$] & 2.10 & --- & ---\\
			2 & 3 & even/odd & [0.02, -0.22], [-0.02$i$, 0.22$i$] & 2.10 & --- & ---\\
			3 & 3 & even/odd & [0.06, -0.24], [-0.06$i$, 0.24$i$] & 2.40 & 2.16 & $\geq99$\%\\
			\hline
		\end{tabular}
		\caption{Optimal settings for the Bell test for each state $\ket{\Psi^{(k,\sigma)}_\textrm{out}}$. $|B|$ is the value of the Bell expression for zero idler losses. $|B|$ (40\kern0.25emdB) is the value for 40\kern0.25emdB idler losses (80\kern0.25emdB losses in the Alice-Bob channel). A value of $|B| > 2$ shows non-locality, with a blank entry signifying that there is no observed non-locality ($|B| \leq 2$). The detector efficiency shown is that required when there are 40\kern0.25emdB idler losses.}
		\label{bell_table}
	\end{center}
\end{table*}

Similarly, in the even/odd test we can express the Bell expression in terms of the Wigner function. The correlation may be written as an expectation value $E(\delta_\alpha, \delta_\beta) = \textrm{Tr}\Big(\rho^{(k,\sigma)}_\textrm{out} \hat{W}(\delta_\alpha, \delta_\beta)\Big)$ of the operator
\begin{equation}
	\hat{W}(\delta_\alpha, \delta_\beta) =  \big[\hat{\Pi}^{(+)}_a(\delta_\alpha) - \hat{\Pi}^{(-)}_a(\delta_\alpha)\Big] \otimes \Big[\hat{\Pi}^{(+)}_b(\delta_\beta) - \hat{\Pi}^{(-)}_b(\delta_\beta)\big],
\end{equation}
where $\hat{\Pi}^{(+)}_a$ and $\hat{\Pi}^{(-)}_a$ are measurement operators for Alice to measure even and odd photon numbers respectively
\begin{align}
	\hat{\Pi}^{(+)}_a(\delta_\alpha) &= \sum_m D(\delta_\alpha)\ket{2m}_a\bra{2m}_a D(\delta_\alpha)^\dagger,\\
	\hat{\Pi}^{(-)}_a(\delta_\alpha) &= \sum_m D(\delta_\alpha)\ket{2m+1}_a\bra{2m+1}_a D(\delta_\alpha)^\dagger,
\end{align}
and similarly for Bob. Using the parity operator
\begin{equation}
	(-1)^{a^\dagger a} = \sum_m \big(\ket{2m}\bra{2m} - \ket{2m+1}\bra{2m+1}\big),
\end{equation}
$\hat{W}(\delta_\alpha, \delta_\beta)$ may be simplified to
\begin{equation}
	\hat{W}(\delta_\alpha, \delta_\beta) = D(\delta_\alpha) D(\delta_\beta) (-1)^{a^\dagger a} (-1)^{b^\dagger b}  D(\delta_\alpha)^\dagger D(\delta_\beta)^\dagger.
\end{equation}
Thus $E(\delta_\alpha, \delta_\beta)$ is the expectation value of this displaced parity operator, which is one of the equivalent definitions of the Wigner function \cite{Banaszek1999}. To make this clear we write $E(\delta_\alpha, \delta_\beta) = W(\delta_\alpha, \delta_\beta)$, which again is true only for perfect photon detection. With zero idler losses, the Wigner function of the state $\ket{\Psi^{(k,\sigma)}_\textrm{out}}$ is
\begin{equation}
	W(\delta_\alpha, \delta_\beta) =
	\begin{aligned}[t]
		&(-1)^\sigma \exp[-2(|\delta_\alpha|^2+|\delta_\beta|^2)]\times{}\\
		&\qquad\times L_k[2(|\delta_\alpha|^2+|\delta_\beta|^2) - 4\textrm{Im}\{\delta_\alpha^* \delta_\beta\}]\times{}\\
		&\qquad\times L_{\sigma-k}[2(|\delta_\alpha|^2+|\delta_\beta|^2) + 4\textrm{Im}\{\delta_\alpha^* \delta_\beta\}],
		\end{aligned}
\end{equation}
where $L_k$ are Laguerre polynomials. This may be calculated by noting that $\ket{\Psi_{\text{out}}^{(k,\sigma)}} = U_\textrm{BS} \ket{k}\ket{\sigma-k}$ (this follows from the symmetry $A_\sigma(n,k) = A_\sigma(k,n)$), and evolving the Wigner function for the two-mode Fock state.

For each state we have numerically maximized the Bell inequality with respect to the four measurement settings $[\delta_{\alpha 1}, \delta_{\alpha 2}], [\delta_{\beta 1}, \delta_{\beta 2}]$. Table II lists the optimal choice of dichotomized variables and settings for each state, and the resulting value of $|B|$. We have also seen how this value is affected by idler losses. Since the entanglement decreases and plateaus after some finite amount of losses (see Fig.~\ref{fig:success_logneg}), the amount of Bell violation displays the same behaviour. For definiteness we have investigated typical Earth-space idler losses of 40\kern0.25emdB (80\kern0.25emdB summed losses in the Alice-Bob channel), with $g = 0.1$, although due to the plateau effect these losses can be considered the limiting behaviour as $r\to \infty$. We have also determined the minimum efficiency of Alice and Bob's local photon detectors required to still observe some violation.

The $\sigma=1$ states have the most robust violation in the presence of losses, $|B| = 2.63$ and require detector efficiencies $> 85\%$. States equivalent to our $\sigma=1$ states in the lossless limit have been explored in ref. \cite{Banaszek1999}, although they used a less optimal choice of settings, obtaining $|B| \approx 2.4$. The higher states $\sigma \geq 2$ also violate these Bell inequalities, but the violation is fragile with respect to signal mode losses and detector efficiencies. In ideal conditions, the states $k=\frac{\sigma}{2}$ have the lowest amount of Bell violation due to the lower amount of entanglement in these states. Some states e.g. $(k=1, \sigma=3)$ and $(k=2, \sigma=3)$ violate inequalities for zero idler losses but fail to do so with arbitrarily high losses. For these specific examples, the maximum amount of losses is around 50\%, or 3\kern0.25emdB. It is expected that the states $\sigma \geq 2$ violate some higher dimensional Bell inequalities beyond the simple CHSH scenario, which will be more loss tolerant, but this is still an active area of research.

\subsection{Device-independent (DI)-QKD}

\begin{figure*}[htbp]\centering
	\includegraphics[width=15cm]{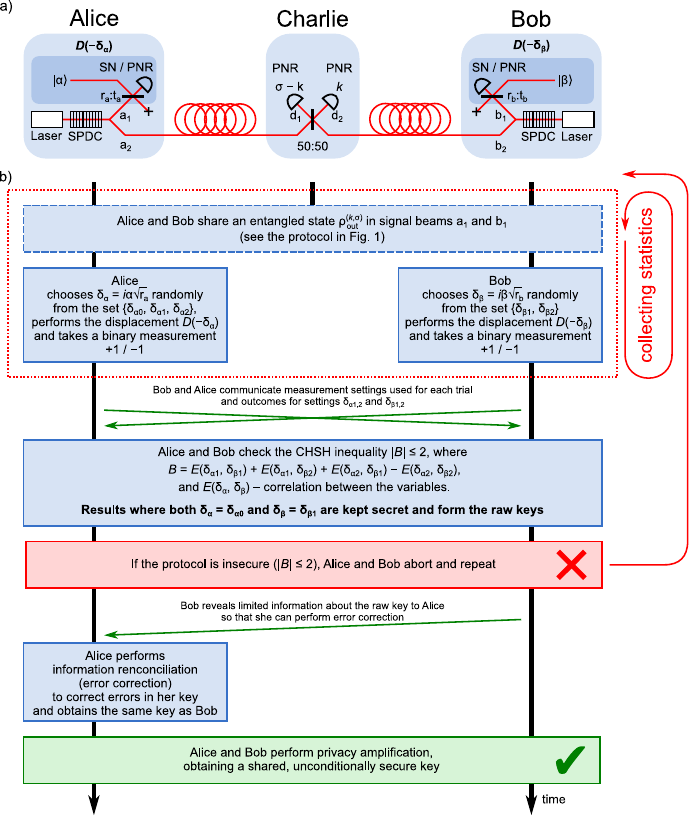}
	\caption{(Color online) Device-independent quantum key distribution (DI-QKD) protocol
			based on the distributed multiphoton bipartite entanglement. a)
		\textit{The setup.} Alice, Bob and Charlie utilize the setup presented in
		Figure~1a but with a few additions: Alice and Bob interfere their signal modes
		$a_1$ and $b_1$ with local oscillators $\ket{\alpha}$ and $\ket{\beta}$,
		respectively. To this end, they use variable beam splitters of ratio $r_a:t_a$
		and $r_b:t_b$. This setup allows them to perform a displacement operator
		$D(-\delta_{\alpha,\beta})$ on their modes, where
		$\delta_{\alpha}=i\alpha\sqrt{r_a}$ and $\delta_{\beta}=i\beta\sqrt{r_b}$. The
		readout obtained with either single-photon detection, such as superconducting
		nanowires (SN), or photon-number-resolved (PNR) detectors allows them to perform
		a Bell test and realize the QKD protocol. b) \textit{A sequence diagram
			of the protocol.} Alice repeatedly chooses $\delta_{\alpha}$ from a set
		$\{\delta_{\alpha0},\delta_{\alpha1},\delta_{\alpha2}\}$ while Bob uses
		$\delta_{\beta}$ from $\{\delta_{\beta1},\delta_{\beta2}\}$. They take binary
		measurements and exchange the values for $\{\delta_{\alpha1},\delta_{\alpha2}\}$
		and $\{\delta_{\beta1},\delta_{\beta2}\}$. Next, they check the CHSH inequality
		$\lvert B\rvert\leq 2$ computed with the readouts. If this inequality is
		fulfilled, then the protocol is insecure -- they should abort and repeat.
		Otherwise, Bob reveals a limited information about his raw key to Alice, she
		performs an information reconciliation (error correction) and, after a privacy
		amplification, they share an unconditionally secure secret key.}
	\label{fig:QKD}
\end{figure*}

If Alice and Bob share entangled states and can violate some Bell inequalities, it is possible for them to extract a shared cryptographic key. If the Bell test is loophole-free, they can place a limit on the amount of information obtained by an eavesdropper Eve and ensure that their key is secure, immune to all possible side channels and eavesdropping strategies. This is called device-independent quantum key distribution (DI-QKD). We follow closely the formulation presented in ref. \cite{Pironio2009}, where we can think of Charlie sending Alice and Bob the state $\rho_\textrm{out}^{(k,\sigma)}$ depending on their measurement, $\ket{k, \sigma-k}$ at the remote station. The protocol is secure even if Charlie is replaced by the adversarial Eve.

As discussed in the main text, we focus on long-distance QKD and thus $\sigma = 1$ states, since they display success rates scaling with transmittance as $O(\sqrt{\eta})$, and are more tolerant to signal mode losses and detector inefficiencies. In the long distance limit, these states form the vast majority of generated states, e.g. for our typical Earth-space parameters with 40\kern0.25emdB losses and $g = 0.1$, $p_{(k, 1)}/(1-p_{(0, 0)}) = 99.9999\%$. In the absence of losses these states are exactly the maximally entangled single-photon states $\frac{1}{\sqrt{2}}(\ket{0,1}_{a_1, b_1} \mp i\ket{1, 0}_{a_1, b_1})$. These have been considered in a single-photon entanglement DI-QKD scheme described in ref. \cite{Kamaruddin2015}, the protocol developed here can be considered equivalent, but using the realistic state $\rho_\textrm{out}^{(k,\sigma)}$ rather than the idealized states, and with improved parameters and key rates.

The setup of the QKD protocol and a sequence diagram are displayed in Fig.~\ref{fig:QKD}. Upon receipt of the entangled state $\rho^{(0,1)}_\textrm{out}$ or $\rho^{(1,1)}_\textrm{out}$, Alice and Bob each perform a measurement as described in the previous section, by performing displacement operations with parameters $\delta_\alpha$ and $\delta_\beta$ respectively, and measuring the output photon numbers $i$ and $j$. Alice randomly chooses her setting $\delta_\alpha$ from three possible choices $[\delta_{\alpha 0}, \delta_{\alpha 1}, \delta_{\alpha 2}]$ while Bob chooses randomly from two choices $[\delta_{\beta 1}, \delta_{\beta 2}]$. They repeat this process a number of times and then communicate publicly which settings they chose for each trial.

The pair $\{\delta_{\alpha 0}$, $\delta_{\beta 1}\}$ are chosen such that $i$ and $j$ are highly correlated for these measurements, and these results are kept secret and used to generate the raw key. The combination of settings $\{\delta_{\alpha 1}, \delta_{\alpha 2}, \delta_{\beta 1}, \delta_{\beta 2}\}$ are chosen such that the measurements violate the CHSH inequality. Alice and Bob communicate through an authenticated channel and reveal the results $i$ and $j$ for all combinations of these settings and ensure that the inequality Eq.~(\ref{B}) is violated. This ensures that these measurement results are random i.e. that a potential eavesdropper Eve cannot predict the results with certainty. The amount of randomness increases with increasing Bell violation, with Eve's probability of guessing the result bounded by $p_\textrm{guess} \leq \frac{1}{2}(1+\sqrt{2-B^2/4})$ \cite{Brunner2014}. Since $p(j|\delta_{\beta 1})$ doesn't depend on Alice's setting by the non-signalling principle, Eve also cannot have complete information about the results for $\{\delta_{\alpha 0}$, $\delta_{\beta 1}\}$ or the raw key generated by them.

Alice and Bob now both have raw keys of $n$ bits, but they may not be perfectly correlated, and Eve may have a small amount of information about them. They thus carry out a process of classical information reconciliation (error correction) and privacy amplification, discarding a number of bits to obtain the final shared key of $\ell$ bits. The raw key rate is defined as the limiting ratio
\begin{equation}
	K = \lim_{n\to\infty} \frac{\ell}{n},
\end{equation}
and depends on the type of attacks we consider from Eve. Eve is assumed to control the source (for us, she is in control of the remote station measurement), and also to have fabricated Alice and Bob's measurement devices. Following \cite{Pironio2009}, we consider collective attacks whereby Eve interacts a fresh system with Alice and Bob's idler modes each time they reach the remote station. She applies the same unitary operation each time, and can perform a coherent measurement on all her systems at any time. The security may be improved to protect against completely general/coherent attacks by using the results from refs. \cite{Vazirani2014, Miller2016}. However this modification will not significantly reduce the key rate, which is dominated by the efficiency of the entanglement distribution, which remains unchanged.

The security proof presented in ref. \cite{Pironio2009} applies provided that the measurements are described by Hermitian operators with eigenvalues $\pm 1$. This is true in our case as we now prove. The event where Alice measures zero transmitted photons after interfering with the coherent state $\ket{\alpha}$ is assigned the value $+1$ and has probability
\begin{align}
	p(+1) &= \bra{\delta_\alpha} \rho_a \ket{\delta_\alpha},\nonumber\\
	&= \Tr_a(\ket{\delta_\alpha} \bra{\delta_\alpha} \rho_a),
\end{align}
where $\rho_a = \Tr_b (\rho_\textrm{out}^{(k,\sigma)})$ is Alice's reduced density matrix. Thus $\ket{\delta_\alpha} \bra{\delta_\alpha}$ is a projection operator (and POVM element) for the outcome $+1$, and the corresponding operator for outcome $-1$ is $1 - \ket{\delta_\alpha} \bra{\delta_\alpha}$. The measurement can then be described by the Hermitian operator
\begin{align}
	A(\delta_\alpha) &= +\ket{\delta_\alpha} \bra{\delta_\alpha} - (1 - \ket{\delta_\alpha} \bra{\delta_\alpha}),\nonumber\\
	&= 2\ket{\delta_\alpha} \bra{\delta_\alpha} - 1,
\end{align}
and the security proof in ref. \cite{Pironio2009} then directly follows.

The raw key rate for collective attacks may be calculated by the Devetak--Winter formula \cite{devetakwinter}
\begin{equation}
	K = I(A_0:B_1) - \chi(B_1:E),
	\label{dwrate}
\end{equation}
where $I(A_0:B_1)$ is the mutual information between Alice and Bob when they use settings $\{\delta_{\alpha 0}, \delta_{\beta 1}\}$, and $\chi(B_1:E)$ is the Holevo quantity which upper-bounds the mutual information between Bob and Eve. This equation assumes the classical communication is performed one way from Bob to Alice. This is because $\chi(A_0:E) > \chi(B_1:E)$ holds for this protocol, i.e. Eve has more information about Alice's results than Bob's, the asymmetry arising from the fact Alice uses three settings while Bob uses only two. The mutual information $\chi(B_1:E)$ is upper-bounded \cite{Pironio2009} by
\begin{equation}
	\chi(B_1:E) \leq h\left(\frac{1+\sqrt{(B/2)^2-1}}{2}\right),
	\label{eq:chi}
\end{equation}
where $h(p)$ is the Binary entropy function. Thus we have a lower bound for the raw key rate
\begin{equation}
	K \geq I(A_0:B_1) - h\left(\frac{1+\sqrt{(B/2)^2-1}}{2}\right),
	\label{K}
\end{equation}
and all that is left to do is calculate this quantity for our protocol and maximize over the five measurement settings.

The mutual information is
\begin{equation}
	I(A_0:B_1) = H(A_0) + H(B_1) - H(A_0, B_1),
\end{equation}
where $H(A_0)$ and $H(B_1)$ are the entropies of Alice and Bob's marginal probabilities
\begin{align}
	H(A_0) &= h[p(i=0|\delta_{\alpha 0})],\\
	H(B_1) &= h[p(j=0|\delta_{\beta 1})],
\end{align}
and $H(A_0, B_1)$ is the entropy of the joint probabilities
\begin{align}
	&H(A_0, B_1) ={}\nonumber\\
	&\qquad-p(i=0,j=0|\delta_{\alpha 0},\delta_{\beta 1}) \log_2[p(i=0,j=0|\delta_{\alpha 0},\delta_{\beta 1})]\nonumber\\
	&\qquad-p(i=0,j\neq 0|\delta_{\alpha 0},\delta_{\beta 1}) \log_2[p(i=0,j\neq 0|\delta_{\alpha 0},\delta_{\beta 1})]\nonumber\\
	&\qquad-p(i\neq 0,j=0|\delta_{\alpha 0},\delta_{\beta 1}) \log_2[p(i\neq 0,j=0|\delta_{\alpha 0},\delta_{\beta 1})]\nonumber\\
	&\qquad-p(i\neq 0,j\neq 0|\delta_{\alpha 0},\delta_{\beta 1}) \log_2[p(i\neq 0,j\neq 0|\delta_{\alpha 0},\delta_{\beta 1})].
\end{align}
In the case of perfect photon detectors, these can be conveniently described in terms of the $Q$ functions using Eqs. (\ref{Q2}) and (\ref{Q}), and writing
\begin{align}
	p(i=0,j=0) &{}= Q(\delta_{\alpha 0}, \delta_{\beta 1}),\\
	p(i=0,j\neq 0) &{}= p(i=0) - p(i=0, j=0) \nonumber\\
	&{}= Q_a(\delta_{\alpha 0}) - Q(\delta_{\alpha 0}, \delta_{\beta 1}),\\
	p(i\neq 0,j= 0) &{}= p(j=0) - p(i=0, j=0) \nonumber\\
	&{}= Q_b(\delta_{\beta 1}) - Q(\delta_{\alpha 0}, \delta_{\beta 1}),\\
	p(i\neq 0,j\neq 0) &{}= 1 + Q(\delta_{\alpha 0}, \delta_{\beta 1}) - Q_a(\delta_{\alpha 0}) - Q_b(\delta_{\beta 1}).
\end{align}
The Bell expression $B$ is given by Eq.~(\ref{B}) as before, which can similarly be expressed in terms of the $Q$ functions.

One can now simply perform a numerical maximization of $K$ as a function of our five measurement settings, but we will take a more instructive approach. As before, we will calculate the optimal parameters for the lossless case and then consider the effect of losses after. The parameters $\{\delta_{\alpha 1}, \delta_{\alpha 2}, \delta_{\beta 1}, \delta_{\beta 2}\}$ must be chosen to maximize the Bell parameter $|B|$ so as to decrease Eve's knowledge about the key $\chi(B_1:E)$. At the same time $\{\delta_{\alpha 0}$, $\delta_{\beta 1}\}$ must be chosen to maximize the correlations between Alice and Bob's key bits $I(A_0: B_1)$. These cannot be done independently since $\delta_{\beta 1}$ appears in both, but we can get a very good approximation by first maximizing $B$ to find $\delta_{\beta 1}$ and then choosing $\delta_{\alpha 0}$. The parameters $\{\delta_{\alpha 1}, \delta_{\alpha 2}, \delta_{\beta 1}, \delta_{\beta 2}\}$ have thus already been calculated and given in Table \ref{bell_table}.

There are two ways for the results for $\{\delta_{\alpha 0}$, $\delta_{\beta 1}\}$ to be perfectly correlated. Either $p(i=0,j=0) + p(i\neq 0, j\neq 0) = 1$, so that Alice and Bob always have equal results, or $p(i=0,j=0) + p(i\neq 0, j\neq 0) = 0$, so that they always have opposite results. Let's first look at the state $(k= 0 , \sigma = 1)$, the probabilities are given by our Q function expressions
\begin{align}
	&p(i=0,j=0|\delta_{\alpha 0}, \delta_{\beta 1}) = \frac{e^{-(|\delta_{\alpha 0}|^2+|\delta_{\beta 1}|^2)}}{2}\times\nonumber\\
	&\qquad\times\Big[|\delta_{\alpha 0}|^2 + |\delta_{\beta 1}|^2 + 2 \textrm{Im}\{\delta_{\alpha 0}^* \delta_{\beta 1}\}\Big],\\
	&p(i\neq0,j\neq0|\delta_{\alpha 0}, \delta_{\beta 1}) = 1 + \frac{e^{-(|\delta_{\alpha 0}|^2+|\delta_{\beta 1}|^2)}}{2}\times\nonumber\\
	&\qquad\times\Big[|\delta_{\alpha 0}|^2 + |\delta_{\beta 1}|^2 + 2 \textrm{Im}\{\delta_{\alpha 0}^* \delta_{\beta 1}\}\Big]+\nonumber\\
	&\qquad{}- \frac{e^{-|\delta_{\alpha 0}|^2}}{2}(1+|\delta_{\alpha 0}|^2) - \frac{e^{-|\delta_{\beta 1}|^2}}{2}(1+|\delta_{\beta 1}|^2).
\end{align}
Thanks to the interference term, the first of these can be set to zero by setting $\delta_{\alpha 0} = i \delta_{\beta 1}$, which also reduces the second probability, simplifying to
\begin{align}
	p(i=0,j=0|\delta_{\alpha 0}, \delta_{\beta 1}) &= 0,\\
	p(i\neq0,j\neq0|\delta_{\alpha 0}, \delta_{\beta 1}) &= 1 - e^{-|\delta_{\beta 1}|^2}(1+|\delta_{\beta 1}|^2) \approx |\delta_{\beta 1}|^4,
\end{align}
where we have expanded around small $|\delta_{\beta 1}|$. Reading from Table \ref{bell_table} we have $\delta_{\beta 1} = 0.17i$, so we should choose $\delta_{\alpha 0} = -0.17$ and thus we find $p(i=0,j=0) + p(i\neq 0, j\neq 0) = 4\times 10^{-4}$. Thus Alice and Bob's results are almost perfectly correlated, if Alice measures $i\neq 0$, Bob always measures $j = 0$ and vice-versa, one party just needs to swap their results to obtain the same key bit. We thus define the quantum bit error rate $q$ as
\begin{equation}
	q = p(i=0,j=0|\delta_{\alpha 0}, \delta_{\beta 1}) + p(i\neq 0, j\neq 0|\delta_{\alpha 0}, \delta_{\beta 1}),
\end{equation}
which will become non-zero once we include losses. Interestingly, Alice and Bob's marginal probabilities are also nearly perfectly balanced between the two outcomes
\begin{align}
	p(i=0|\delta_{\alpha 0}) &{}= p(j=0|\delta_{\beta 1})\nonumber\\
	&{}= \frac{e^{-|\delta_{\beta 1}|^2}}{2}(1+|\delta_{\beta 1}|^2) \approx \frac{1}{2}(1-|\delta_{\beta 1}|^4),
\end{align}
so that they have near maximum entropy $H(A_0) = H(B_1) = 1 - 10^{-7}$. The joint entropy and mutual information are $H(A_0, B_1) = 1.005$ and $I(A_0:B_1) = 0.995$. Inserting our value for the Bell expression $B = 2.688$ into Eq.~(\ref{eq:chi}) we find the Holevo bound $\chi(B_1:E) \leq 0.290$ and thus raw key rate for the lossless case
\begin{equation}
	K \geq 0.705.
\end{equation}
The key rate per pulse $R$ (number of secure key bits per second divided by the repetition rate of the pump $f_\textrm{rep}$) is found by multiplying the dimensionless key rate $K$ by the number of raw key bits generated per pulse. The probabilities of using the settings $\{\delta_{\alpha 1}, \delta_{\alpha 2}, \delta_{\beta 2}\}$ can be made very small such that the majority of Alice and Bob's measurements create raw key bits using the settings $\{\delta_{\alpha 0}$, $\delta_{\beta 1}\}$. This leaves a small fraction of other measurements, but still enough to reliably calculate the Bell inequality. Thus the number of raw key bits generated per pulse is approximately equal to the protocol efficiency $(1 - p_{(0, 0)})$. Thus $R$ is given by
\begin{equation}
	R = K \left(1 - \left[\frac{1-\tanh^2(g)}{1-r\tanh^2(g)}\right]^2 \right),
\end{equation}
with $K$ given by Eq.~(\ref{K}), and we have inserted $p_{(0,0)}$ from Eq.~(\ref{efficiency2}). In the lossless case we obtain
\begin{equation}
	R \geq 1.39\times 10^{-2}.
\end{equation}
The calculation for the state $(k=1,\sigma=1)$ proceeds in the same way except that we take $\delta_{\alpha 0} = -i \delta_{\beta 1}$. The results are summarized in Table \ref{qkd_table}.

\begin{table*}[h]
	\begin{center}
		\begin{tabular}{ |c|c||c|c|c|c|c| } 
			\hline
			$k$ & $\sigma$ & $[\delta_{\alpha 0}, \delta_{\alpha 1}, \delta_{\alpha 2}], [\delta_{\beta 1}, \delta_{\beta 2}]$ & $|B|$ & Bit error rate, $q$ & Raw key rate, $K$ & Key rate per pulse, $R$\\
			\hline
			0 & 1 & [-0.17, 0.17, -0.56], [0.17$i$, -0.56$i$] & 2.688 & $4\times 10^{-4}$ & $\geq 0.556$ & $\geq 1.39\times 10^{-2}$\\ 
			1 & 1 & [-0.17, 0.17, -0.56], [-0.17$i$, 0.56$i$] & 2.688 & $4\times 10^{-4}$ & $\geq 0.556$ & $\geq 1.39\times 10^{-2}$\\
			\hline
		\end{tabular}
		\caption{Optimal settings for the QKD protocol in the lossless case.}
		\label{qkd_table}
	\end{center}
\end{table*}
\begin{table*}[h]
	\begin{center}
		\begin{tabular}{ |c|c||c|c|c|c|c| } 
			\hline
			$k$ & $\sigma$ & $[\delta_{\alpha 0}, \delta_{\alpha 1}, \delta_{\alpha 2}], [\delta_{\beta 1}, \delta_{\beta 2}]$ & $|B|$ & Bit error rate, $q$ & Raw key rate, $K$ & Key rate per pulse, $R$\\
			\hline
			0 & 1 & [-0.17, 0.17, -0.56], [0.17$i$, -0.56$i$] & 2.635 & $0.01$ & $\geq 0.705$ & $\geq 1.12\times 10^{-6}$\\ 
			1 & 1 & [-0.17, 0.17, -0.56], [-0.17$i$, 0.56$i$] & 2.635 & $0.01$ & $\geq 0.705$ & $\geq 1.12\times 10^{-6}$\\
			\hline
		\end{tabular}
		\caption{Optimal settings for the QKD protocol in the case of 40\kern0.25emdB idler losses (80\kern0.25emdB losses in the Alice-Bob channel).}
		\label{qkd_table2}
	\end{center}
\end{table*}

We will see that the raw key rate $K$ is not significantly affected by idler losses, but the true rate $R$ decreases with the same scaling as the protocol efficiency, $O(\sqrt{\eta}) = O(1-r)$. Although we have calculated the optimal parameters $[\delta_{\alpha 0}, \delta_{\alpha 1}, \delta_{\alpha 2}], [\delta_{\beta 1}, \delta_{\beta 2}]$ assuming the lossless scenario, we find that they remain very close to optimal when losses are included.

Inserting idler losses of 40 dB, Alice and Bob's marginals are still nearly perfectly balanced $H(A_0) = H(B_1) = 1 - 10^{-4}$, while the joint entropy and mutual information are only slightly modified $H(A_0, B_1) = 1.073$, $I(A_0: B_1) = 0.927$. This corresponds to a quantum bit error rate of $q = 0.01$. We already calculated $|B|$ for 40 dB losses and found $|B| = 2.635$, which yields a Holevo bound $\chi(B_1:E) \leq 0.371$. Thus for 40 dB idler losses and perfect photon detectors we have a raw key rate
\begin{equation}
	K \geq 0.556,
\end{equation}
and multiplying by the protocol efficiency, $1-p_{(0,0)} = 2\times10^{-6}$, the key rate per pulse is
\begin{equation}
	R \geq 1.12 \times 10^{-6}.
\end{equation}
The results are summarized in Table \ref{qkd_table2}.

\begin{figure}[hbp]\centering
	\includegraphics[width=\columnwidth]{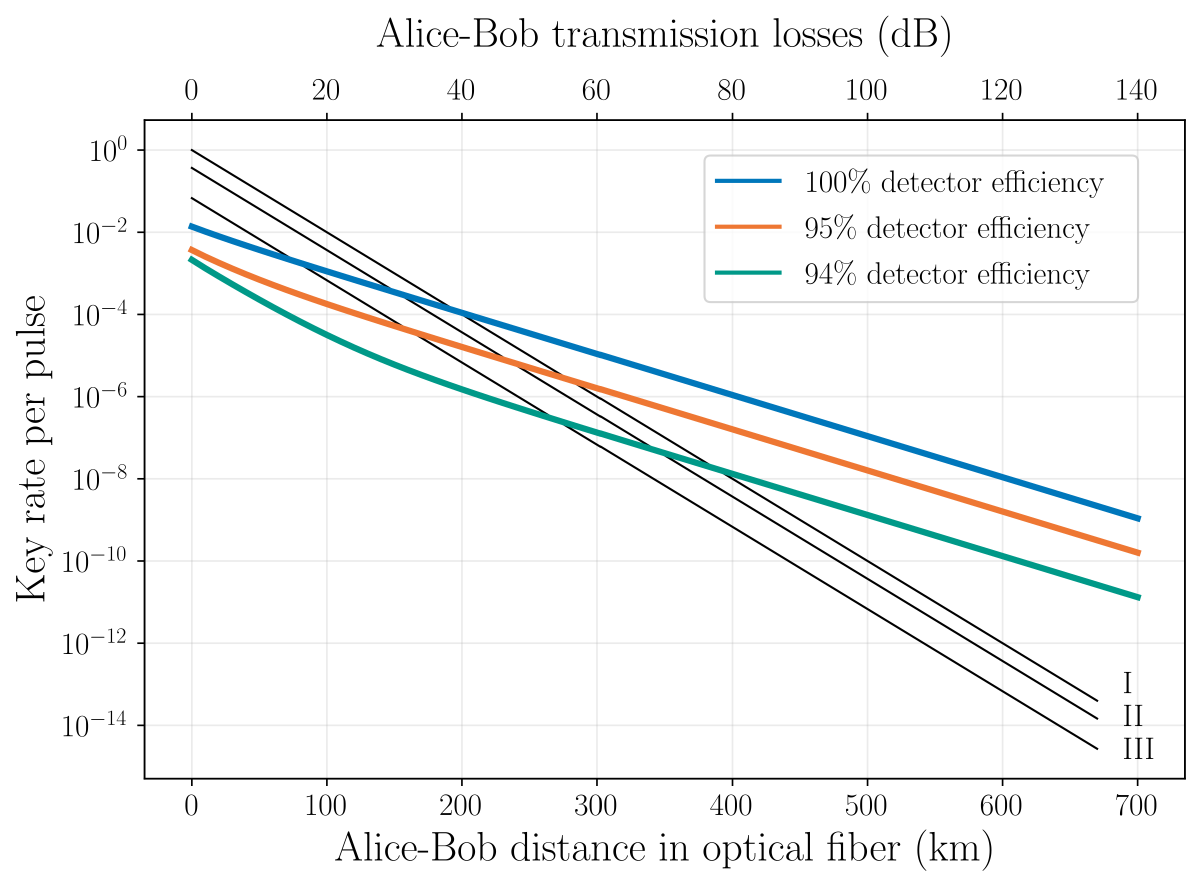}
	\caption{(Color online) Key rate of our DI-QKD protocol as a function of transmission losses
			and detector efficiencies. Lines I-III are three QKD protocols with key rates
		scaling as $O(\eta)$. I: Ideal BB84, $R=\eta$, II: Decoy state QKD,
		$R=\frac{\eta}{e}$, III: Decoy-state measurement-device-independent (MDI)-QKD,
		$R = \frac{\eta}{2e^2}$~(32).}
	\label{fig:key_rate_eff}
\end{figure}

So far we have assumed perfect efficiency of Alice and Bob's local detectors. Non-zero detector losses reduce $K$ by the reduction of the Bell violation (increasing $\chi(B_1:E)$), and by an increase of the quantum bit error rate $q$ (decreasing $I(A_0:B_1)$). Assuming 95\% efficiency, we obtain marginal entropies $H(A_0) = H(B_1) = 0.999$, joint entropy $H(A_0,B_1) = 1.290$ and mutual information $I(A_0:B_1) = 0.708$, corresponding to a quantum bit error rate $q = 0.055$. The Bell violation remains large $|B| = 2.430$, yielding $\chi(B_1:E) \leq 0.622$. Thus we find, in the presence of 40 dB idler losses (80 dB Alice-Bob losses) and 95\% detector efficiency
\begin{align}
	K &\geq 0.086,\\
	R &\geq 1.72\times 10^{-7}.
\end{align}
Fig.~\ref{fig:key_rate_eff} shows how the key rate decreases as a function of transmission losses and detector efficiencies.

\section{Bell test with post-selection and SEPM-QKD}
\label{app:SEPM-QKD}

\subsection{Bell test with post-selection}

If high detection efficiencies are not available for Alice and Bob's local detectors, a different Bell test may be carried out using post-selection \cite{Li2019}. The setup is the same as before (Fig.~\ref{fig:key_rate_eff}a), with Alice and Bob interfering with coherent states but they now measure both the transmitted and reflected photon numbers with detectors $A_1, A_2$ and $B_1, B_2$. We again focus on the $\sigma = 1$ states due to the increased efficiency of their generation.

After measuring $\sigma = 1$ at the remote station, Charlie has effectively distributed the following state to Alice and Bob
\begin{equation}
	\ket{\Psi_{\text{out}}^{(k,1)}} = \frac{1}{\sqrt{2}}(\ket{0,1}_{a,b} + e^{i\theta}\ket{1,0}_{a,b}),
\end{equation}
where $\theta = \mp \pi/2$ for $k=0/1$. In our analytical results we will use this state generated with zero idler losses, but in our numerical results we will use the full state $\rho_\textrm{out}^{(k,1)}$, with the understanding that in the limit $g^2 \ll 1$ they become identical. Alice interferes her mode with a coherent state $\ket{\alpha}$, where $\alpha = |\alpha|e^{i\theta_a}$, on a 50/50 beam splitter $U_A$. Bob similarly interferes his mode with a coherent state $\ket{\beta}$, where $\beta = |\beta|e^{i\theta_b}$ on a 50/50 beam splitter $U_B$. After the interactions they then share the state
\begin{equation}
	\ket{\chi} = U_A U_B \ket{\Psi_{\text{out}}^{(k,1)}} \ket{\alpha}\ket{\beta}.
\end{equation}
They each assign a result $+1$ to events where photons are detected exclusively in the transmitted channel (detector $A_1$/$B_1$ clicks for Alice/Bob respectively), and a result $-1$ to events where photons are detected exclusively in the reflected channel (detector $A_2$/$B_2$ clicks). They ignore outcomes where neither or both detectors click i.e., they apply the fair-sampling assumption in the usual way.

Let's consider the event where $A_1$ and $B_1$ click. The primary contribution to the probability $p(A_1,B_1)$ is from single photon events, and assuming $|\alpha| = |\beta|$ this probability takes the simple form
\begin{equation}
	\big\lvert\bra{1,0}_a \bra{1,0}_b \ket{\chi} \big\rvert^2 = \frac{e^{-2|\alpha|^2} |\alpha|^2}{4}[1+\cos(\theta_a-\theta_b-\theta)].
\end{equation}
If Alice and Bob used PNR detectors they could post-select on these single photon outcomes and this would be the total probability $p(A_1,B_1)$. If however they use non-photon-number-resolving detectors, such as superconducting nanowires, there will be extra contributions from multi-photon events. Retaining terms up to $|\alpha|^4$ the extra contributions come from
\begin{align}
	\big\lvert\bra{2,0}_a \bra{1,0}_b \ket{\chi} \big\rvert^2 &{}= \big\lvert\bra{1,0}_a \bra{2,0}_b \ket{\chi}\big\rvert^2\nonumber\\&{}= \frac{e^{-2|\alpha|^2}|\alpha|^4}{8}\left[\frac{5}{4}+\cos(\theta_a-\theta_b-\theta)\right],
\end{align}
so that the total probability is
\begin{equation}
	p(A_1, B_1) = \begin{aligned}[t]&\frac{e^{-2|\alpha|^2}}{4}(|\alpha|^2+|\alpha|^4)[1+\cos(\theta_a-\theta_b-\theta)]+{}\\&\qquad + \frac{e^{-2|\alpha|^2} |\alpha|^4}{16}.\end{aligned}
\end{equation}

Similar calculations follow for the other three events $(A_1, B_2)$, $(A_2, B_1)$ and $(A_2, B_2)$. In general, the probability that detectors $A_i$ and $B_j$ click is
\begin{equation}
	p(A_i, B_j) =\begin{aligned}[t]& \frac{e^{-2|\alpha|^2}}{4}(|\alpha|^2+|\alpha|^4)\times{}\\
		&\qquad\times[1+(-1)^{i+j}\cos(\theta_a-\theta_b-\theta)] +{}\\&\qquad+ \frac{e^{-2|\alpha|^2} |\alpha|^4}{16}.\end{aligned}
	\label{paibj_noloss}
\end{equation}
The term proportional to $[1+(-1)^{i+j}\cos(\theta_a-\theta_b-\theta)]$ allows for perfect correlations while the remaining term introduces a small error due to multi-photon events.

We now consider the effect of losses. In contrast to ref \cite{Li2019}, our states $\rho_\textrm{out}^{(k,\sigma)}$ are heralded at Alice and Bob via a pre-selection measurement rather than being sent directly through free space. As we've seen, this means the idler losses have a negligible effect on the state so that $\rho_\textrm{out}^{(k,\sigma)} \approx \ket{\Psi_{\text{out}}^{(k,\sigma)}}\bra{\Psi_{\text{out}}^{(k,\sigma)}}$, which comes at the cost of reduced efficiency of the protocol. In contrast, signal mode losses $r_s$, including detector inefficiencies of Alice and Bob's detectors, lower the amount of entanglement. With non-perfect signal mode transmittance $\eta_s = 1-r_s$, the state reaching the local beam splitters is then
\begin{equation}
	\rho = \eta_s \ket{\Psi_{\text{out}}^{(k,1)}}\bra{\Psi_{\text{out}}^{(k,1)}} + (1-\eta_s)\ket{0,0}_{a,b}\bra{0,0}_{a,b}.
\end{equation}
Adding the extra contribution from the vacuum interfering with the coherent states
\begin{equation}
	\big\lvert\bra{1,0}_a \bra{1,0}_b U_A U_B \ket{0,0}_{a,b}\ket{\alpha}\ket{\beta} \big\rvert^2 = \frac{e^{-2|\alpha|^2} |\alpha|^2}{4},
\end{equation}
Eq.~(\ref{paibj_noloss}) is modified to
\begin{equation}
	p(A_i, B_j) = \begin{aligned}[t]
		&\frac{\eta_s e^{-2|\alpha|^2}}{4}(|\alpha|^2+|\alpha|^4)\times{}\\
		&\qquad{}\times\left[1+(-1)^{i+j}\cos(\theta_a-\theta_b-\theta)\right] +{}\\
		&\qquad{}+ \frac{e^{-2|\alpha|^2} |\alpha|^4}{4}\left(1-\frac{3}{4}\eta_s\right).
		\end{aligned}
	\label{paibj}
\end{equation}
A slightly different result is obtained in ref. \cite{Li2019} due to them truncating the coherent states at the single photon level, but this doesn't affect their conclusions.

The expectation value of the product of their outcomes can then be calculated
\begin{align}
	E(\theta_a, \theta_b) &= \frac{p(A_1,B_1) - p(A_1,B_2) - p(A_2,B_1) + p(A_2,B_2)}{p(A_1,B_1) + p(A_1,B_2) + p(A_2,B_1) + p(A_2,B_2)}\nonumber\\ &= \left[\frac{\eta_s (1+|\alpha|^2)}{|\alpha|^2 + \eta_s +\frac{|\alpha|^2 \eta_s}{4}}\right] \cos(\theta_a - \theta_b - \theta),
\end{align}
and using the measurement settings $\theta_a = \{0, \frac{\pi}{2}\}$ and $\theta_b = \{\frac{\pi}{4}-\theta, -\frac{\pi}{4}-\theta\}$, one can evaluate the CHSH expression
\begin{align}
	B &= E(\theta_{a1},\theta_{b1}) + E(\theta_{a1},\theta_{b2}) + E(\theta_{a2},\theta_{b1}) - E(\theta_{a2},\theta_{b2})\nonumber\\
	&= \left[\frac{\eta_s (1+|\alpha|^2)}{|\alpha|^2 + \eta_s +\frac{|\alpha|^2 \eta_s}{4}}\right] 2\sqrt{2}.
\end{align}
In the limit $|\alpha|^2 \to 0$ this becomes a maximal violation of $B=2\sqrt{2}$, however the combination of multi-photon events and non-perfect transmittance lowers the violation. The transmittance required to observe $|B| > 2$ is
\begin{equation}
	\eta_s > \frac{|\alpha|^2}{\sqrt{2} - 1 + |\alpha|^2(\sqrt{2}-\frac{1}{4})},
\end{equation}
which for e.g. $|\alpha| = 0.35$, is $\eta_s > 22\%$, much better than the $85\%$ efficiency required in the loophole-free case. This of course comes at the cost of having to post-select on a small amount of events. The fraction of events that are kept is given by
\begin{equation}
	\sum_{ij=1}^2 p(A_i,B_j) = e^{-2|\alpha|^2} |\alpha|^2 \left(|\alpha|^2 + \eta_s + \frac{|\alpha|^2\eta_s}{4}\right),
\end{equation}
equal to $4\%$ for $|\alpha|=0.35$ and $\eta_s = 30\%$. Note that this probability does not take into account the probability of generating the starting state $\ket{\Psi_{\text{out}}^{(k,1)}}$, which is approximately equal to the entanglement distribution protocol efficiency $1-p_{(0,0)}$. This is considered a pre-selection rather than a post-selection, since if Charlie measures $\sigma = 0$ and communicates this, Alice and Bob don't share an entangled state and don't perform any measurements.

\subsection{Single-photon entanglement-based phase-matching (SEPM)-QKD}

The above Bell inequality is no longer loophole-free so we do not have DI levels of security, but it can nevertheless be used to limit the information leaked to Eve in a MDI-QKD protocol described in ref. \cite{Li2019}. This protocol is called single-photon entanglement-based phase-matching (SEPM)-QKD. Applying this protocol to our case requires slight modifications due to the states $\rho_\textrm{out}^{(k,\sigma)}$ being heralded at Alice and Bob via a pre-selection measurement rather than being sent directly through free space.

The coherent state phases are expressed as $\theta_a = \phi_a + k_a \pi$ and $\theta_b = -\theta + \phi_b + k_b\pi$. The phases $\phi_{a(b)}$ are both randomly chosen from the set $\{-\frac{\pi}{4}, 0, \frac{\pi}{4}, \frac{\pi}{2}\}$ while $k_{a(b)}$ are random bits $\{0, 1\}$. The settings used above for the Bell test are contained within this set of measurements, but there are also additional events where the phases $\phi_a$ and $\phi_b$ match. The shared key bits are extracted from these events. Setting $\phi_a = \phi_b$ one has $\cos(\theta_a-\theta_b-\theta) = (-1)^{k_a + k_b}$, so that if Alice and Bob's key bits are equal $k_a = k_b$, their measurement probabilities $p(A_i,B_j)$ are
\begin{align}
	p(A_1,B_1) &{}= p(A_2,B_2)\nonumber\\
	&{}= \frac{\eta_s e^{-2|\alpha|^2}}{2}(|\alpha|^2+|\alpha|^4) + \frac{e^{-2|\alpha|^2} |\alpha|^4}{4}\left(1-\frac{3}{4}\eta_s\right)
	\nonumber\\
	&\approx \sum_{ij}p(A_i,B_j),\\
	p(A_1,B_2) &{}= p(A_2,B_1)\nonumber\\
	&{}= \frac{e^{-2|\alpha|^2} |\alpha|^4}{4}\left(1-\frac{3}{4}\eta_s\right)\nonumber\\
	&{}\approx 0.
\end{align}
Similarly, if their key bits are different $k_a \neq k_b$ we have the opposite scenario
\begin{align}
	p(A_1,B_1) &{}= p(A_2,B_2)\nonumber\\
	&{}= \frac{e^{-2|\alpha|^2} |\alpha|^4}{4}\left(1-\frac{3}{4}\eta_s\right)\nonumber\\
	&{} \approx 0, \\
	p(A_1,B_2) &{}= p(A_2,B_1)\nonumber\\
	&{}= \frac{\eta_s e^{-2|\alpha|^2}}{2}(|\alpha|^2+|\alpha|^4) + \frac{e^{-2|\alpha|^2} |\alpha|^4}{4}\left(1-\frac{3}{4}\eta_s\right)\nonumber\\
	&{}\approx \sum_{ij}p(A_i,B_j).
\end{align}
After the measurements are complete they both publish the phases $\phi_{a(b)}$ and measurement results $i,j$ in each trial but keep the key bits $k_{a(b)}$ secret. From the above equations we see that for events where they obtain the same outcome $i = j$, their key bits are the same with high probability, and for events where they obtain different outcomes $i \neq j$, their key bits are different with high probability, so Bob simply flips his key. The error rate is small and given by (e.g. for $k_a = k_b$)
\begin{equation}
	q = \frac{p(A_1,B_2) + p(A_2,B_1)}{\sum_{i,j}p(A_i,B_j)} = \frac{|\alpha|^2(1-\frac{3}{4}\eta_s)}{2(|\alpha|^2 + \eta_s + \frac{|\alpha|^2 \eta_s}{4})}.
\end{equation}

From the published information, Eve can determine $|k_a - k_b|$ for each trial but cannot determine the exact values $k_a, k_b$. The raw key rate considering attacks by Eve is
\begin{equation}
	K = I(A:B) - \chi(AB:E).
\end{equation}
The mutual information between Alice and Bob is $I(A:B) = 1 - h(q)$ while the Holevo quantity $\chi(AB:E)$ is the maximum amount of information obtainable by Eve. In ref. \cite{Li2019} they determine the optimal collective attack by Eve and find that Eve's information gain is limited to twice the quantum bit error rate $\chi(AB:E) \leq 2q$. They also consider a beam splitting attack which must be treated separately to collective attacks, however we find that the information gain is negligible, being proportional to $|\alpha|^4$, whereas $q$ is proportional to $|\alpha|^2$.

The measurement results for the other combinations of settings may be used to evaluate the CHSH inequality and, since $B \approx (1-2q) 2\sqrt{2}$, the amount of violation can be used to estimate Eve's information gain and the quantum bit error rate. After performing error correction, they then have a shared secure key. Since the fair sampling approximation was used to evaluate the Bell inequality, the security is no longer device-independent. Alice and Bob must now trust their local sources and measurement apparatus, but still need not trust the measurement at the remote station performed by Charlie/Eve, placing the security within the framework of measurement-device-independent (MDI)-QKD.

The key rate per pulse is found by multiplying $K$ by the number of raw key bits generated per pulse. There is the probability of post-selection $\sum_{ij} p(A_i, B_j)$ but in our case there is also the probability of distributing the entangled state, i.e. pre-selection, $1-p_{(0,0)}$. We then obtain
\begin{equation}
	R = (1-f h(q)-2q) (1-p_{(0,0)}) \sum_{ij} p(A_i, B_j),
\end{equation}
where we have allowed for an inefficiency $f$, for the error correction. Due to the pre-selection, the term $\sum_{ij} p(A_i, B_j)$ depends only on the signal mode transmittance, with the idler mode transmittance affecting only the term $1-p_{(0,0)}$. This is a key benefit, since in ref. \cite{Li2019} they were restricted to the regime $|\alpha|^2 < 1-r$ where the key rates were much lower than in state-of-the-art protocols. Our pre-selection means we instead have the less strict limit $|\alpha|^2 < \eta_s$, allowing higher values of $|\alpha|$ and higher key rates.

We have numerically simulated this SEPM-QKD protocol and compared it to the case of realistic TF-QKD, with the results being presented in Fig.~\ref{fig:keyrate}. The simulation parameters were chosen to match those in ref. \cite{Lucamarini2018}: 30\% detector efficiencies at Alice and Bob's local detectors and an error correction coefficient $f = 1.15$. The coherent state amplitude was chosen as $|\alpha| = 0.35$, which we found to be close to optimal. For a more fair comparison we have also included a small probability of false detection events $p_\textrm{dark} = 10^{-8}$, which could be due to dark counts or background (stray) photons. The false detection rate per unit time is then $p_\textrm{dark} f_\textrm{rep}$ where $f_\textrm{rep}$ is the repetition rate of the pump. The main effect of these false detections is that sometimes Charlie incorrectly declares that the state $\rho_\textrm{out}^{(k,1)}$ has been distributed due to a false detection of $\sigma = 1$. If the protocol efficiency becomes comparable or lower than the false detection efficiency, $(1-p_{(0,0)}) < p_\textrm{dark}$, the state can then no longer be distributed and the key rate drops to zero. This is responsible for the maximum distance shown in Fig.~\ref{fig:keyrate} The other key rates plotted on this figure are obtained with equations presented in \cite{Lucamarini2018}. Lines I-III are three QKD protocols with key rates scaling as $O(\eta)$. I: Ideal BB84, $R=\eta$, II: Decoy state QKD, $R=\frac{\eta}{e}$, III: Decoy-state measurement-device-independent (MDI)-QKD, $R = \frac{\eta}{2e^2}$

Finally we note the similarity between this protocol and the ideal entanglement-based one presented in ref. \cite{Curty2019} (protocol one). There Alice and Bob each prepare a state $\ket{\Phi} = \sqrt{q} \ket{0,0}_{a1,a2} + \sqrt{1-q} \ket{1,1}_{a1,a2}$ where $0 \leq q \leq 1$ and send the idler beams to Charlie who interferes them on a beam splitter and measures the output photon numbers. Alice and Bob then share an entangled state, and perform measurements either in the $z$-basis, $\ket{0}, \ket{1}$, or the $x$-basis, $\ket{\pm} = (\ket{0} \pm \ket{1})/\sqrt{2}$. The QKD protocol presented here can be considered equivalent, with realistic interference measurements replacing the ideal $x$-basis qubit measurement, and the two-mode squeezed vacuum state approximating $\ket{\Phi}$.

\section{Software}
\label{app:software}

The software that we have developed and used was written in Python 3 using the NumPy package, which was employed for matrix algebra. The program performs operations described by the following algorithm
\begin{enumerate}
	\item for both SPDC sources belonging to Alice and Bob, compute $\rho_\Psi=\ket{\Psi}\bra{\Psi}$ where, $\ket{\Psi}=\sum_{n=0}^{n_{\text{max}}}\sqrt{\lambda_n}\,\lvert n,n\rangle$, and $n_{\text{max}}$ is a sum cutoff found by solving $\lambda_{n_{\text{max}}}/\lambda_0<10^{-15}$ for a given $g$ and $10^{-15}$ results from the precision of the numeric data type; usually $n_{\text{max}}\geq 6$.
	\item for all modes leaving the sources, add losses and then trace out unwanted modes, i.e.\ perform the following operation
	\begin{displaymath}
		\rho_{\Psi'} = \Tr_{3,4}\left\{U_{\text{BS}}^{(r_s)}\,U_{\text{BS}}^{(r_i)}\ket{\Psi}\bra{\Psi} \left(U_{\text{BS}}^{(r_i)}\right)^\dagger\left(U_{\text{BS}}^{(r_s)}\right)^\dagger  \right\},
	\end{displaymath}
	where $r_s$ and $r_i$ are losses in signal and idler modes, respectively,
	\item construct the density operator matrix for  $\rho_{\text{in}}=\rho_{\Psi'}\otimes\rho_{\Psi'}$,
	\item apply the $50:50$ BS operation, $\rho_{\text{BS}}= U_{\text{BS}}\,\rho_{\text{in}}\,\left(U_{\text{BS}}\right)^\dagger$,
	\item add losses $r_{d_1}$ and $r_{d_2}$ to the modes leaving the BS
	\begin{displaymath}
		\rho_{\text{out}} = \Tr_{5,6}\left\{U_{\text{BS}}^{(r_{d_1})}U_{\text{BS}}^{(r_{d_2})}\rho_\text{BS} \left(U_{\text{BS}}^{(r_{d_2})}\right)^\dagger\left(U_{\text{BS}}^{(r_{d_1})}\right)^\dagger\right\},
	\end{displaymath}
	\item for all $k=0,\dots,\sigma$ perform the following operations:
	\label{itm:final}
	\begin{enumerate}
		\item project $\rho_{\text{out}}$ onto $\ket{k,\sigma-k}$: $\rho_{\text{out}}^{(k,\sigma)}=\bra{k,\sigma-k}\rho_{\text{out}}\ket{k,\sigma-k}$ and renormalize the result,
		\item compute partial transposition of $\rho_{\text{out}}^{(k,\sigma)}$: $\left(\rho_{\text{out}}^{(k,\sigma)}\right)^\Gamma$,
		\item compute eigenvalues $\{\alpha_k\}$ of $\left(\rho_{\text{out}}^{(k,\sigma)}\right)^\Gamma$ using routines built into the NumPy package,
		\item compute the logarithmic negativity
		\begin{displaymath}
			E_{\mathcal{N}} = \log_2\left(1+2\sum_k\frac{\lvert \alpha_k\rvert-\alpha_k}{2}\right).
		\end{displaymath}
	\end{enumerate}
\end{enumerate}
The advantage of the above algorithm is relatively fast operation at a cost of huge memory requirements because of the large size of the matrices. In the worst case, $\rho_{\text{out}}$ contains $n_{\text{max}}^8$ double-precision values (ca.\ 13\kern.25emMB for $n_{\text{max}}=6$, 800\kern.25emMB for $n_{\text{max}}=10$) and therefore, computations for $n_{\text{max}}$ as large as 20 or 50 would be impossible. This has been solved by noticing that the matrix contains mostly zeros and that the sum of the number of photons in all modes cannot be simultaneously higher than $4n_{\text{max}}$, which allowed us to apply denser packing. Once the final density operator matrix is computed, subsequent computations (step~\ref{itm:final}) of the algorithm can be performed for all interesting values of $S$ and $k$.

\end{document}